\documentclass{jfm}
\usepackage{graphicx,pdflscape,rotating,psfrag}
\usepackage{epstopdf, epsfig, placeins,upgreek,url}
\usepackage{amsmath,mathtools,mathptmx,tikz}
\usepackage{amsfonts,oldgerm}
\usepackage{amssymb} 
\usepackage[mathlines]{lineno}
\usepackage{bm}
\usepackage{authblk}
\usepackage{tabularx}
\usepackage{float}
\usepackage{caption}
\usepackage{subcaption}

\ifCUPmtlplainloaded \else
\checkfont{msam10}
\iffontfound
\IfFileExists{amssymb.sty}
{\typeout{^^JFound AMS Symbol fonts on the system, using the
		'amssymb' package.^^J}%
	\usepackage{amssymb}%
	\let\le=\leqslant  \let\leq=\leqslant
	  
}{}
\fi
\fi

\ifCUPmtlplainloaded \else
\IfFileExists{amsbsy.sty}
{\typeout{^^JFound the 'amsbsy' package on the system, using it.^^J}%
	\usepackage{amsbsy}}
{\providecommand\boldsymbol[1]{\mbox{\boldmath $##1$}}}
\fi

\usepackage{caption}
\usepackage{subcaption}

\def\*#1{\mathbf{#1}}
\def\$#1{\hat{\mathbf{#1}}}
\def\9#1{{\boldsymbol{#1}}}
\def\##1{{\tilde{#1}}}
\def\%#1{{\hat{#1}}}
\shorttitle{The dipole--multipole transition in rapidly rotating dynamos}
\shortauthor{D. Majumder, B. Sreenivasan and G. Maurya}

\title{Self-similarity 
of the dipole--multipole transition in rapidly rotating
dynamos}

\author[1]{Debarshi Majumder}
\author[1]{Binod Sreenivasan
  \corresp{\email{bsreeni@iisc.ac.in}}}
\author[1]{Gaurav Maurya}

\affil[1]{Centre for Earth Sciences, 
Indian Institute of Science, Bangalore 560012, India}

\begin{document}
	\maketitle
\begin{abstract}
The dipole--multipole transition in rapidly rotating
dynamos is investigated through the analysis of
forced magnetohydrodynamic waves in an
unstably stratified fluid. The focus of this
study is on the inertia-free limit
applicable to planetary cores, where the
Rossby number is small not only on the core
depth but also on the length scale
of columnar convection. By progressively
increasing the buoyant forcing in a linear
magnetoconvection model, the slow
{{Magnetic-Archimedean-Coriolis}} (MAC) waves are
significantly attenuated so that their kinetic
helicity decreases to zero; the fast MAC wave helicity,
on the other hand, 
is practically unaffected. In turn, polarity reversals
in low-inertia spherical dynamos are
shown to occur when the
slow MAC waves disappear under strong forcing. Two
dynamically similar regimes are identified --
the suppression of slow waves in a strongly
forced dynamo and the excitation 
of slow waves in a moderately forced
dynamo starting from a small seed
field. While the former regime
results in polarity reversals, the latter regime produces
the axial dipole from a chaotic multipolar
state. For either polarity transition,
a local Rayleigh number based on the mean wavenumber
of the energy-containing scales bears the same linear
relationship with the square of the
peak magnetic field measured
at the transition. The self-similarity of
the dipole--multipole transition can place
a constraint on the Rayleigh number for
polarity reversals in the Earth.
\end{abstract}
\section{Introduction}
\label{intro}
The dynamo operating in the Earth's outer core
generates a predominantly North--South 
dipole magnetic field. Occasionally, 
the magnetic dipole axis flips its orientation
and retains its approximate alignment with the Earth's rotation
axis. The last such polarity reversal occurred
nearly 0.78 million years ago \citep{merrill2011}.
Geomagnetic excursions, 
the periods during which the magnetic axis 
wanders up to 45$^\circ$ 
from the rotation axis 
before returning to its original state, have
been more frequent in the Earth's past. 
As reversals and excursions are likely to
result from similar 
convective states of the core \citep{gubbins1999,valet2005},
{ it is possible that the geodynamo operates
in a strongly driven state below the
threshold for reversals.}

The first polarity reversals 
in numerical
dynamo models were obtained by
\citeauthor{glatz1995a} 
\citeyearpar{glatz1995a,glatz1995b}. Since then,
 several other
studies reported reversals in comparable parameter regimes
\citep{sarson1999,kutzner2002,wicht2004,olsdris2009,
sreeni2014},
and it is now 
well accepted that dynamo reversals occur
under strong buoyancy-driven convection.
Any explanation for polarity reversals
must follow from an explanation for
the preference for the axial dipole
in planetary dynamos.
The kinetic helicity $\bm{u}\cdot \bm{\zeta}$
(where $\bm{u}$ is the velocity and
$\bm{\zeta}$ is the vorticity) generated
in convection columns that arise in
rapid rotation is thought to
be essential for dipolar dynamo
action \citep{moffatt1978,olson1999numerical}. Consequently,
the loss of columnar helicity can lead
to collapse of the dipole \citep[e.g.][]{soderlund2012}.
A number of numerical dynamo models show a 
dipole--multipole transition for $Ro_\ell \approx 0.1$,
where $Ro_\ell$ is a \lq local' Rossby number that
measures the ratio of nonlinear
inertial to Coriolis
forces on the length scale of columnar
convection \citep{chraub2006}. { Since
it has been hypothesized that
the columnar vortices
in rotating turbulence are formed by
the propagation of linear
inertial waves} \citep{06davidstaple}, 
the above polarity 
transition may result from 
the suppression of
inertial waves in the dynamo above
a critical value of $Ro_\ell$ \citep{mcdermott2019}. 
Multipolar solutions may also be found
when the ratio of nonlinear inertia to Lorentz forces
exceeds a critical value of $O(1)$ 
{\citep{tassin2021,zaire2022}.}
That having been said,
in the rapidly rotating
limit of zero nonlinear inertia, the helicity
of columnar vortices aligned with the
rotation axis is considerably enhanced
by the magnetic field in the dynamo \citep{jfm11}. 
It is thought that this field-induced helicity
is essential for the formation of the
axial dipole in planetary dynamos.
Under strong buoyancy-driven convection,
the magnetically enhanced columnar flow
breaks down, likely causing
collapse of the axial dipole { \citep{jfm11}.} Given the
spatial inhomogeneity of the magnetic
field  \citep[e.g.][]{schaeff2017}, an analysis of
 volume-averaged
 forces in the dynamo cannot possibly reveal how
buoyancy offsets the
Lorentz--Coriolis force balance 
in isolated columns. 
The analysis of helical dynamo waves at
progressively increasing forcing can, on the 
other hand, provide an insight into the
role of buoyancy in dipole collapse. 
{ This study distinguishes
between polarity-reversing and multipolar
dynamos in the inertia-free limit,
where the Rossby
number is small not only on the depth of
the planetary core but also on the 
columnar length scale transverse to the
rotation axis. The analysis
of dynamo waves in this limit in turn places
 a constraint on the convective
state of the dynamo that admits polarity reversals. }

In convectively-driven 
planetary cores, isolated density disturbances
generate fast and slow 
Magnetic-Archimedean-Coriolis (MAC)
waves under the combined influence of background
rotation, magnetic field and unstable
stratification. The fast waves are inertial
waves weakly modified by the magnetic field
and buoyancy while the slow waves are
magnetostrophic waves produced by localized
balances between the magnetic, Coriolis and buoyancy
forces { \citep[pp.165--168]{brag1967,07bussechapter}. } 
MAC waves in the stably stratified regions of
the Earth's core, where gravity acts as a restoring force,
 have been an active
area of research { \citep[e.g.][]{nicolas2023} }.
The present study, on the other hand,
focuses on MAC waves in an unstably
stratified medium that supports 
convection in a planetary
dynamo.
 In a rapidly
rotating dynamo, the fast MAC waves of frequency 
$\sim \omega_C$, the frequency of linear inertial waves,
exist even in 
the strong field state where
the square of the scaled mean
magnetic field, $B^2/2 \varOmega \rho \mu \eta
= O(1) $, where $\varOmega$ is the angular velocity
 of rotation, $\rho$ is the fluid density, 
 $\eta$ is the magnetic diffusivity and $\mu$
is the magnetic permeability. The intensity
of slow MAC wave motions becomes comparable to that of
the fast waves when the ratio of Alfv\'en to
inertial wave frequencies,
\begin{equation}
\left(\frac{\omega_M}{\omega_C}\right)_{\!0} 
\sim \frac{V_M}{2 \varOmega \delta} \sim 10^{-2},
\label{leh1}
\end{equation}
where $V_M$ is the Alfv\'en wave velocity based on
the peak magnetic field, $\delta$
is the length scale of the buoyancy disturbance and the subscript
\lq 0' refers to the initial state of the disturbance as it is
released into the flow \citep{aditya2022}. Because of the
anisotropy of the convection, the instantaneous value
of $\omega_M/\omega_C$ for parity between fast and slow wave
motion would be higher than its initial value, and
inferred to be $\sim 0.1$ from the spherical shell dynamo models
of Varma \& Sreenivasan. In the energy-containing scales, given by the
range of spherical harmonic degrees lesser than the mean value at
energy injection, the 
 absolute kinetic helicity in the nonlinear dynamo
is nearly twice that in the equivalent 
nonmagnetic
convection problem. This
result implies that the helicity produced
by the slow MAC waves
in the nonlinear dynamo would be of the same magnitude as
that produced by the inertial waves  in nonmagnetic
convection. Further, we infer that 
the slow waves are essential
for dipole formation since the hydrodynamic dynamo at the
same parameters, where the Lorentz force is zero,
 does not generate the 
axial dipole \citep[see also][]{prf18}.
Although the magnetic diffusion frequency $\omega_\eta$
is the lowest frequency in the dynamo, 
small but finite magnetic diffusion can place a lower bound on
the length scale that supports slow MAC
waves in the energy-containing scales
of the dynamo. Linear magnetoconvection analysis of
a forced damped system \citep{jfm21} indicates that, for
\begin{equation}
\left(\frac{\omega_\eta}{\omega_C}\right)_{\!0}  
\sim \frac{\eta}{2 \varOmega \delta^2} 
\lesssim 10^{-5},
\end{equation}
the energy of the fast and
slow wave motions are comparable. (Here, $\omega_M/\omega_C \sim 0.1$).
Therefore, slow wave motions at length scales $ \delta \sim$ 10 km can
be influential in the generation of the dipole field. At
higher values of the ratio
 $(\omega_\eta/\omega_C)_0$, the slow wave energy falls below the
fast wave energy, implying that scales smaller than
$\sim$ 10 km are rapidly damped by magnetic diffusion.

The role of buoyancy in a rapidly rotating 
dynamo is not merely in the excitation
of MAC waves.
In a convective dynamo that evolves from a small
seed magnetic field, the slow MAC waves are
first excited when  $|\omega_M| > |\omega_A|$,
where $\omega_A$ is a measure of the strength of
buoyancy in an unstably stratified fluid
and has the same magnitude as that of the frequency
of internal gravity waves in a stably stratified fluid
\citep{aditya2022}. As the dynamo field intensity 
increases from
its starting seed value, $|\omega_M|$ progressively
increases and 
exceeds $|\omega_A|$, after which the axial dipole
eventually forms
from a chaotic multipolar state.
Thus, the inequality $|\omega_C| > |\omega_M| >|\omega_A|
> |\omega_\eta|$ represents a large region of the parameter
space where dipole-dominated dynamos exist.
If a rotating dynamo is subject to progressively increasing
buoyant forcing, the larger self-generated fields would result
in progressively larger $\omega_M$ until a state is
reached where $|\omega_A| \sim |\omega_M|$ as
the field attains its highest intensity for a given rotation
rate. Here, the slow MAC waves disappear, likely causing
collapse of the dipole. This transition from dipole
to polarity-reversing states is dynamically
similar to the transition
from a multipolar state to the dipole that occurs in the growth phase
of a dynamo starting from a seed field. Since $\omega_C$
remains the dominant frequency while forcing
is increased, the dynamo reverses polarity in a 
rotationally dominant regime where slow magnetostrophic
waves are suppressed. Although increased forcing 
may result in enhanced nonlinear inertia
in numerical dynamo models, we consider it 
unlikely that inertia has any role in polarity transitions 
in the Earth
even for the smallest buoyancy disturbances that
support MAC waves.
For disturbances of size $L_\perp \approx$ 15 km transverse
to the rotation axis, the actual ratio of nonlinear inertia
to Coriolis forces,
\begin{equation}
\dfrac{|\nabla \times (\bm{u} \cdot \nabla) \bm{u}|}
{|2 (\bm{\varOmega} \cdot \nabla) \bm{u}|} 
\sim \dfrac{u_\star L}{2 \varOmega \, L_\perp^2} \approx 0.03,
\end{equation}
taking $L=$ 2260 km and $u_\star= 5 \times 10^{-4}$ 
ms$^{-1}$ \citep{starjones2002}. Low-inertia numerical dynamos must have a small
Rossby number based on the length scale of convection.
The use of magnetic Prandtl number $Pm =\nu/\eta \sim$ 1--10,
where $\nu$ is the kinematic viscosity,
is useful in realizing strong magnetic fields in 
numerical simulations \citep{sreenivasan2006,07willis,teed2015,16dormy} 
in the inertia-less regime relevant to rotating 
planetary cores. Although the choice of a large $Pm$
at moderate Ekman number $E=\nu/2\varOmega L^2
\sim 10^{-4}$--$10^{-6}$
has the unphysical consequence of the
viscous dissipation being at least as high as the Ohmic dissipation,
the advantage derived in terms of reproducing the MAC force balance
in the energy-containing scales
is crucial in the understanding of wave motions in both dipole-dominated
and reversing regimes. The present study analyses polarity
transitions in strongly driven, low-inertia dynamos.

In \S 2, we consider the evolution of
a buoyancy disturbance in an unstably stratified rotating fluid
subject to a magnetic field. {In Cartesian geometry, the axes 
parallel to gravity, rotation, and the magnetic field are 
chosen to be mutually orthogonal, 
which is referred to as the ``equatorial toroidal configuration" by
\cite{loper2003}.}
At times much shorter than the
time scale for exponential increase of the perturbation,
the relative intensities of the fast and slow MAC wave motions
generated by the perturbation are studied. Apart from the
dipole-dominated regime given by $|\omega_C|>
|\omega_M| > |\omega_A| > |\omega_\eta|$, the regime 
thought to be relevant to polarity transitions,
$|\omega_C|> |\omega_M| \approx |\omega_A| > |\omega_\eta|$,
 is analysed. { Since the relative magnitudes
of the frequencies in these inequalities
do not depend on the precise orientation of the gravity,
rotation and magnetic field axes, the Cartesian
linear model serves as the basis for
the study of the role of wave motions in 
polarity transitions
in nonlinear dynamo models}, given in \S 3. Here, we find that
the formation of the axial dipole from a chaotic multipolar state
and the collapse of the axial dipole into a polarity-reversing
state are dynamically similar phenomena, in that they both 
occur at $|\omega_A/\omega_M| \sim 1$. While dipole
formation requires the excitation of slow MAC waves as a dynamo
evolves from a small seed magnetic field, polarity reversals
occur when the slow waves are suppressed in a strongly driven
dynamo. The self-similarity of
the dipole--multipole transition in the inertia-free
regime places a constraint on the
Rayleigh number for reversals. In \S 4, we discuss the implications
of our results for planetary cores and future work.

\section{Evolution of a density disturbance under 
rapid rotation and a magnetic field}
\label{linmodel}
\subsection{Problem set-up and governing equations}
\label{psetup}
\begin{figure}
	\centering
	\includegraphics[width=0.45\linewidth]
	{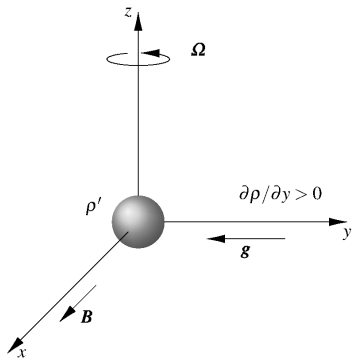}\\
	\caption{{The initial state of a density perturbation
	 $\rho^{\prime}$ that evolves in an unstably stratified 
fluid subject to a uniform magnetic
	field $\bm{B} = B \hat{\bm{e}}_x$, background rotation 
	$\bm{\varOmega} = \varOmega \hat{\bm{e}}_z$ and gravity 
	$\bm{g} = -g \hat{\bm{e}}_y$ in
	Cartesian coordinates $(x,y,z)$.}}
	\label{coordinate}
\end{figure}
A localized density disturbance $\rho^\prime$ that occurs in an unstably stratified rotating fluid layer threaded by a
uniform magnetic field is considered. Since  
$\rho^\prime$ is related to a temperature perturbation
$\Theta$ by $\rho^\prime = -\rho \alpha \Theta$, 
where $\rho$ is the ambient density and $\alpha$ is the coefficient of thermal
expansion, an initial temperature perturbation is chosen in the form
\begin{equation}
\Theta_0=C \, \exp \left[-(x^2+y^2+z^2)/\delta^2 \right],
\label{pert}
\end{equation}
where $C$ is a constant and $\delta$ is the length scale 
of the perturbation. Figure \ref{coordinate} shows the initial
perturbation which subsequently evolves under gravity 
$\bm{g} = -g \hat{\bm{e}}_y$, background rotation 
$\bm{\varOmega} = \varOmega \hat{\bm{e}}_z$ 
and a uniform magnetic field $\bm{B} = B \hat{\bm{e}}_x$ in
Cartesian coordinates $(x,y,z)$. In an otherwise
quiescent medium, the initial 
temperature perturbation \eqref{pert} gives rise
to a velocity field $\bm{u}$, which in turn interacts with
 $\bm{B}$ to generate the induced magnetic field $\bm{b}$.
The initial velocity $\bm{u}_0$ and induced field $\bm{b}_0$
are both zero. In the Boussinesq approximation, the following
magnetohydrodynamic (MHD) equations give the evolution of
$\bm{u}$, $\bm{b}$ and $\Theta$:
\begin{eqnarray}
& & \frac{\partial{\bm{u}}}{\partial{t}} =
-\frac{1}{\rho}\nabla{p^*} -2 \bm{\varOmega}
 \times \bm{u} +
\frac{1}{\mu \rho}\left(\bm{B} \cdot \nabla \right) \bm{b} - 
\bm{g} \alpha\Theta +\nu\nabla^2\bm{u},
\label{mom}\\
& &\frac{\partial{\bm{b}}}{\partial{t}} =\left(\bm{B} \cdot \nabla \right) {\bm{u}}
+\eta \nabla^2 \bm{b},
\label{ind}\\
&& \frac{\partial{\Theta}}{\partial{t}}=
-\gamma \bm{\hat{e}}_y \cdot \bm{u}+ \kappa\nabla^2\Theta,
\label{temp}\\
&& \nabla \cdot \bm{u} = \nabla \cdot \bm{b}=0,
\label{divcond}
\end{eqnarray}
where $\nu$ is the kinematic viscosity,
$\kappa$ is the thermal diffusivity, $\eta$ is the magnetic diffusivity,
$\mu$ is the magnetic permeability, $\bm{\varOmega}= 
\varOmega \hat{\bm{e}}_z$, $p^*=p -(\rho/2) |{\bm{\Omega}} \times
{\bm{x}}|^2+ (\bm{B} \cdot \bm{b})/\mu$ and 
$\gamma=\partial T_0/\partial y <0$ is the mean temperature gradient
in the unstably stratified fluid.

\subsection{Solutions for the velocity field}
\label{velsol}
Taking the  curl of equations \eqref{mom} and
the \eqref{ind} and eliminating the
electric current density between them, we obtain
for the velocity field,
\begin{equation}
\begin{aligned}
\bigg[\bigg(\frac{\partial}{\partial{t}}&
-\nu \nabla^2\bigg)\left(\frac{\partial}{\partial{t}}-\eta \nabla^2\right)
-V_M^2 \frac{\partial^2{}}{\partial{x^2}}\bigg]^2
(-\nabla^2 \bm{u})\\
&=4\Omega^2\left(\frac{\partial}{\partial{t}}
-\eta\nabla^2\right)^2\frac{\partial^2{\bm{u}}}{\partial{z^2}}
-2\Omega\alpha\frac{\partial}{\partial{z}}\left(\frac{\partial}{\partial{t}}
-\eta\nabla^2\right)^2(\nabla\times\Theta\bm{g})\\
&-\alpha\left(\frac{\partial}{\partial{t}}
-\eta \nabla^2\right)\bigg[\left(\frac{\partial}{\partial{t}}
-\nu \nabla^2\right)\left(\frac{\partial}{\partial{t}}-\eta \nabla^2\right)
- V_M^2 \frac{\partial^2{}}{\partial{x^2}}\bigg]
(\nabla\times\nabla\times\Theta\bm{g}),
\label{curlthree}
\end{aligned}
\end{equation}
where $V_M= B/\sqrt{\mu \rho}$ is the Alfv\'en velocity.
The assumption of an unbounded domain facilitates 
the use of Fourier transforms along each coordinate direction,

\begin{equation}
\mathcal{F}(\mathbf{A}) = \hat{\mathbf{A}}=  
\int_{-\infty}^{\infty}\int_{-\infty}^{\infty}\int_{-\infty}^{\infty}\mathbf{A}
\mbox{e}^{- \mathrm{i} \, \bm{k}\cdot\bm{x}}\,\mathrm{d}x\,
\mathrm{d}y\,\mathrm{d}z,
\label{fourier}
\end{equation}

and 
\begin{equation}
\mathcal{F}^{-1}\left(\hat{\mathbf{A}}\right)=\mathbf{A}= 
{ \frac{1}{(2\pi)^3} } 
\int_{-\infty}^{\infty}\int_{-\infty}^{\infty}\int_{-\infty}^{\infty}\hat{\mathbf{A}}
\mbox{e}^{\mathrm{i} \,\bm{k}\cdot\bm{x}}\, \mathrm{d}k_x\,
\mathrm{d}k_y\,\mathrm{d}k_z,
\label{invfourier}
\end{equation}
where $\bm{k}=(k_x,k_y,k_z)$ represents the wave vector such that $|\bm{k}|=k=\sqrt{k_x^2+k_y^2+k_z^2}$. Application
of the Fourier transform \eqref{fourier} to the Cartesian components of equation \eqref{curlthree} gives
\begin{equation}
\begin{aligned}
\bigg[\bigg(\bigg(\frac{\partial}{\partial{t}}
&+\omega_\nu\bigg)\left(\frac{\partial}{\partial{t}}
+\omega_\eta\right)+\omega_M^2\bigg)^2\left(\frac{\partial}{\partial{t}}
+\omega_\kappa\right)+\omega_C^2\left(\frac{\partial}{\partial{t}}
+\omega_\eta\right)^2\left(\frac{\partial}{\partial{t}}
+\omega_\kappa\right)\bigg]\hat{u}_x\\
&=\bigg[-\omega_C\omega_A^2\frac{k_zk}{k_x^2+k_z^2}
\left(\frac{\partial}{\partial{t}}+\omega_\eta\right)^2\\
&+\omega_A^2 \frac{k_xk_y}{k_x^2+k_z^2}\left(\frac{\partial}{\partial{t}}
+\omega_\eta\right) \bigg(\left(\frac{\partial}{\partial{t}}
+\omega_\nu\right)\left(\frac{\partial}{\partial{t}}
+\omega_\eta\right)+\omega_M^2\bigg)\bigg]\hat{u}_y,
\label{ux}
\end{aligned}
\end{equation}
\begin{equation}
\begin{aligned}
\bigg[\bigg(\bigg(\frac{\partial}{\partial{t}}
&+\omega_\nu\bigg)\left(\frac{\partial}{\partial{t}}
+\omega_\eta \right)+\omega_M^2\bigg)^2\left(\frac{\partial}{\partial{t}}
+\omega_\kappa\right)+\omega_C^2\left(\frac{\partial}{\partial{t}}
+\omega_\eta\right)^2\left(\frac{\partial}{\partial{t}}
+\omega_\kappa\right)\\
&+\omega_A^2\left(\frac{\partial}{\partial{t}}
+\omega_\eta\right) \bigg(\left(\frac{\partial}{\partial{t}}
+\omega_\nu\right)\left(\frac{\partial}{\partial{t}}
+\omega_\eta\right)+\omega_M^2\bigg)\bigg]\hat{u}_y=0,
\label{uy}
\end{aligned}
\end{equation}
\begin{equation}
\begin{aligned}
\bigg[\bigg(\bigg(\frac{\partial}{\partial{t}}
&+\omega_\nu\bigg)\left(\frac{\partial}{\partial{t}}
+\omega_\eta\right)+\omega_M^2\bigg)^2\left(\frac{\partial}{\partial{t}}
+\omega_\kappa\right)+\omega_C^2\left(\frac{\partial}{\partial{t}}
+\omega_\eta\right)^2\left(\frac{\partial}{\partial{t}}
+\omega_\kappa\right)\bigg]\hat{u}_z\\
&=\bigg[\omega_C\omega_A^2\frac{k_x k}{k_x^2+k_z^2}
\left(\frac{\partial}{\partial{t}}
+\omega_\eta\right)^2\\
&+\omega_A^2\frac{k_yk_z}{k_z^2+k_x^2}\left(\frac{\partial}{\partial{t}}
+\omega_\eta\right) \bigg(\left(\frac{\partial}{\partial{t}}
+\omega_\nu\right)\left(\frac{\partial}{\partial{t}}
+\omega_\eta\right)+\omega_M^2\bigg)\bigg]\hat{u}_y,
\label{uz}
\end{aligned}
\end{equation}
where we have combined the transformed temperature equation 
\eqref{temp} with the transform of \eqref{curlthree}.
 In equations \eqref{ux}--\eqref{uz}, 
\begin{equation}
\omega_C^2=4 \varOmega^2 k_z^2/k^2, \quad
 \omega_A^2=g\alpha \gamma \left(k_x^2+k_z^2 \right)/k^2 , \quad
 \omega_M^2=V_M^2 k_x^2,
\label{linom}
\end{equation}
 represent the squares of the frequencies
 of linear inertial, buoyancy and Alfv\'en waves respectively \citep{brag1967,07bussechapter}. 
 In an unstably stratified medium, wherein
 $\omega_A^2 <0$, $|\omega_A|$ is a measure of the strength
 of buoyancy. As the present study focuses on a system where
 the viscous and thermal diffusion are much smaller than
 magnetic diffusion, the frequencies $\omega_\nu = \nu k^2$ and 
$\omega_\kappa = \kappa k^2$ in \eqref{ux}--\eqref{uz} are small compared 
$\omega_\eta = \eta k^2$. In this limit, the solution of the form
$\hat{u}_y \sim \mbox{e}^{\mathrm{i} \lambda t}$ for
the homogeneous equation \eqref{uy} gives the 
following quintic equation in $\lambda$: 
\begin{equation}
\begin{aligned}
\lambda^5- 2 \mathrm{i}\omega_\eta \lambda^4&
- (\omega_A^2+\omega_C^2+\omega_\eta^2+2\omega_M^2) \lambda^3
+2 \mathrm{i} \omega_\eta(
\omega_A^2+\omega_C^2+\omega_M^2) \lambda^2\\&
+(\omega_A^2\omega_\eta^2+\omega_C^2\omega_\eta^2+
\omega_A^2\omega_M^2+\omega_M^4) \lambda-
\mathrm{i} \omega_A^2\omega_\eta\omega_M^2=0,
\label{maineqn}
\end{aligned}
\end{equation}
the approximate roots of which were discussed in
\cite{jfm21} for known relative orders of magnitudes of
$\omega_M$, $\omega_A$, $\omega_C$ and $\omega_\eta$.
In the solution for \eqref{uy},
\begin{equation}
\hat{u}_y = \sum_{m=1}^{5}D_{m} \mbox{e}^{\mathrm{i} \lambda_{m}t},
\label{solutionyhat}
\end{equation}
the coefficients 
$D_{m}$ are determined using the initial conditions for
$\hat{u}_y$ and its time derivatives
 (\S \ref{coeff}). Of the five terms in the
expansion on the right-hand side of \eqref{solutionyhat}, 
two terms represent oppositely travelling fast MAC waves, 
two other terms represent oppositely travelling slow MAC
waves, and the fifth term represents the overall growth of the velocity perturbation. 

By substituting \eqref{solutionyhat}
  in \eqref{ux} and \eqref{uz}, the following
   solutions for \eqref{ux} and \eqref{uz} are obtained:
\begin{equation}
\hat{u}_{x} =\hat{u}^H_{x}+\hat{u}^P_{x}=\sum_{m=1}^{5}A_{m}
\mathrm{e}^{\mathrm{i}\lambda_m^H t}+\sum_{m=1}^{5}M_me^{\mathrm{i} \lambda_m^P t},
\label{solutionxhat}
\end{equation}
\begin{equation}
\begin{aligned}
\hat{u}_{z} =\hat{u}^H_{z}+\hat{u}^P_{z}=\sum_{m=1}^{5}C_{m}
\mathrm{e}^{\mathrm{i}\lambda_m^H t}
+\sum_{m=1}^{5}N_me^{\mathrm{i}\lambda_m^P t}
\label{solutionzhat}
\end{aligned}
\end{equation}
In \eqref{solutionxhat} and \eqref{solutionzhat}, $\hat{u}_x^H$
 and $\hat{u}_z^H$ are the homogeneous solutions of 
  \eqref{ux} and \eqref{uz} respectively, $\hat{u}_x^P$ 
  and $\hat{u}_z^P$ are the particular solutions,
   $\lambda_m^P$ are the roots of 
    \eqref{maineqn} and
    $\lambda_m^H$ are the roots of \eqref{maineqn} with 
    $\omega_A=0$,
\begin{equation}
\begin{aligned}
\lambda^H_1&=\frac{1}{2}\left(\omega_C+ \mathrm{i} \omega_\eta+\sqrt{\omega_C^2-2 \mathrm{i} \omega_C\omega_\eta-\omega_\eta^2+4\omega_M^2}\right),\\
\lambda^H_2&= \frac{1}{2} \left(- \omega_C+ \mathrm{i} 
\omega_\eta - \sqrt{\omega_C^2 + 2 \mathrm{i} \omega_C\omega_\eta-\omega_\eta^2+4\omega_M^2}\right),\\
\lambda^H_3&=\frac{1}{2}\left(\omega_C+ \mathrm{i} \omega_\eta-\sqrt{\omega_C^2-2 \mathrm{i}\omega_C\omega_\eta-\omega_\eta^2+4\omega_M^2}\right),\\
\lambda^H_4&=\frac{1}{2} \left(- \omega_C+ \mathrm{i} \omega_\eta
 + \sqrt{\omega_C^2 +2 \mathrm{i}\omega_C\omega_\eta 
 -\omega_\eta^2+4\omega_M^2}\right),\\
\lambda^H_5&=0,
\label{homogroots}
\end{aligned}
\end{equation}
which give the frequencies of the fast and slow Magneto-Coriolis
(MC) waves in the absence of buoyancy \citep{jfm17b}.
The coefficients $A_m$, $C_m$, $M_m$ and $N_m$ in 
\eqref{solutionxhat} and \eqref{solutionzhat} are
evaluated as in \S\ref{coeff} below.
      
The solutions for the induced magnetic
 field transforms $\hat{b}_x$, $\hat{b}_y$ and $\hat{b}_z$  
are obtained following a similar approach.

\subsection{Evaluation of spectral coefficients}
\label{coeff}
From \eqref{solutionyhat}, the initial conditions 
for $\hat{u}_y$ and its time derivatives are given by
\begin{equation}
\begin{aligned}
\mathrm{i}^n \sum_{m=1}^{5}D_m \lambda_{m}^n
=\left(\frac{\partial^n\hat{u}_y}{\partial t^n}\right)_{t=0}=a_{n+1}, \ n=0,1,2,3,4.
\end{aligned}
\label{axzn}
\end{equation}
Algebraic simplifications give the right-hand sides of \eqref{axzn} in the
limit of $\nu=\kappa=0$,  as follows:
\begin{equation}
\begin{aligned}
a_1&=\hat{u}_y|_{t=0}=0,\\
a_2&=\frac{\partial{\hat{u}_y}}{\partial{t}}|_{t=0}=
\alpha g \left(\frac{k_z^2+k_x^2}{k^2}\right)\hat{\Theta}_0,\\
a_3&=\frac{\partial^2 \hat{u}_y}{\partial{t^2}}|_{t=0}=0,\\
a_4&=\frac{\partial^3{\hat{u}_y}}{\partial{t^3}}|_{t=0}=
-(\omega_M^2+\omega_C^2+\omega_A^2)~a_2,\\
a_5&=\frac{\partial^4{\hat{u}_y}}{\partial{t^4}}|_{t=0}=
\omega_M^2\omega_\eta ~a_2.
\label{smalla5v}
\end{aligned}
\end{equation}
The coefficients $D_m$ may now be obtained using the roots of equation \eqref{maineqn}.
For example, we obtain,
\begin{eqnarray}
D_{1}&=\dfrac{a_{5}- \mathrm{i} \, a_{4}(\lambda_2+\lambda_3+\lambda_4+\lambda_5)
	+\mathrm{i} \, a_{2}(\lambda_2\lambda_4\lambda_5+\lambda_3\lambda_4\lambda_5
	+\lambda_2\lambda_3\lambda_4+\lambda_2\lambda_3\lambda_5)}{(\lambda_1-\lambda_2)
(\lambda_1-\lambda_3)(\lambda_1-\lambda_4)(\lambda_1-\lambda_5)},
\label{d1coeff}\\
D_{3}&=\dfrac{a_{5}- \mathrm{i} \, a_{4}(\lambda_1+\lambda_2+\lambda_4+\lambda_5)
	+ \mathrm{i} \,a_{2}(\lambda_1\lambda_4\lambda_5+\lambda_2\lambda_4\lambda_5
	+\lambda_1\lambda_2\lambda_4+\lambda_1\lambda_2\lambda_5)}{(\lambda_3-\lambda_1)
	(\lambda_3-\lambda_2)(\lambda_3-\lambda_4)(\lambda_3-\lambda_5)},
\label{d3coeff}
\end{eqnarray}
for the forward-travelling fast and slow wave solutions respectively.
The coefficients  $A_m$ and $C_m$ are determined in a similar way.
The coefficients $M_m$ and $N_m$ in equations 
\eqref{solutionxhat} and \eqref{solutionzhat}
 are obtained using the method of undetermined coefficients, as follows:
\begin{equation}
\begin{aligned}
M_m &= \frac{D_{m}}{T_m}\bigg[-\omega_C\omega_A^2
\left(\frac{kk_z}{k_x^2+k_z^2}\right)
\left(-(\lambda^P_m)^2+\omega_\eta^2
+2 \mathrm{i} \omega_\eta\lambda^P_m\right)\\
&+\omega_A^2\frac{k_xk_y}{k_x^2+k_z^2}
\left(- \mathrm{i}(\lambda^P_m)^3-2\omega_\eta(\lambda^P_m)^2
+ \mathrm{i}(\omega_\eta^2+\omega_M^2)\lambda^P_m
+\omega_M^2\omega_\eta\right)\bigg],
\end{aligned}
\end{equation}
and 
\begin{equation}
\begin{aligned}
N_m& = \frac{D_{m}}{T_m}\bigg[\omega_C\omega_A^2
\left(\frac{kk_x}{k_x^2+k_z^2}\right)
\left(-(\lambda^P_m)^2+\omega_\eta^2
+2 \mathrm{i}\omega_\eta\lambda^P_m\right)\\
&+\omega_A^2\frac{k_zk_y}{k_x^2+k_z^2}
\left(- \mathrm{i}(\lambda^P)^3-2\omega_\eta(\lambda^P_m)^2
+ \mathrm{i}(\omega_\eta^2+\omega_M^2)\lambda^P_m
+\omega_M^2\omega_\eta\right)\bigg],
\end{aligned}
\end{equation}
with
\begin{equation}
\begin{aligned}
T_m = \mathrm{i} (\lambda^P_m)^5+2(\lambda^P_m)^4\omega_\eta
- \mathrm{i}(\lambda^P_m)^3(\omega_C^2+\omega_\eta^2
+2\omega_M^2)-2\omega_\eta(\lambda^P_m)^2(2\omega_C^2+\omega_M^2)\\
+ \mathrm{i}\lambda^P_m(\omega_C^2\omega_\eta^2+\omega_M^4),
\quad m=1,2...5.
\label{mi}
\end{aligned}
\end{equation}

\subsection{Fast and slow MAC waves in unstable stratification} 
\label{linresults}

{From the solution of the initial value 
problem, the velocity and induced magnetic field 
can be obtained at discrete points in time.} The analysis
of the solutions is limited to times much shorter than the 
time scale for the exponential increase
of the perturbations. When the buoyancy force is small
compared with the Lorentz force ($|\omega_A/\omega_M| \ll 1$),
the parameter regime is determined
by the Lehnert number $Le$ and the magnetic Ekman number $E_\eta$,

\begin{subequations}\label{pars}
\begin{gather}
Le= \dfrac{V_M}{2 \varOmega \delta}, \quad 
E_\eta= \dfrac{\eta}{2 \varOmega
\delta^2},
\tag{\theequation a,b}
\end{gather}
\end{subequations}
both based on the length scale $\delta$ of the initial
buoyancy perturbation \eqref{pert}. 

\begin{figure}
	\centering
	\hspace{-1.4 in}	(a)  \hspace{1.5 in} (b) \hspace{1.4 in} (c) \\
	\includegraphics[width=0.28\linewidth,height=.47\linewidth]{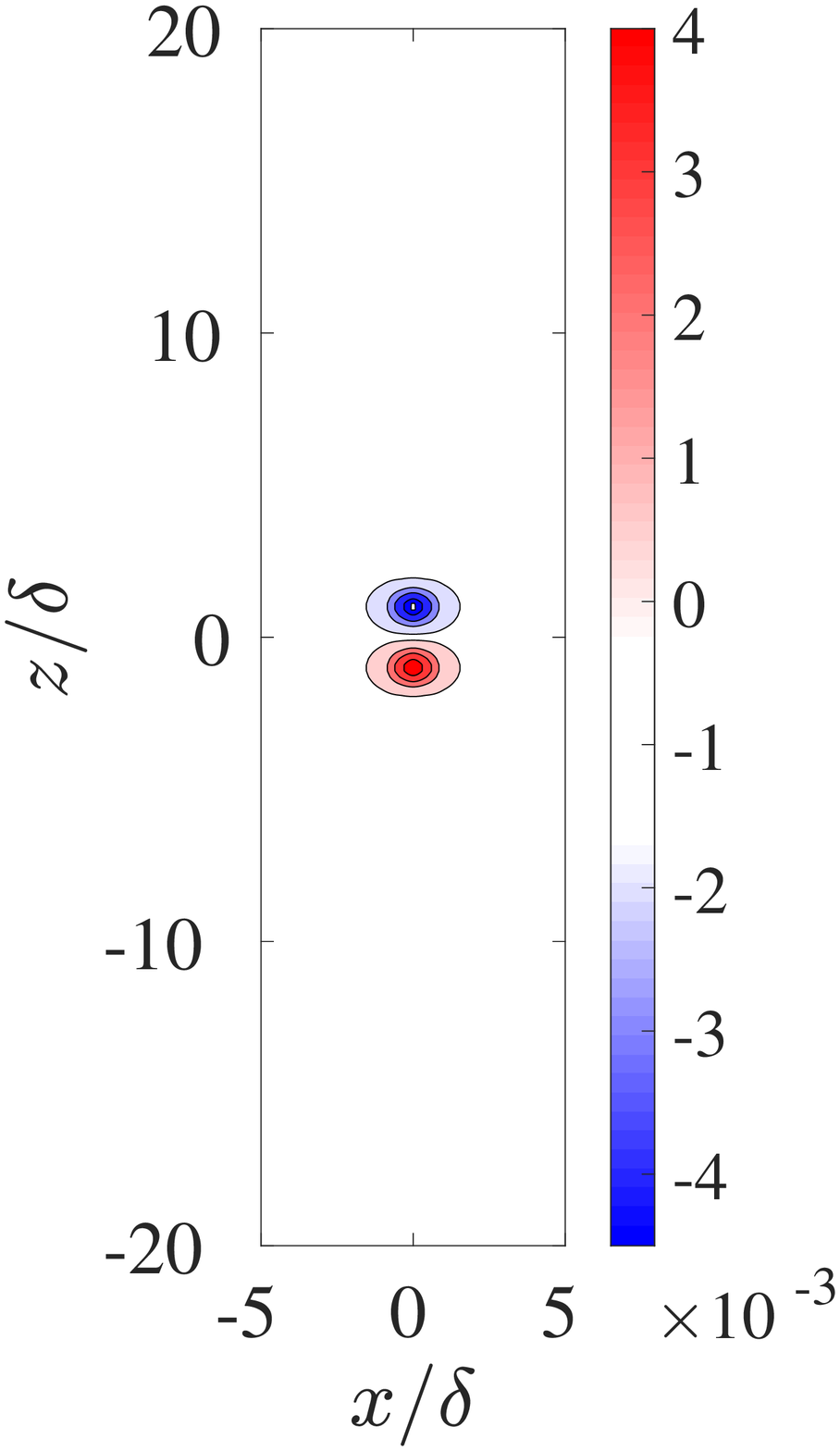}	
	\includegraphics[width=0.28\linewidth,height=.47\linewidth]{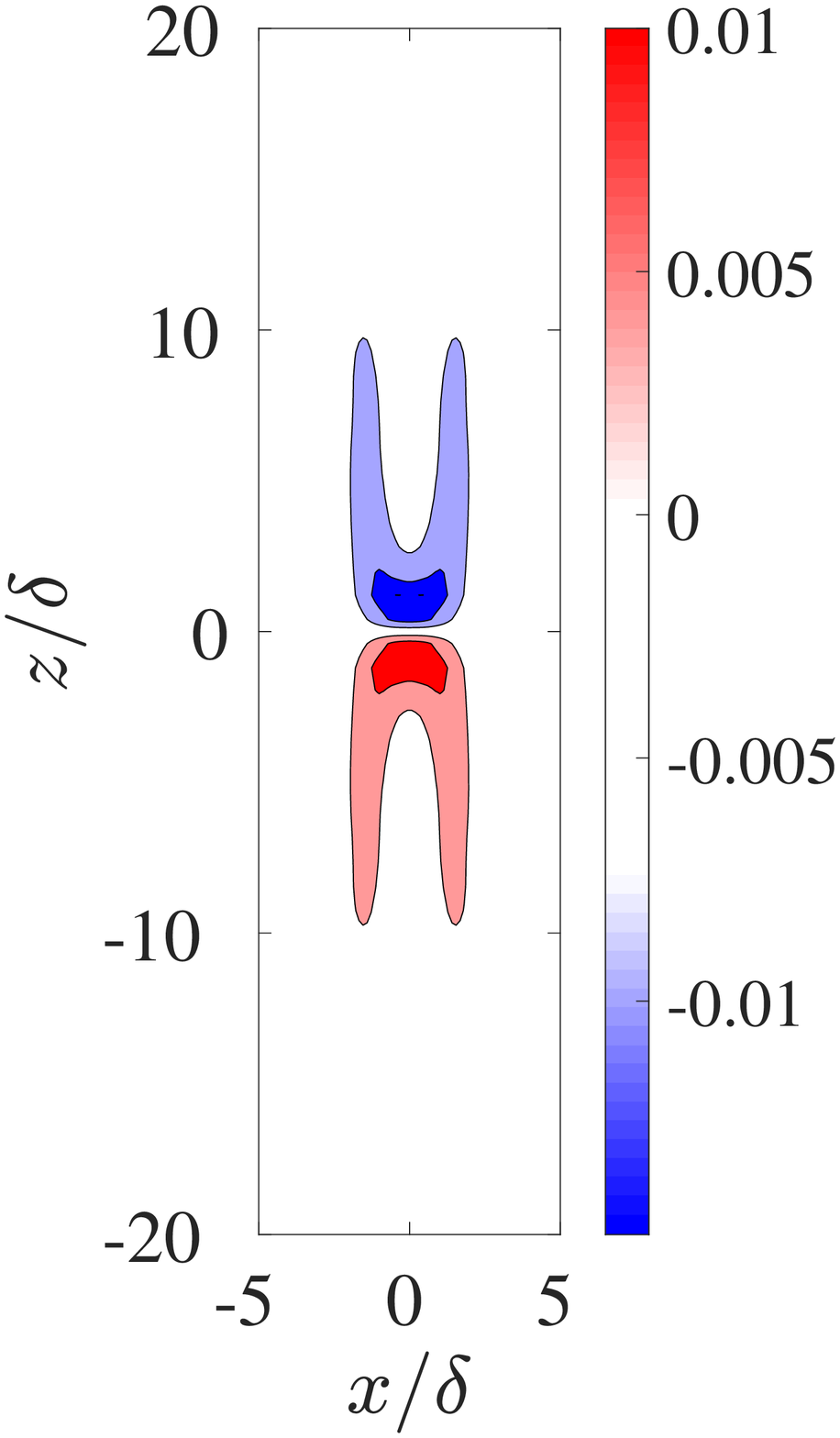}	
	\includegraphics[width=0.28\linewidth,height=.49\linewidth]{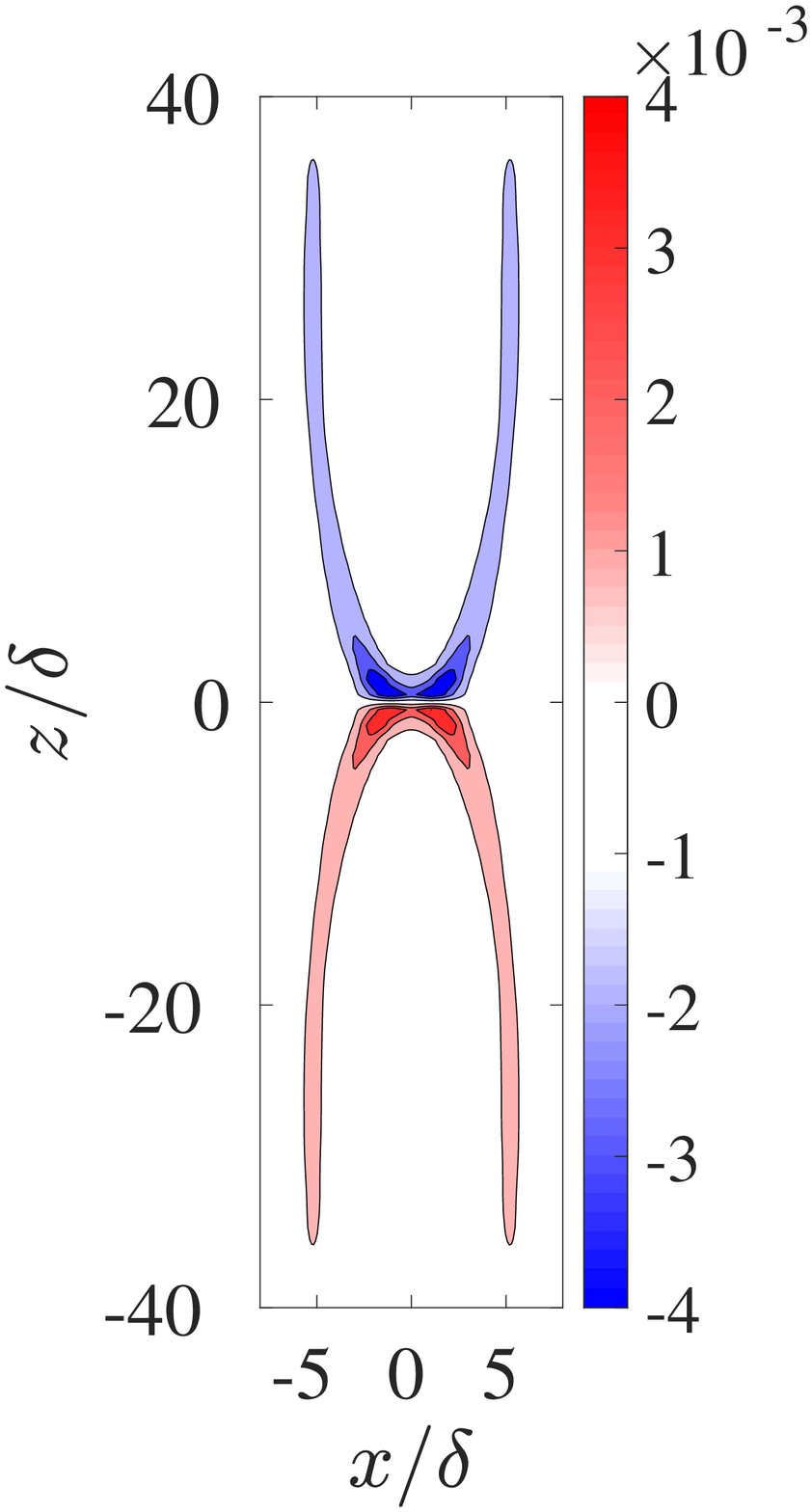}\\
	\caption{Evolution of the kinetic helicity 
on the $x$-$z$ plane at $y=0$
with time (measured in units of the magnetic diffusion time $t_\eta$)
for $Le=0.03$ and $E_\eta=2\times 10^{-5}$. The snapshots are
at (a) $t/t_\eta =1\times10^{-4}$, (b) $t/t_\eta=2.5\times10^{-3}$
and (c) $t/t_\eta = 1\times10^{-2}$. The ratio
$|\omega_A/\omega_M| =0.05$ at times after the formation of
the waves.}
	\label{evolution}
\end{figure}

Figure \ref{evolution} shows the evolution of the kinetic
helicity $\bm{u} \cdot \bm{\zeta}$ at $y=0$. The
real-space fields are obtained from
the transforms $\hat{\bm{u}}$ and $\hat{\bm{\zeta}}$ via
the inverse Fourier transform \eqref{invfourier}. Here, a truncation
value of $\pm 10/\delta$ is used for
the three wavenumbers in the integrals
since the initial wavenumber $k_0= \sqrt{3}/\delta$ (Appendix \ref{k0}).
 Apart from the segregation of oppositely
signed helicity 
between the two halves about the 
mid-plane $z=0$,
the evolution of blobs into columnar structures through
the propagation of damped waves is evident. 

\begin{figure}
	\centering
	\hspace{-2.5 in}	(a)  \hspace{2.5 in} (b) \\
	\includegraphics[width=0.48\linewidth]{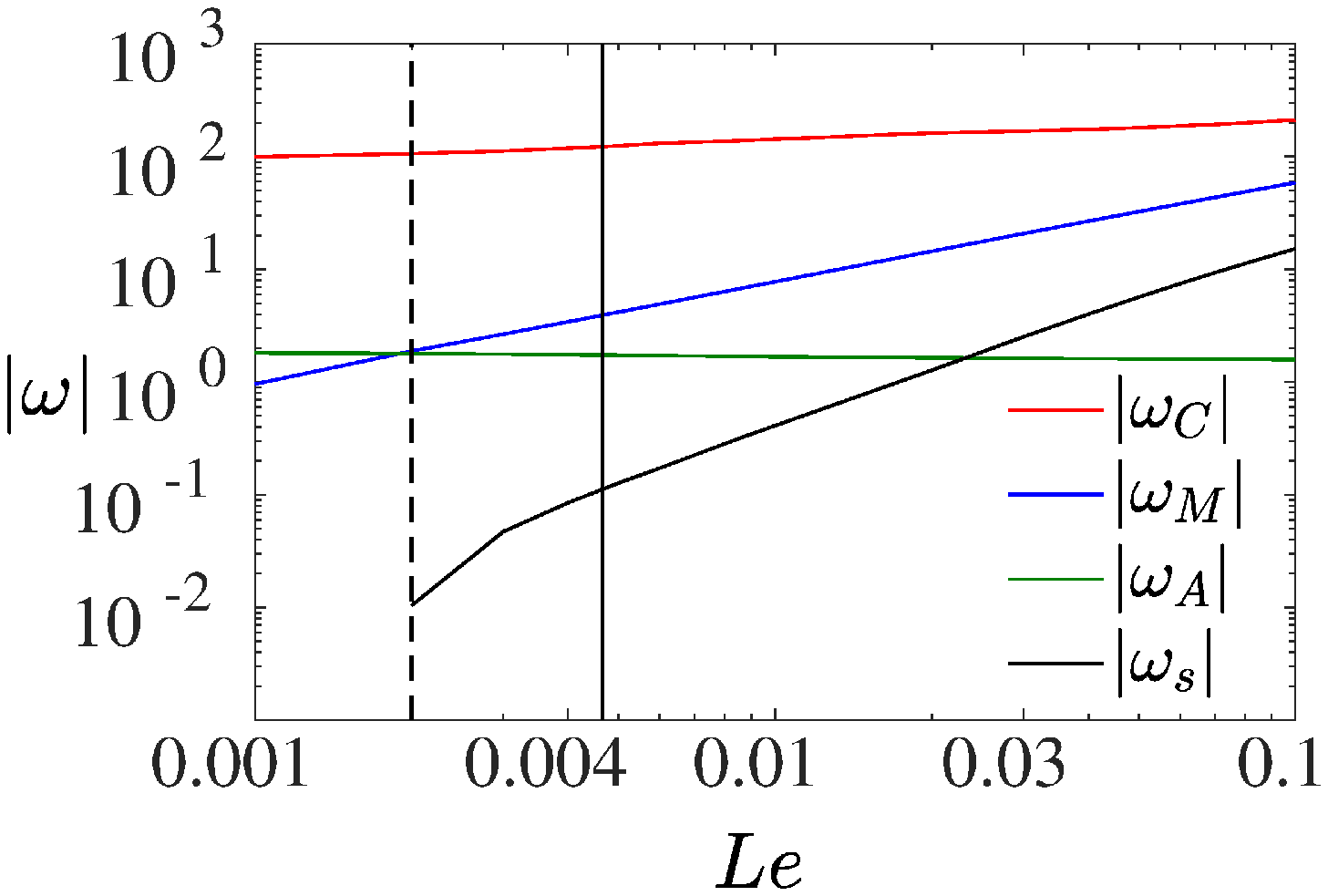}			
\includegraphics[width=0.48\linewidth]{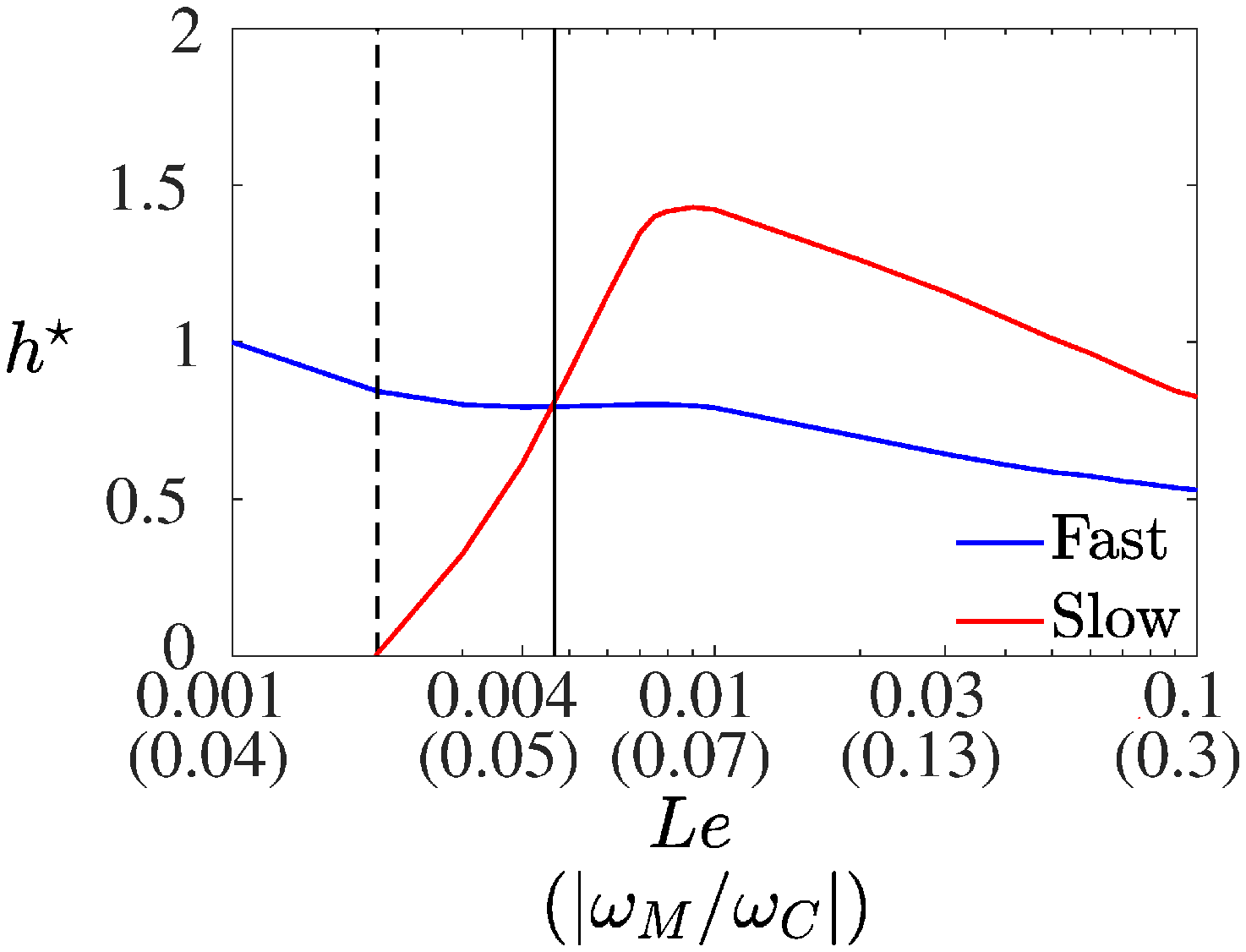}\\

	\caption{(a) Variation of absolute values of frequencies with
 $Le$. (b) Variation with $Le$ of the total kinetic helicity in the region $z<0$ 
normalized by the nonmagnetic helicity. { The values of $|\omega_M/\omega_C|$ that
correspond to the values of $Le$ are given in brackets 
below the horizontal axis.} 
All calculations
are performed for $E_\eta = 2 \times 10^{-5}$. The slow wave frequency, 
$\omega_s$ takes non-zero values for $Le > 2 \times 10^{-3}$,
when $|\omega_M| > |\omega_A|$. }
	\label{cmahel}
\end{figure}

As we seek to understand the role of
MAC waves in the dipolar and multipolar dynamo regimes, 
we may separate the fast and slow MAC wave parts of the 
general solution, which is a linear superposition of
the two wave solutions \citep[see][]{jfm21}. For example,
\begin{equation}
\begin{aligned}
\hat{u}_{x,f}&=M_1 \mathrm{e}^{\mathrm{i}\lambda^P_1 t}+
M_2 \mathrm{e}^{\mathrm{i}\lambda^P_2 t} + A_{1} \mathrm{e}^{\mathrm{i}{\lambda}_1^H t}+
A_{2} \mathrm{e}^{\mathrm{i} {\lambda}_2^H t},\\
\hat{u}_{y,f}& = D_1 \mathrm{e}^{\mathrm{i}\lambda_1 t}+ D_2 \mathrm{e}^{\mathrm{i}\lambda_2 t},\\
\hat{u}_{z,f}&=N_1 \mathrm{e}^{\mathrm{i} \lambda^P_1 t}+ N_2 \mathrm{e}^{\mathrm{i} \lambda^P_2 t}
+C_{1} \mathrm{e}^{ \mathrm{i} {\lambda}_1^H t}+C_{2} \mathrm{e}^{\mathrm{i}{\lambda}_2^H t},
\label{fast}
\end{aligned}
\end{equation}
and
\begin{equation}
\begin{aligned}
\hat{u}_{x,s}&= M_3 \mathrm{e}^{\mathrm{i} \lambda^P_3 t} + M_4 \mathrm{e}^{\mathrm{i}\lambda^P_4 t}
+A_{3} \mathrm{e}^{\mathrm{i}{\lambda}_3^H t} + A_{4} \mathrm{e}^{\mathrm{i}{\lambda}_4^H t},\\
\hat{u}_{y,s}& = D_3 \mathrm{e}^{\mathrm{i}\lambda_3 t}+D_4 \mathrm{e}^{\mathrm{i}\lambda_4 t},\\
\hat{u}_{z,s}&=N_3 \mathrm{e}^{\mathrm{i}\lambda^P_3 t}+
N_4 \mathrm{e}^{\mathrm{i}\lambda^P_4 t}+C_{3} \mathrm{e}^{\mathrm{i}{\lambda}_3^H t}+C_{4}
\mathrm{e}^{\mathrm{i} {\lambda}_4^H t},
\label{slow}
\end{aligned}
\end{equation}
where the subscripts $f$ and $s$ in the left-hand sides
of \eqref{fast} and \eqref{slow} denote the fast and slow
wave parts of the solution.
Figure \ref{cmahel}(a) shows the variation of fundamental 
frequencies with $Le$. { To compute
the frequencies, the mean wavenumbers are first calculated
 through ratios of $L^2$ norms; e.g.
 \begin{equation}
\bar{k}_x = \frac{|\!|k_x \, \hat{u}|\!| }{|\!| \hat{u}|\!|},
\quad \bar{k} = \frac{|\!| k \, \hat{u}|\!|}{|\!| \hat{u}|\!|},
\label{l2norm}
\end{equation}
which are based on the velocity field.}
The values of $|\omega_M/\omega_C|$, given
in brackets below
the horizontal axis of figure \ref{cmahel}(b), 
are systematically higher than that
of $Le$, which is essentially the initial value of this
ratio. The enhanced instantaneous value of
$|\omega_M/\omega_C|$ is due to the
anisotropy of the columnar flow,
and would not be evident if $\bm{\varOmega}$ were aligned
with $\bm{B}$ \citep{jfm21}. For $E_\eta= 2 \times 10^{-5}$,
all calculations for $Le > 2 \times 10^{-3}$ satisfy 
the inequality $|\omega_C| > |\omega_M| > |\omega_A|
>|\omega_\eta|$, thought to be essential for axial
dipole formation in convective dynamos. 
Figure \ref{cmahel}(b) shows the variation
of the dimensionless helicity $h^*$ of the fast and slow
MAC waves, obtained by summing the helicity at all
points in $(x,z)$ for $z<0$ and $y=0$ and then normalizing
this value by the nonmagnetic helicity. For $Le > 2 \times 10^{-3}$
($|\omega_M/\omega_C| > 0.045$), 
the slow wave helicity increases dramatically, and for
$Le \sim 10^{-2}$ ($|\omega_M/\omega_C| \sim 0.1$), the slow
wave helicity is greater than the fast wave helicity.
This result is consistent  with the higher
intensity of the slow wave motions relative to that of the
fast wave motions
for $|\omega_M/\omega_C| \sim 0.1$, inferred from the
fast Fourier transform (FFT) spectra in 
dynamo simulations \citep{aditya2022}.

\begin{figure}
	\centering
	\hspace{-1.5 in}	(a)  \hspace{1.5 in} (b) \\
	\includegraphics[width=0.3\linewidth,height=.5\linewidth]{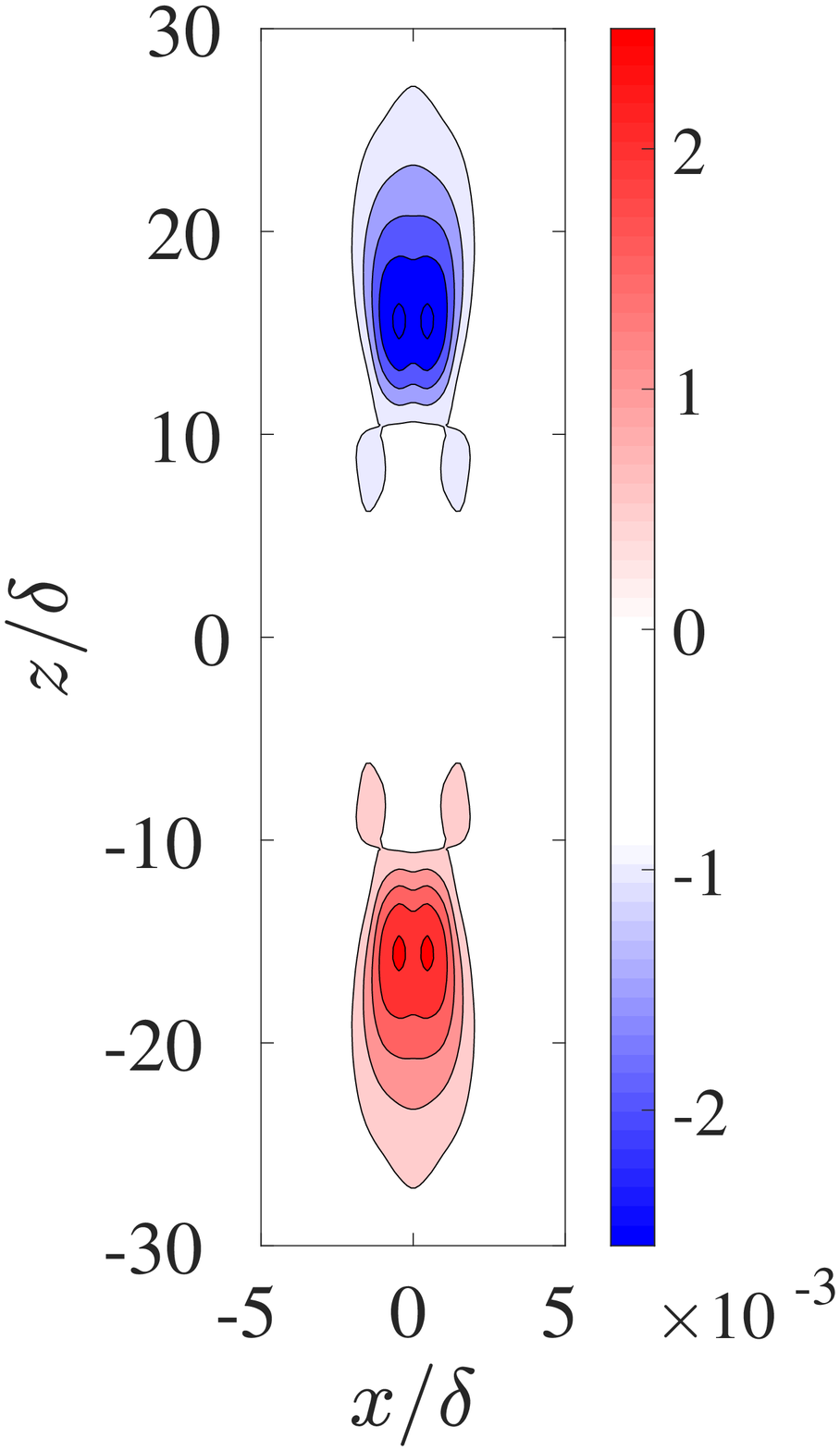}		
	\includegraphics[width=0.3\linewidth,height=.5\linewidth]{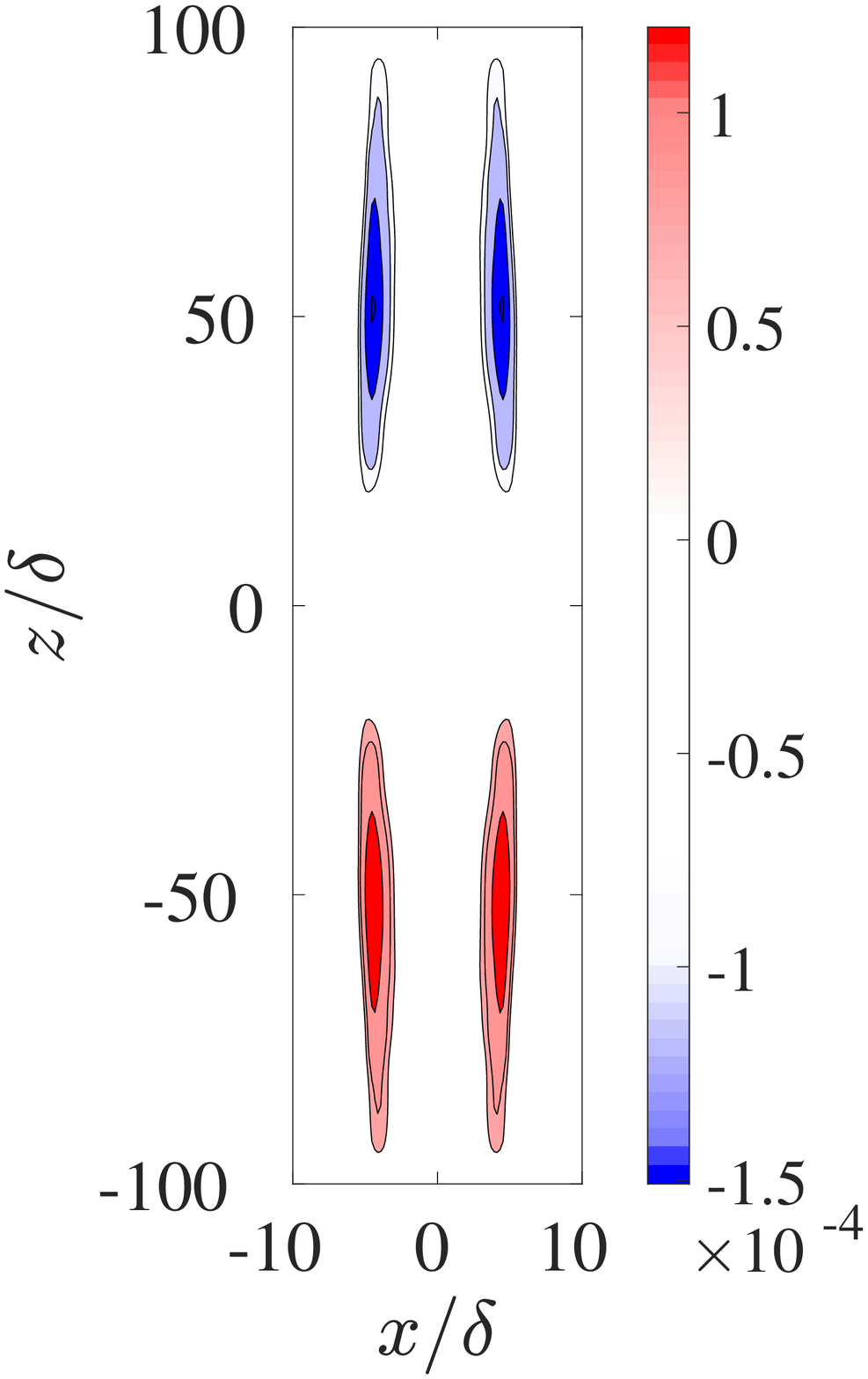}\\
	\hspace{-1.5 in}	(c)  \hspace{1.5 in} (d) \\
	\includegraphics[width=0.3\linewidth,height=.5\linewidth]{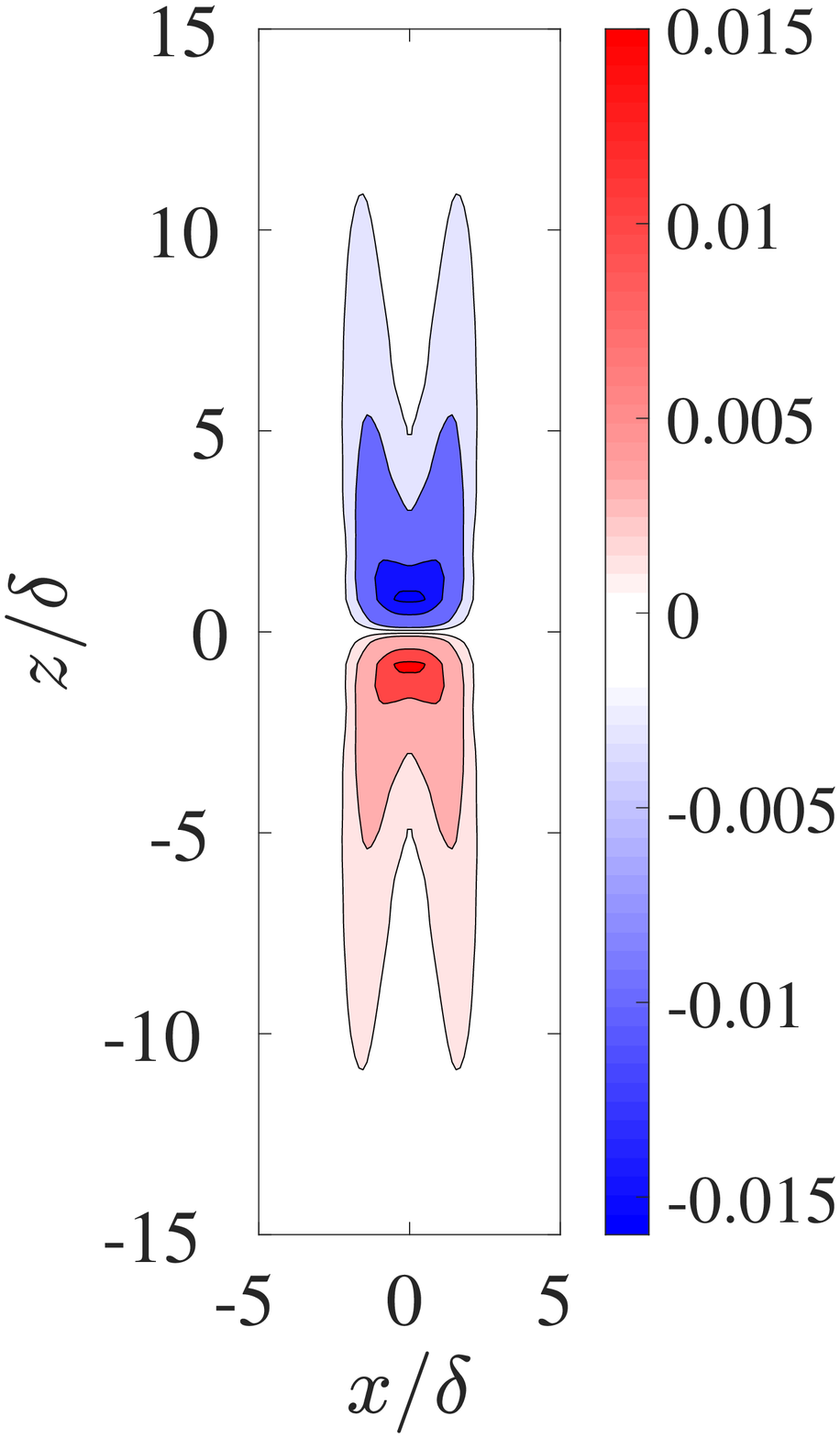}		
	\includegraphics[width=0.3\linewidth,height=.5\linewidth]{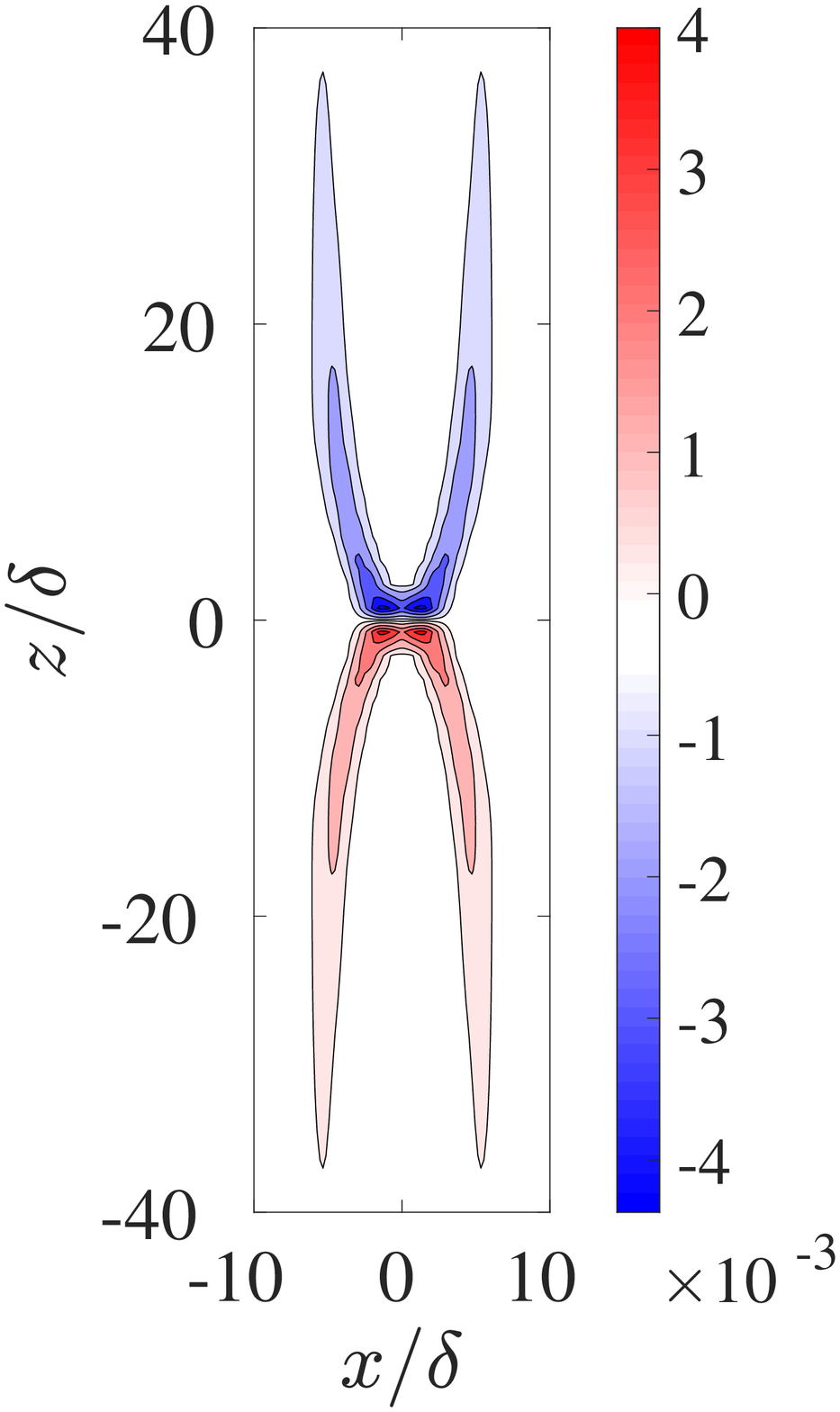}
	\caption{Helicity on the section $y=0$ of fast (a, b) and slow 
(c,d) MAC waves at two times, $t/t_\eta=2.5\times10^{-3}$ (left panels)
and $t/t_\eta=1 \times10^{-2}$ (right panels). 
Here, $E_\eta = 2 \times 10^{-5}$        and 
$Le=0.03$ ($|\omega_M/\omega_C|=0.13$).}
	\label{le03}
\end{figure}

In figure \ref{le03}, the contours of the fast
and slow MAC wave helicities are shown at two times for
$|\omega_M/\omega_C|=0.13$, which lies in the region
of slow wave dominance in figure \ref{cmahel}(b). 
The fast waves split in two and propagate rapidly along $z$.
The slow waves do not propagate
as far as the fast waves at the same time due to
their lower group velocity. Yet, as indicated
by the colour bars, the slow waves are markedly
more intense than the fast waves. Both fast and
slow wave columns propagate along $x$ at the
Alfv\'en velocity (\S \ref{buoy1} below). 

The effect of progressively increasing the buoyant
forcing on the fast and slow waves is examined in the
following section.

\subsection{Effect of progressively increasing buoyancy}
\label{buoy1}
For the regime given by $|\omega_C| \gg |\omega_M|
\gg |\omega_A| \gg |\omega_\eta|$, the roots of
the homogeneous equation
\eqref{maineqn} are approximated by \citep{jfm21}
\begin{eqnarray}
\lambda_{1,2} &\approx& \pm \left(\omega_C+ \frac{\omega_M^2}{\omega_C} \right) +
\mathrm{i} \, \frac{\omega_M^2\omega_\eta}{\omega_C^2}, \label{l12approx}\\ 
\lambda_{3,4} &\approx& \pm \left(\frac{\omega_M^2}{\omega_C}+
\frac{\omega_A^2}{2\omega_C}\right) + \mathrm{i} \, \omega_\eta \,
\left(1-\frac{\omega_A^2}{2\omega_M^2}\right), \label{l34approx} \\
\lambda_{5} &\approx& \mathrm{i}
 \, \frac{\omega_A^2\omega_\eta}{\omega_M^2}. \label{l5approx}
\end{eqnarray}
As $|\omega_A|$ increases relative to $|\omega_M|$, the fast
MAC waves, given by frequencies $\lambda_{1,2}$, { are 
nearly unaffected.}
However, the slow waves, whose real frequencies are approximated by
\begin{equation}
\mbox{Re}(\lambda_{3,4}) \approx \pm \left(\frac{\omega_M^2}{\omega_C}+
\frac{\omega_A^2}{2\omega_C}\right) \approx \pm \frac{\omega_M^2}{\omega_C}
\, \left(1+ \frac{\omega_A^2}{\omega_M^2} \right)^{1/2}
\end{equation}
for $|\omega_C| \gg |\omega_M|, |\omega_A|$ \citep{brag1967}, would be
significantly attenuated
 in an unstably stratified fluid {with 
$\omega_{A}^2<0$ }  as
 $|\omega_A|$ nears $|\omega_M|$.
We see
below that the decrease of the slow wave frequency
translates into the marked 
decrease of the slow wave helicity
relative to the fast wave helicity.

\begin{figure}
	\centering
	\hspace{-2.5 in}	(a)  \hspace{2.2 in} (b) \\
	\includegraphics[width=0.46\linewidth]{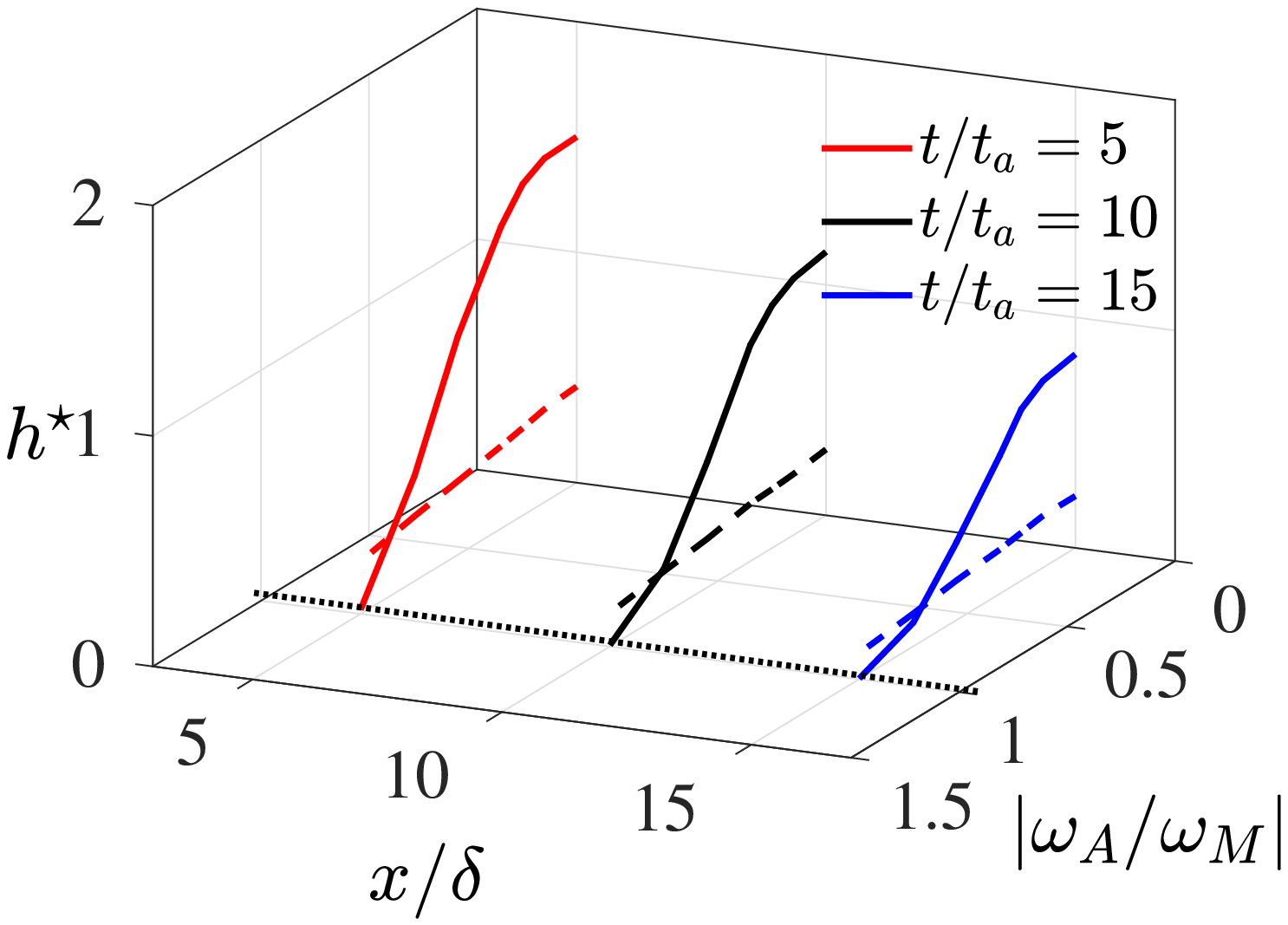} \includegraphics[width=0.46\linewidth]{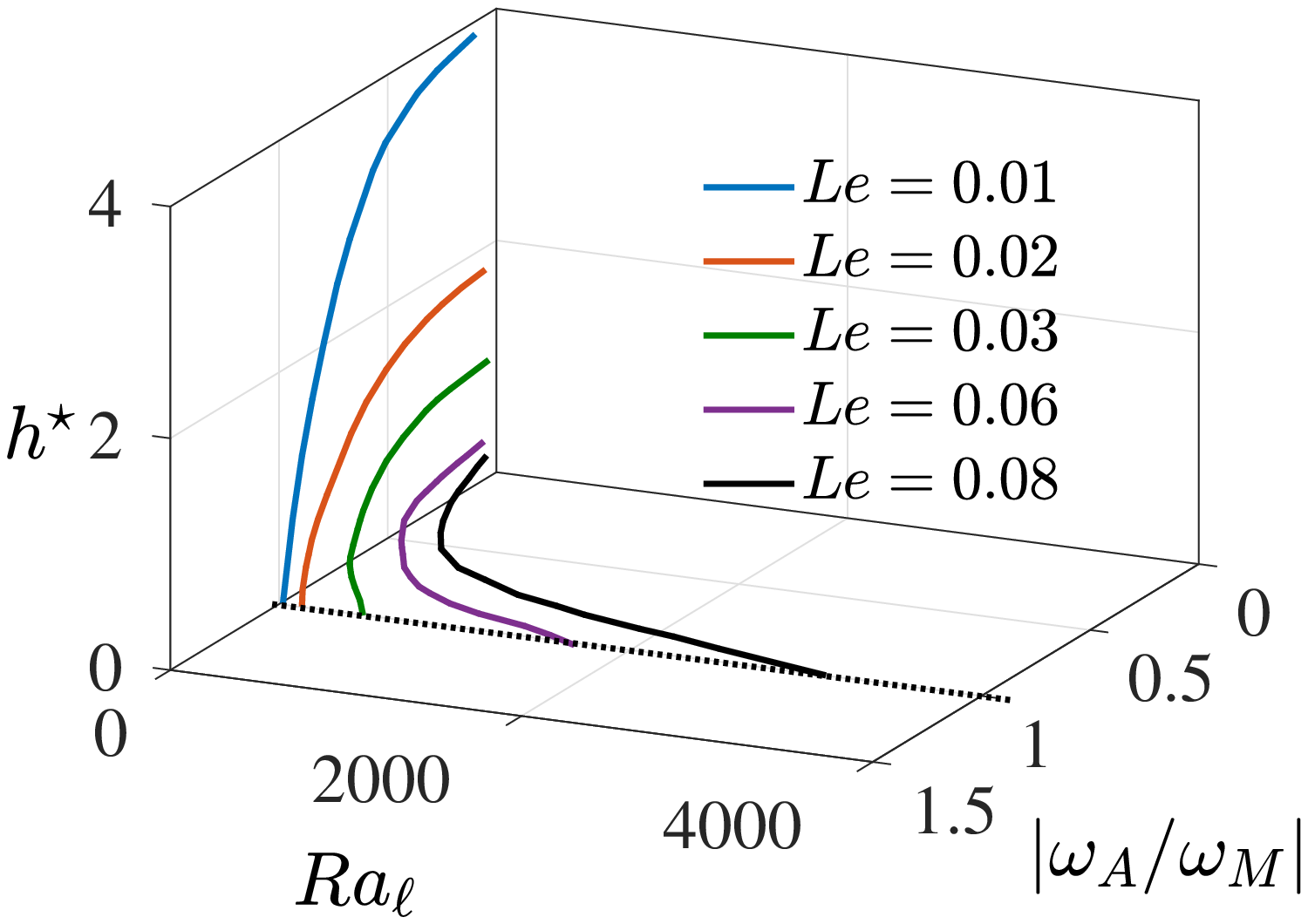}\\
	\caption{(a) Variation of the fast wave (dashed line)
and slow wave (solid line) helicity 
for $z<0$,
normalized by the non-magnetic helicity, with $x/\delta$ and
$|\omega_A/\omega_M|$ at different times (measured in units
of the Alfv\'en wave travel time $t_a$). Here, $E_\eta=2\times 10^{-5}$
and $Le=0.03$. (b) Variation of the normalized slow
wave helicity with the local
Rayleigh number $Ra_\ell$ (defined in \eqref{radelta}) and
$|\omega_A/\omega_M|$ for different $Le$.}
	\label{linrev}
\end{figure}

\begin{figure}
	\centering
	\hspace{-1.4 in}	(a)  \hspace{1.5 in} (b) \hspace{1.4 in} (c) \\
	\includegraphics[width=0.28\linewidth,
height=.45\linewidth]{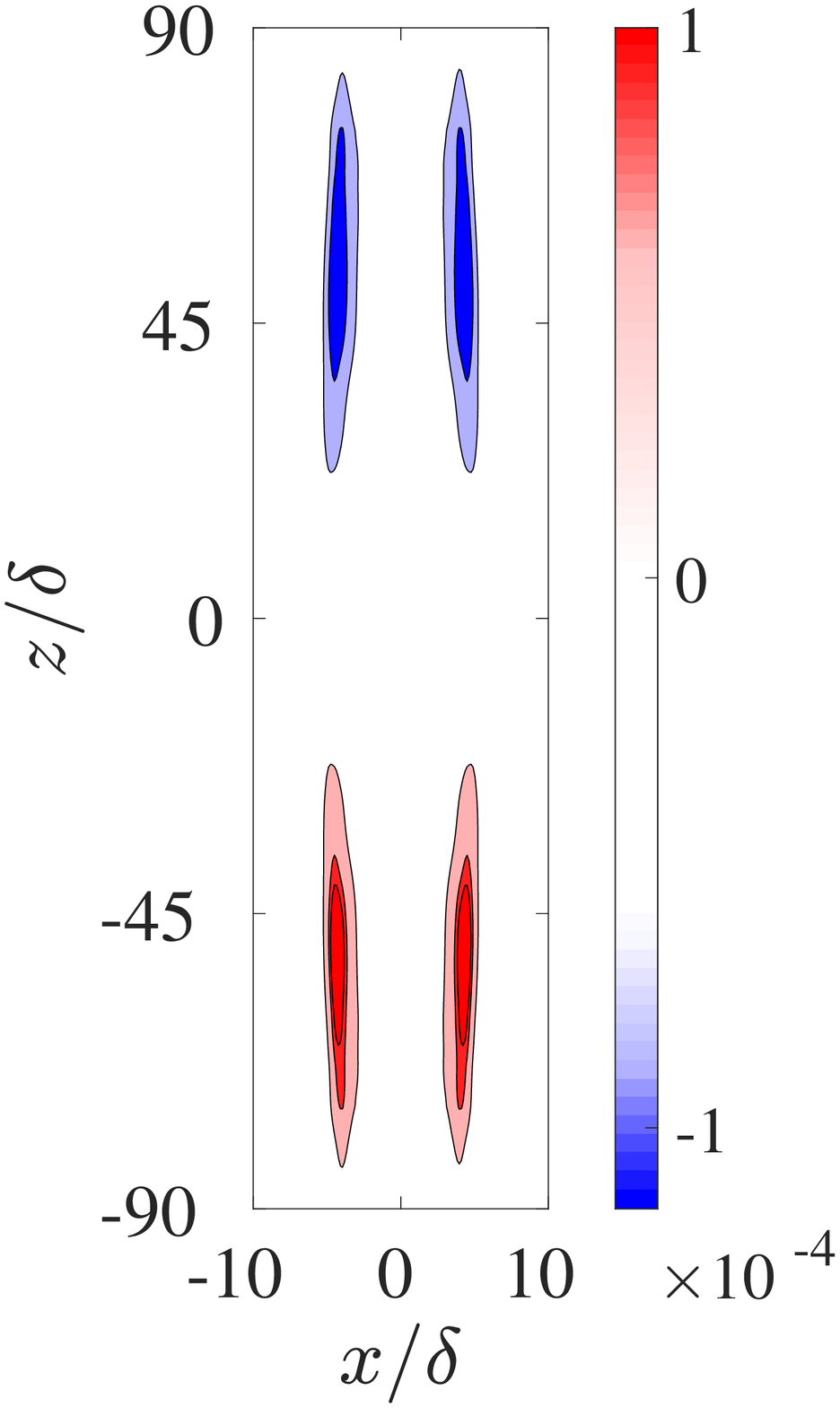}	
	\includegraphics[width=0.28\linewidth,
height=.45\linewidth]{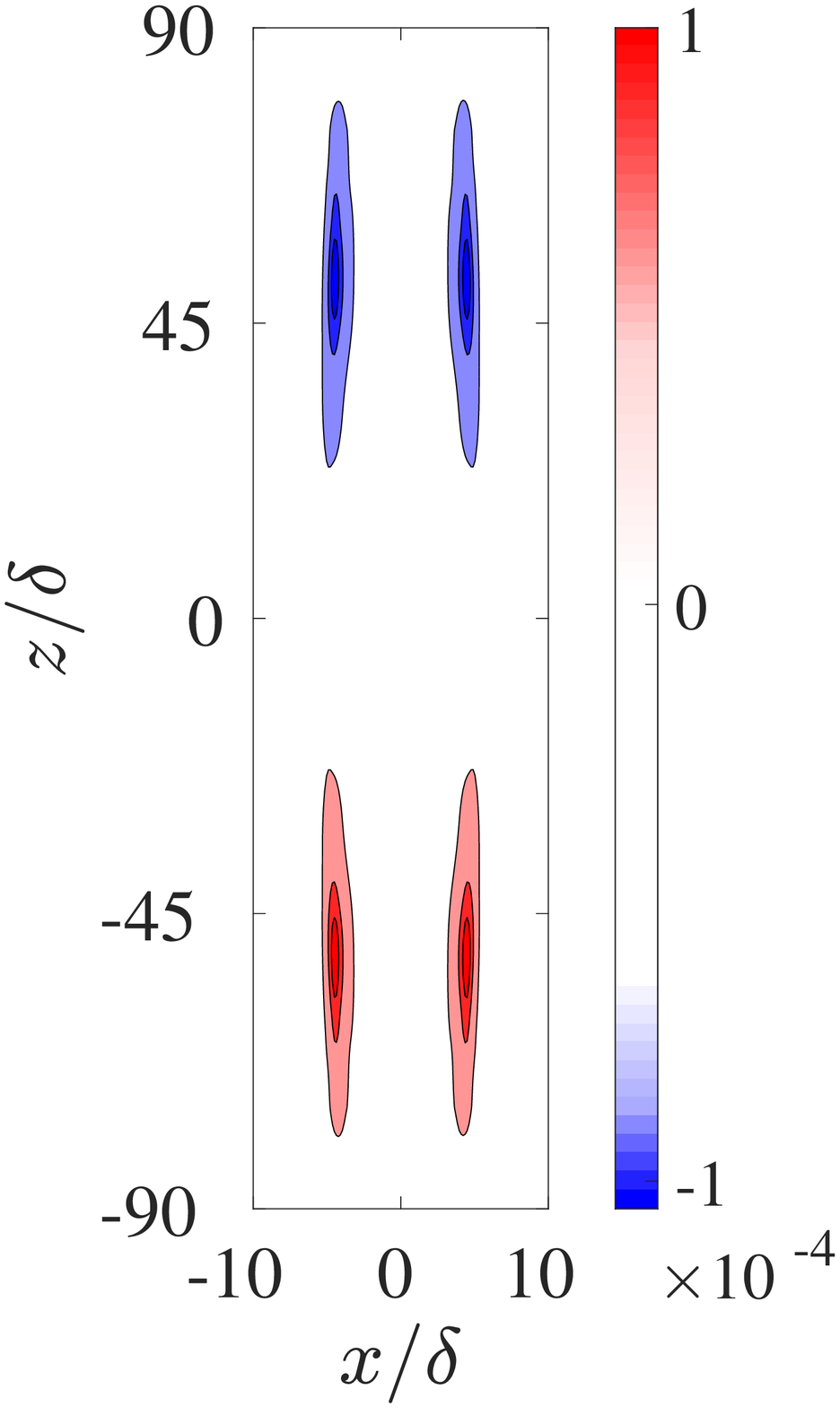}	
	\includegraphics[width=0.28\linewidth,
height=.45\linewidth]{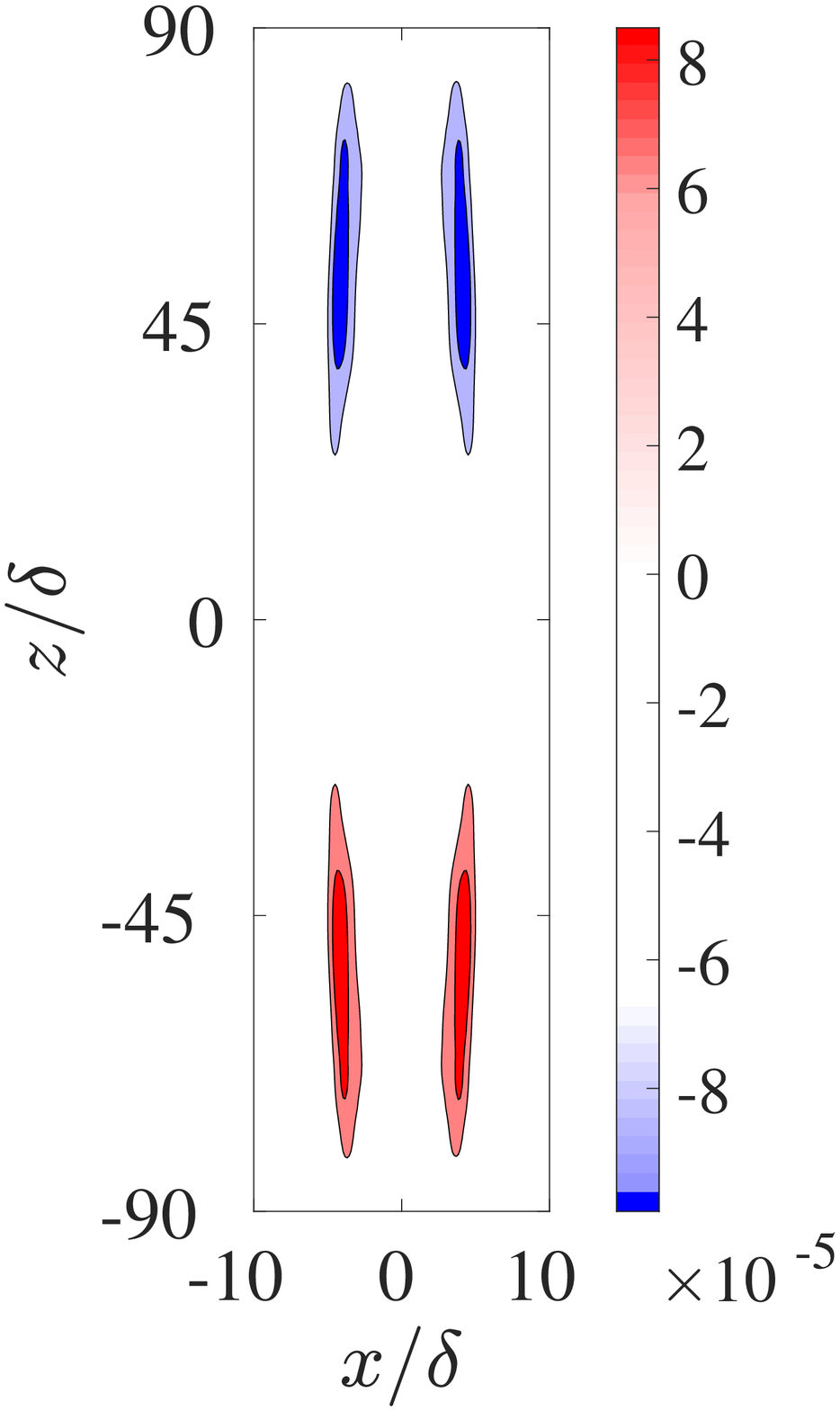}	\\
	\hspace{-1.4 in}	(d)  \hspace{1.5 in} (e) \hspace{1.4 in} (f) \\
	\includegraphics[width=0.28\linewidth,
height=.45\linewidth]{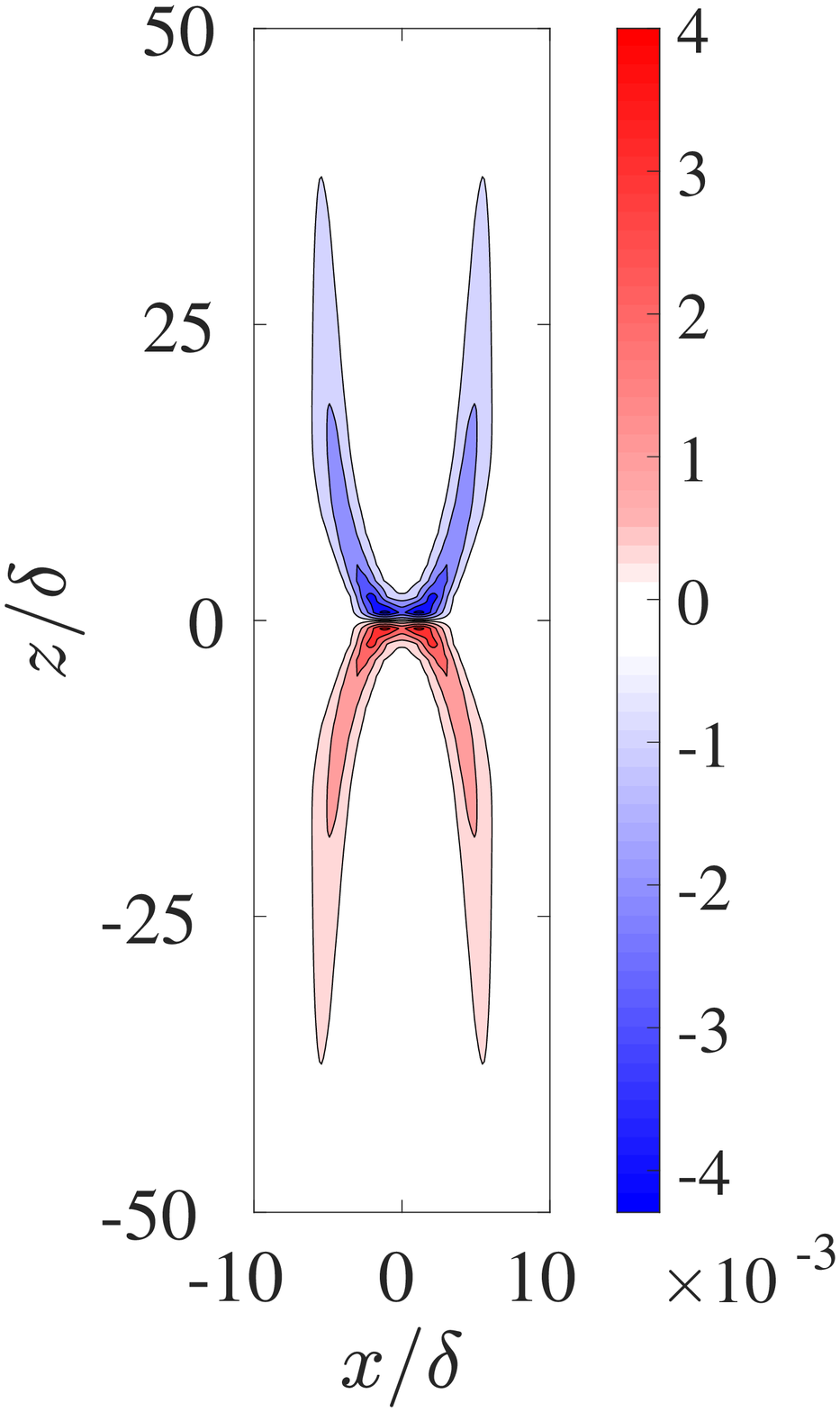}	
	\includegraphics[width=0.28\linewidth,
height=.45\linewidth]{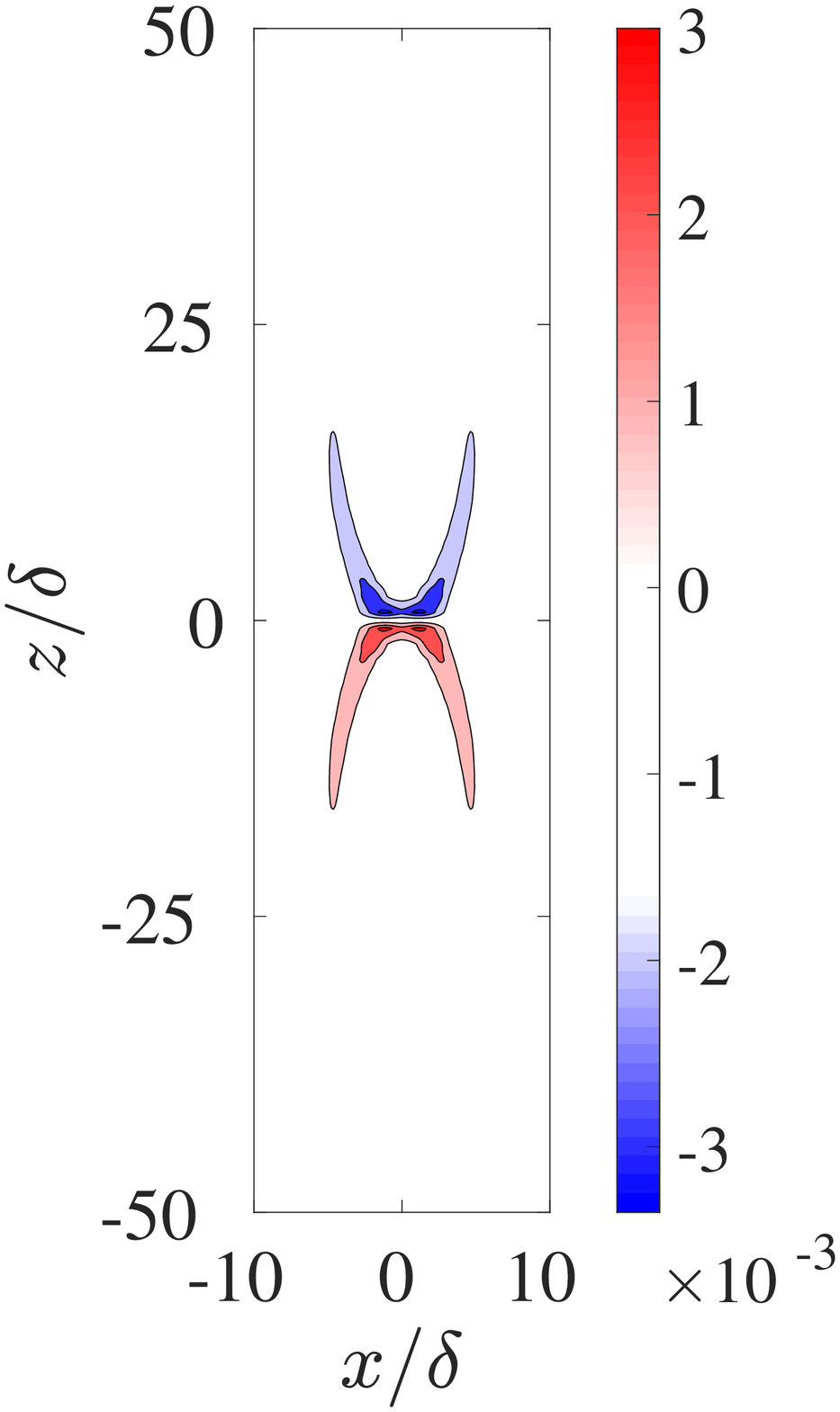}	
	\includegraphics[width=0.28\linewidth,
height=.45\linewidth]{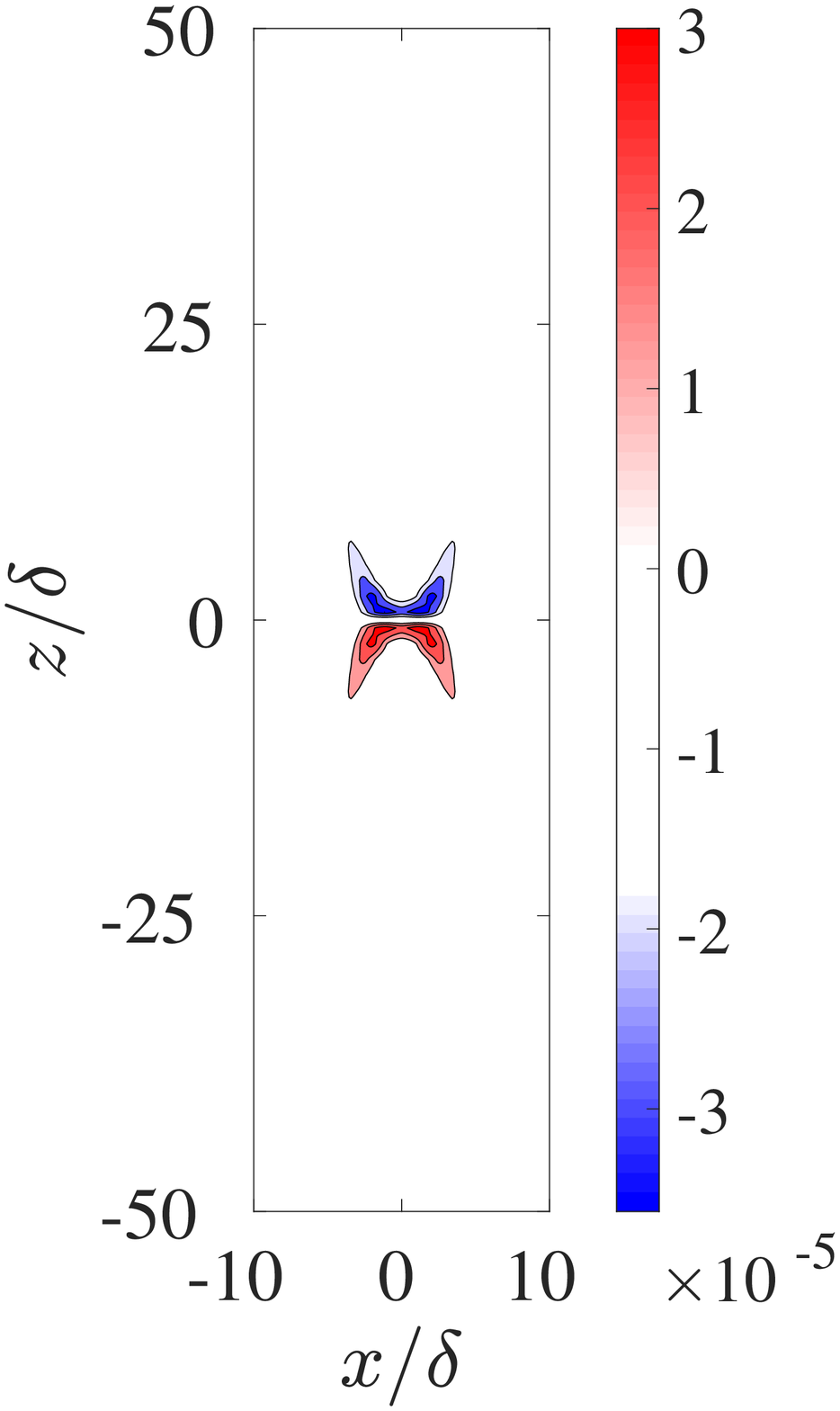}\\
	\hspace{-1.4 in}	(g)  \hspace{1.5 in} (h) \hspace{1.4 in} (i) \\
	\includegraphics[width=0.28\linewidth,
height=.45\linewidth]{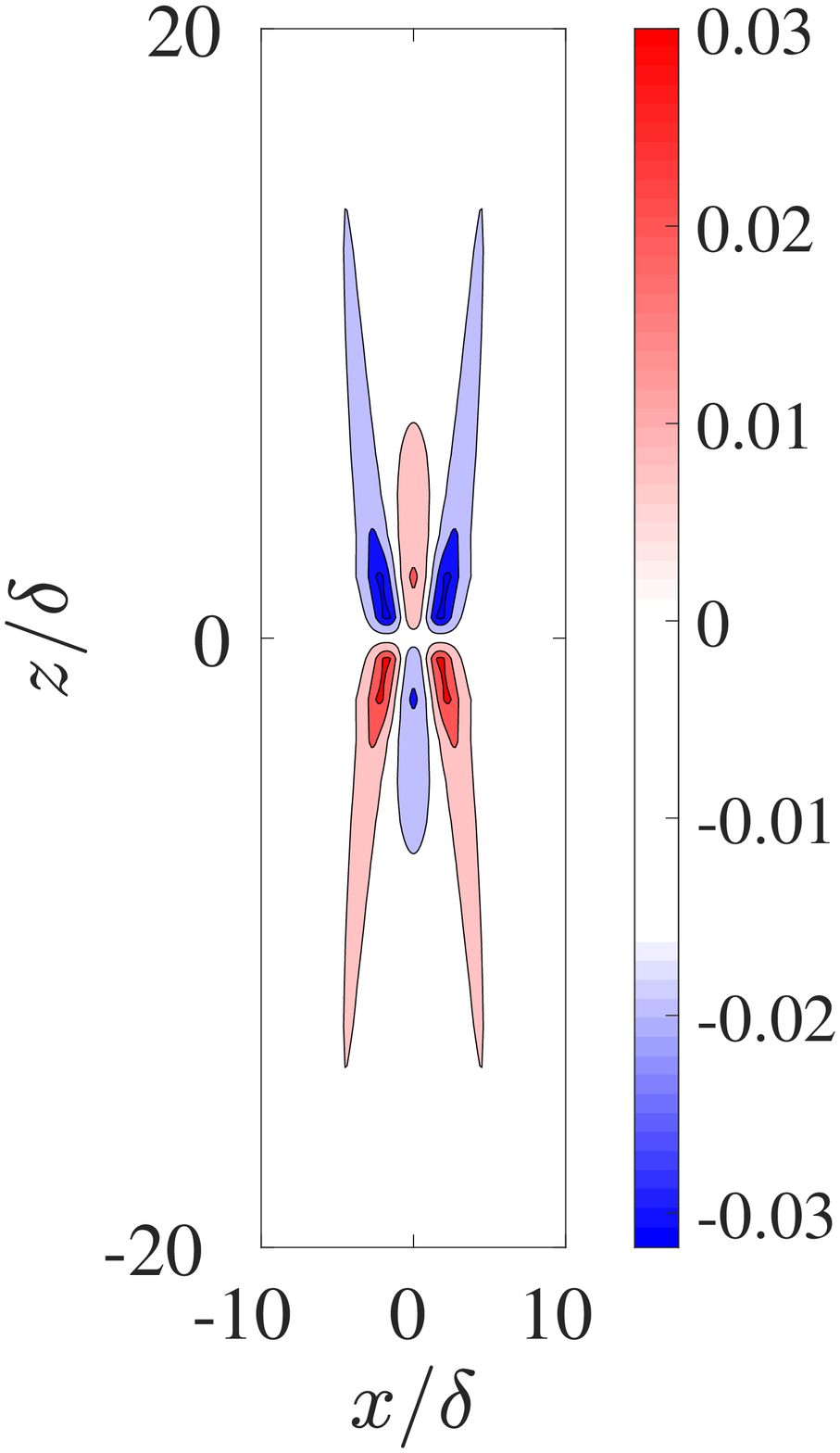}	
	\includegraphics[width=0.28\linewidth,
height=.45\linewidth]{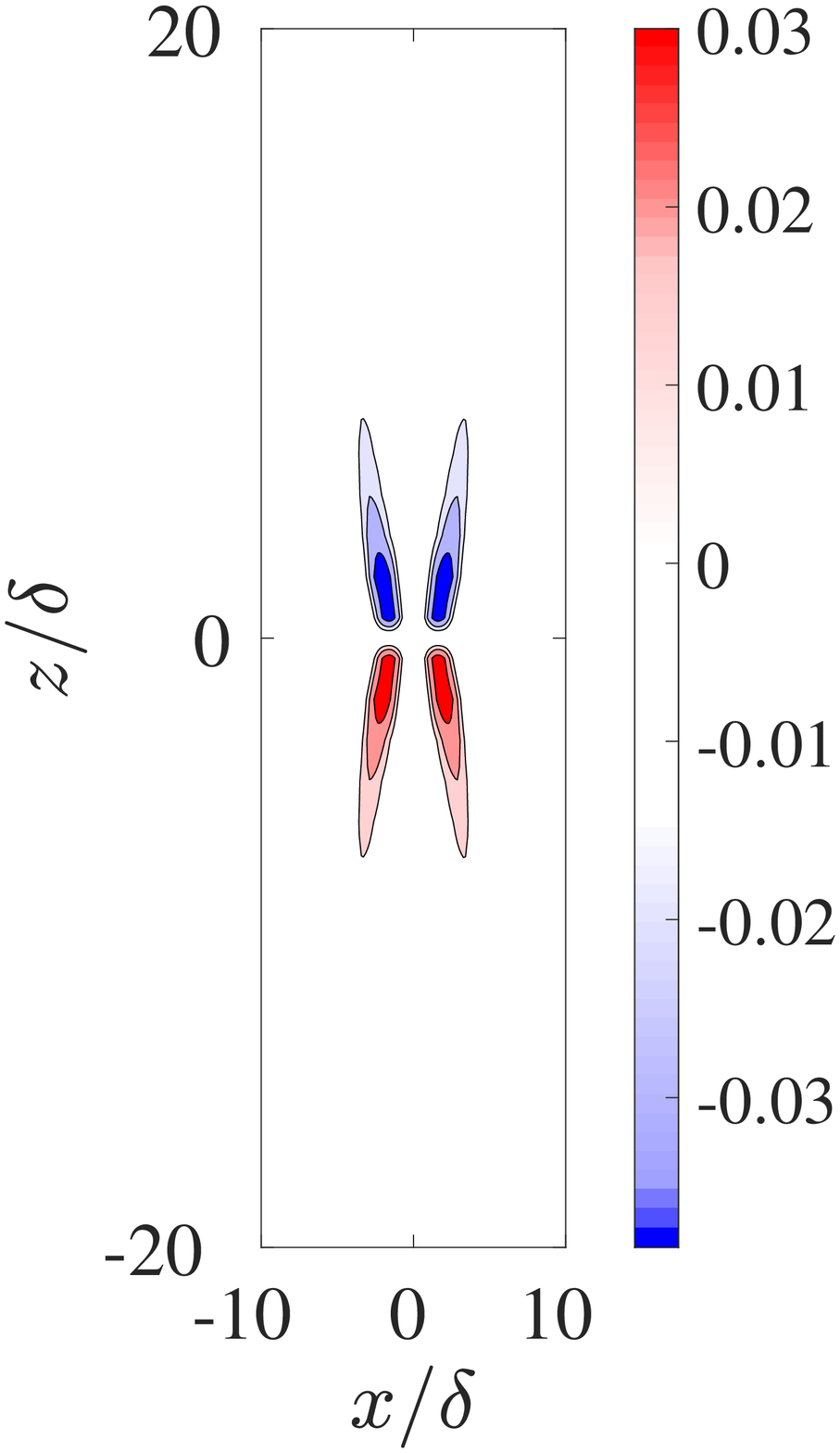}	
	\includegraphics[width=0.28\linewidth,
height=.45\linewidth]{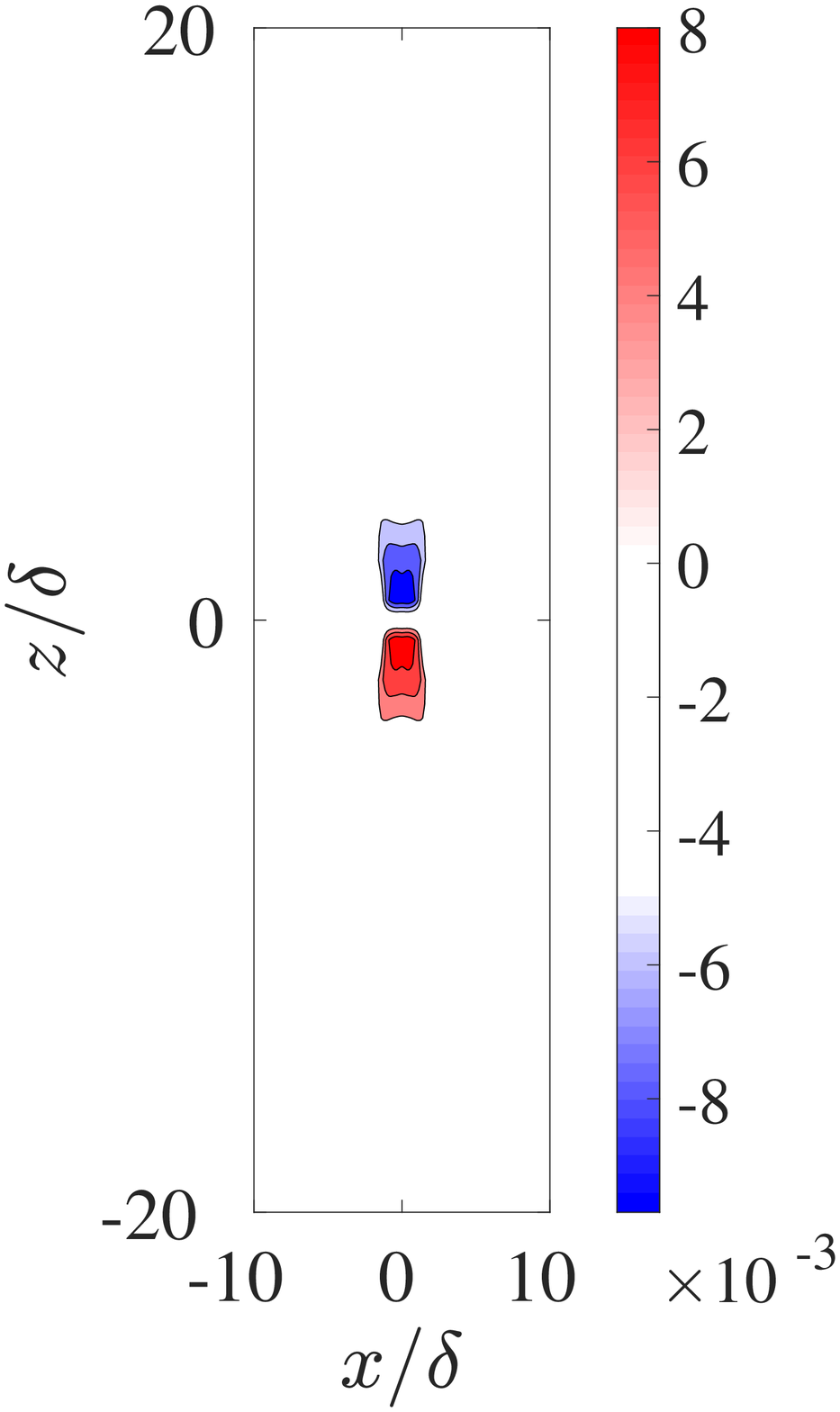}\\
	\caption{Fast MAC wave helicity (a--c), slow MAC wave
helicity (d--f) and $z$-component of the induced magnetic field, $b_z$ 
(g--i) for $|\omega_A/\omega_M| =$ 0.1, 0.6, 0.95 (left to right). 
The plots are generated at time $t/t_\eta=0.01$ for the parameters
 $Le=0.03$ and $E_\eta=2 \times 10^{-5}$.}
	\label{fsbz} 
\end{figure}

Figure \ref{linrev}(a) indicates that both fast and
slow MAC waves propagate along the mean-field direction $x$
such that $x/\delta= t/t_a$, where $t_a$ is the 
Alfv\'en wave travel time. 
{ Contrary
to that in the inertial--Alfv\'en  wave system
discussed in \cite{bardsley2017}, the Alfv\'en waves
propagating in the direction orthogonal to the rotation
axis are simply the degenerate form of the MAC waves
\citep[see also][]{aditya2022}. }
For small $|\omega_A/\omega_M|$,
the helicity of slow wave motions is greater than that
of fast waves, but as $|\omega_A/\omega_M|$ approaches
unity, the slow wave helicity weakens considerably and
falls below that of the fast wave. The effect of increasing
buoyancy forcing on the fast and slow wave helicity is shown
graphically in figure \ref{fsbz}. The fast waves are
practically unaffected by the strength of forcing as
their intensity and 
$z$ propagation rate are nearly invariant for
$|\omega_A/\omega_M|$ in the range 0.1--1 
(figure \ref{fsbz}(a--c)). The slow
wave helicity, on the other hand, is substantially weakened
as $|\omega_A/\omega_M|$ increases
in the same range (figure \ref{fsbz}(d--f)).
For $|\omega_A/\omega_M| \approx 1$, a state
is reached where the slow wave helicity is nearly
zero.
The induced magnetic field $b_z$ also
weakens considerably with increasing
 $|\omega_A/\omega_M|$ (figure \ref{fsbz}(g--i)), which
indicates that only the slow waves have a direct bearing
on field generation. 

\begin{figure}
		\centering
	\includegraphics[width=0.55\linewidth]{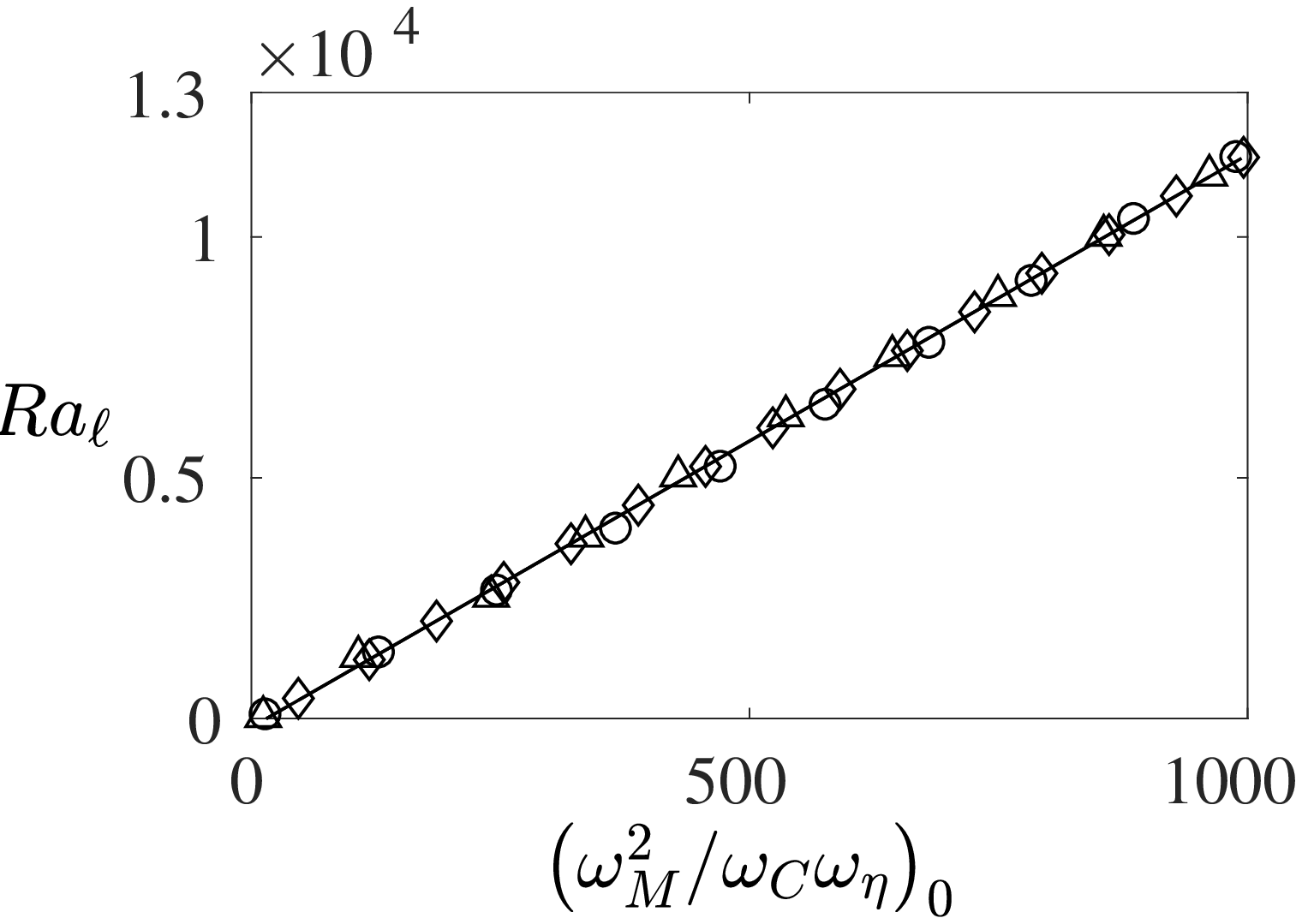}
	\caption{Variation of the local Rayleigh number $Ra_\ell$,
defined in \eqref{radelta}, shown against 
$\varLambda$, defined in \eqref{varl}, for the state of
approximately zero slow MAC wave helicity. Three values
of $E_\eta$ are considered -- 
the diamonds represent $E_\eta=6\times 10^{-7}$,
circles represent $E_\eta=6\times 10^{-6}$ and triangles 
represent  $E_\eta=2\times 10^{-5}$.}
	\label{trans1}
\end{figure}

Figure \ref{linrev}(b) shows the normalized slow wave
helicity against a \lq local' Rayleigh number based on the 
length scale of the initial perturbation,

{{
\begin{equation}
Ra_\ell = \dfrac{g \alpha |\gamma| \delta^2}{2 \varOmega \eta}
\label{radelta}
\end{equation}
}}
as well as $|\omega_A/\omega_M|$. Evidently, the forcing
needed to suppress the slow waves increases with $Le$,
although the total suppression of these waves occurs
universally at $|\omega_A/\omega_M| \approx 1$. This result
prompts us to look at the condition for
vanishing slow wave helicity through a relation between $Ra_\ell$
and a parameter $\varLambda$ defined by
\begin{equation}
\varLambda=\left(\dfrac{\omega_M^2}{\omega_C \omega_\eta}\right)_{\!0}
\sim \dfrac{V_M^2}{2 \varOmega \eta},
\label{varl}
\end{equation}
which measures the initial ratio of the slow MC wave frequency to
the magnetic diffusion frequency. Figure \ref{trans1}
shows that the same linear relation between $Ra_\ell$ and $\varLambda$,
whose values are tabulated in table \ref{lintable},
holds for any $E_\eta$.  
Both $Ra_\ell$ and $\varLambda$
are measurable in dynamo models, the latter being of the same
order of magnitude as the Elsasser number -- the square of the
scaled magnetic field -- in many models.
{ If the state of vanishing slow wave helicity is
taken as a proxy for polarity transitions,
then we may expect a self-similar relationship
between $Ra_\ell$ and $\varLambda$
in low-inertia dynamos, where the nonlinear
inertial force is small compared with the Coriolis force.}
This
idea is explored further in \S \ref{nonlinear}.

\begin{table}
	\centering
	\begin{tabular}{lll|lll|lll}
		\multicolumn{3}{c|}{ $E_\eta=6\times10^{-7}$}
		&\multicolumn{3}{c|}{ $E_\eta=6\times10^{-6}$ }
		&\multicolumn{3}{c}{ $E_\eta=2\times10^{-5}$ }\\
&&&&&&&&\\
		$Le$ & $\varLambda$ & $Ra_\ell$ & $Le$ & $\varLambda$ & $Ra_\ell$ & $Le$ &$ \varLambda$ & $Ra_\ell$\\
		&&$\times10^{4}$&&&$\times10^{4}$&&&$\times10^{4}$\\
		0.0071&47&0.0380&0.0120  & 13.67  & 0.0097 &0.0200 &11.87&0.0039   \\
		0.0111&118&0.1227&0.0366  & 127.49 & 0.1383 &0.0601 &107.33&0.1258  \\
		0.0139&185&0.2029&0.0509  & 246.04 & 0.2669 &0.0901 &240.68&0.2538  \\
		0.0162&253&0.2832&0.0620  & 365.39 & 0.3956 &0.1063 &335.34&0.3787   \\	
		0.0201&338&0.4437&0.0704  & 470.60 & 0.5242 &0.1201 &428.41&0.5036   \\	
		0.0233&523&0.6042&0.0779  & 575.81 & 0.6528 &0.1344 &536.45&0.6286   \\
		0.0262&658&0.7647&0.0846  & 680.00 & 0.7814 &0.1472 &653.35&0.7535   \\	
		0.0287&793&0.9252&0.0908  & 782.69 & 0.9100 &0.1589 &749.42&0.8784  \\
		0.0311&928&1.0862&0.0966  & 885.38 & 1.0386 &0.1698 &858.50&1.0035   \\	
		0.0322&996&1.1335&0.1020  & 988.07 & 1.1672 &0.1800 &961.57&1.1283   \\
	\end{tabular}
	\caption{Values of $Ra_\ell$, defined in \eqref{radelta}, at
different $\varLambda$, defined in \eqref{varl}, for suppression
of slow MAC waves in the linear magnetoconvection calculations. 
The dimensionless parameters $Le$ and $E_\eta$ are defined
in (\ref{pars}a,b).}
	\label{lintable}
\end{table}

\section{Nonlinear dynamo simulations}
\label{nonlinear}

We consider a convection-driven dynamo
operating in a spherical shell, the boundaries
of which  correspond to the inner core
boundary (ICB) and the core--mantle boundary (CMB). 
The ratio of inner to outer radius is
0.35. Fluid motion is driven by 
thermal buoyancy, although our formulation can also 
study thermochemical buoyancy using
the codensity formulation \citep{95brag}. The other
body forces acting on the fluid are the 
electromagnetic Lorentz force 
and the Coriolis force.
Lengths are scaled by the thickness of the spherical shell $L$ 
and time is scaled by magnetic diffusion time 
$L^2/\eta$.
 The velocity $\bm{u}$ and magnetic field 
$\bm{B}$ are scaled by $\eta/L$ and $(2\Omega\rho\mu\eta)^{1/2}$
 respectively. 
The temperature is scaled by $\beta L$, where $\beta$ is the 
radial
temperature gradient at the outer boundary.
In the Boussinesq approximation, the non-dimensional 
MHD equations for the velocity, 
magnetic field and temperature are given by,	

	\begin{align}
E Pm^{-1}  \Bigl(\frac{\partial {\bm u}}{\partial t} + 
(\nabla \times {\bm u}) \times {\bm u}
\Bigr)+  {\hat{\bm{z}}} \times {\bm u} = - \nabla p^\star +
Ra \, Pm Pr^{-1} \, T \, {\bm r} \,  \nonumber\\ +  (\nabla \times {\bm B})
\times {\bm B} + E\nabla^2 {\bm u}, \label{momentum} \\
\frac{\partial {\bm B}}{\partial t} = \nabla \times ({\bm u} \times {\bm B}) 
+ \nabla^2 {\bm B},  \label{induction}\\
\frac{\partial T}{\partial t} +({\bm u} \cdot \nabla) T =  Pm Pr^{-1} \,
\nabla^2 T,  \label{heat1}\\
\nabla \cdot {\bm u}  =  \nabla \cdot {\bm B} = 0,  \label{div}
\end{align}
	
The modified pressure $p^*$  in equation \eqref{momentum}
 is given by $p+ \frac1{2} E \, Pm^{-1} \, |\bm{u}|^2$.
The dimensionless parameters in the above equations 
are the Ekman number $E=\nu/2\varOmega L^2$,
the Prandtl number, $Pr=\nu/\kappa$, 
the magnetic Prandtl number, $Pm=\nu/\eta$ and
the modified Rayleigh number $g \alpha \beta L^2/2 \Omega \kappa$. 
Here, $g$ is the gravitational acceleration, 
$\nu$ is the kinematic viscosity, 
$\kappa$ is the thermal diffusivity and $\alpha$ 
is the coefficient of thermal expansion.

The
basic-state temperature profile represents a basal heating
given by $T_0(r) = r_i r_o/r$, where $r_i$ and $r_o$ are the
inner and outer radii of the spherical shell.
The velocity and magnetic fields satisfy the no-slip and electrically 
insulating conditions respectively at the two boundaries. The inner
boundary is isothermal while the outer boundary has constant heat flux.
The calculations are performed by a pseudospectral 
code that uses
spherical harmonic expansions in the angular coordinates $(\theta,\phi)$
and finite differences in radius $r$ \citep{07willis}.

As in recent studies \citep{aditya2022}, the 
dynamo simulations
begin from a dipole seed magnetic field of small
volume-averaged intensity $\bar{B} =$ 0.01.
 The runs are performed for at least 2 magnetic
diffusion times, well into the saturated state of the dynamo.
The main output parameters of the dynamo
simulations, given in table \ref{parameters},
are time-averaged values in the saturated state.
For three values of the Ekman number $E$, a series
of simulations at progressively increasing Rayleigh number 
$Ra$ are performed, 
spanning the dipole-dominated regime up to
the onset of polarity transitions. 
The mean spherical harmonic degrees for convection and
	energy injection are defined by
	\begin{equation}
	l_{c}= \dfrac{\Sigma \hspace{1pt} l \hspace{1pt}
	E_{k}(l)}  {\Sigma \hspace{1pt} E_{k}(l)}; \hspace{5pt}
	l_{E}= \dfrac{\Sigma \hspace{1pt} l \hspace{1pt}
	E_{T}(l)}{\Sigma \hspace{1pt} E_{T}(l)},
	\label{elldef}
	\end{equation}
	where $E_{k}(l)$ is the kinetic energy spectrum and
	$E_{T}(l)$ is the spectrum obtained from the product of 
the transform of $u_{r}T$
	and its conjugate. 
{  The total kinetic and magnetic energies in the saturated
dynamo are given by the volume integrals
\begin{equation}
E_k= \dfrac{1}{2} \int \bm{u}^2 \mbox{d}V; 
\quad E_m= \dfrac{Pm}{2E} \int \bm{B}^2 \mbox{d}V.
\label{energies}
\end{equation}
The relative dipole field strength $f_{dip}$, 
which is the ratio of the mean dipole field
strength to the field strength in harmonic degrees 
$l =$ 1--12 at the outer boundary
\citep{chraub2006}, takes values $>0.5$ in 
all the dipole-dominated runs
(table \ref{parameters}.) 
The distinction between multipolar and
polarity-reversing runs, 
however, is well understood from the evolution of the
dipole colatitude, presented in \S \ref{revmult}. }

For each
$E$, the value of $Pm=Pr$
is chosen such that the local Rossby number $Ro_\ell$,
which gives the ratio of the inertial to Coriolis
forces on the characteristic length scale of
convection \citep{chraub2006} is $<0.1$
(table \ref{parameters}). Therefore,
our dynamo simulations lie in the rotationally
dominant, or low-inertia, regime. {  While low-$Pm$
simulations starting from a seed field can result in
multipolar fields even for $Ro_\ell <0.1$ \citep{petitdemange2018}, our choice
of $Pm$ ensures that the low-inertia dynamos are dipole dominated
regardless of whether one begins from a seed field
or a strong field. The dipole dominance exists
for a wide range of forcing, up to $Ra/Ra_c \sim 10^3$, where
$Ra_c$ is the critical Rayleigh number at the onset of nonmagnetic
convection. The only exception here is the last run in each
Ekman number series (marked by an asterisk in table \ref{parameters}),
where a multipolar or polarity-reversing dynamo is obtained
depending on the initial field strength (see \S \ref{revmult} below).

The role of the magnetic field in helicity generation is well
understood by comparing a dynamo calculation with its
equivalent nonmagnetic convection calculation. For $Pm=Pr$,
the dynamo obtained by solving equations \eqref{momentum}--\eqref{div}
is compared with its nonmagnetic counterpart, obtained by solving
equations \eqref{mom1}--\eqref{div1}, Appendix \ref{nmeqns}. }
\subsection{Dipolar, multipolar and polarity-reversing dynamos}
\label{revmult}
\begin{figure}
	\centering
	\hspace{-2 in}	(a)  \hspace{1.6 in} (b)  \hspace{1.6 in} (c) \\%
	\hspace{-0.26 in} 	\includegraphics[width=0.32\linewidth]
	{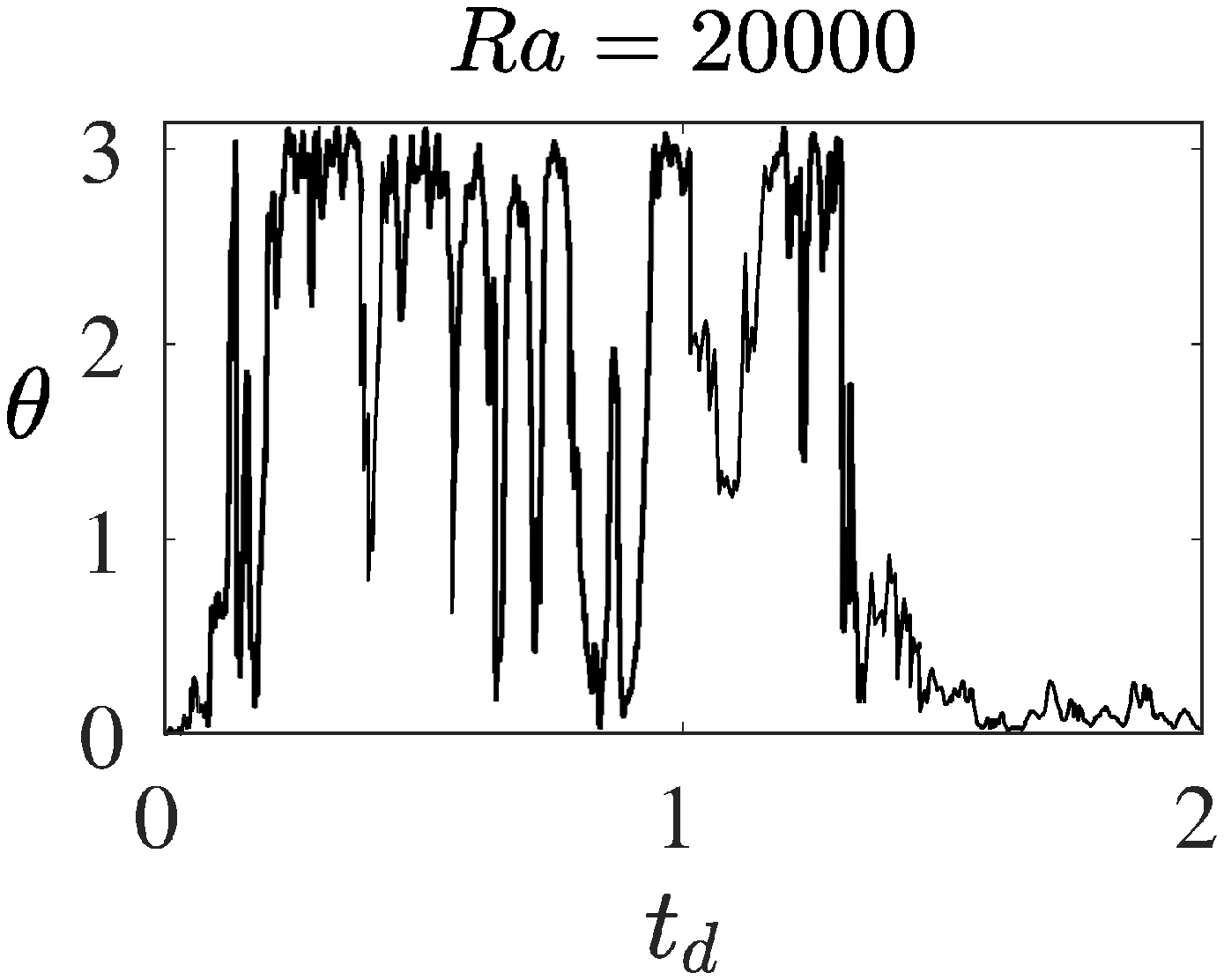}\hspace{-0.05 in} 
	\includegraphics[width=0.32\linewidth]{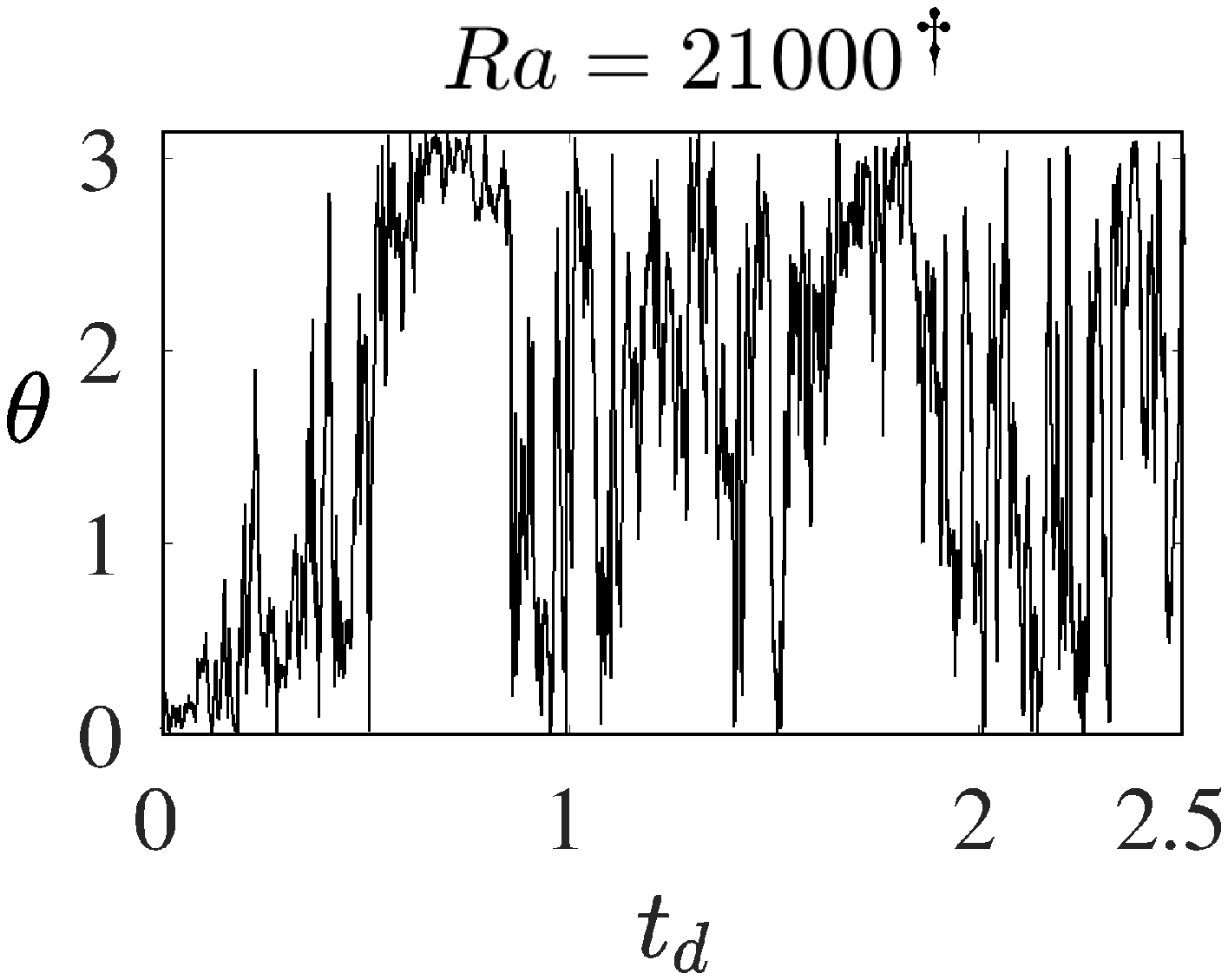}\hspace{-0.05 in} 
	\includegraphics[width=0.32\linewidth]{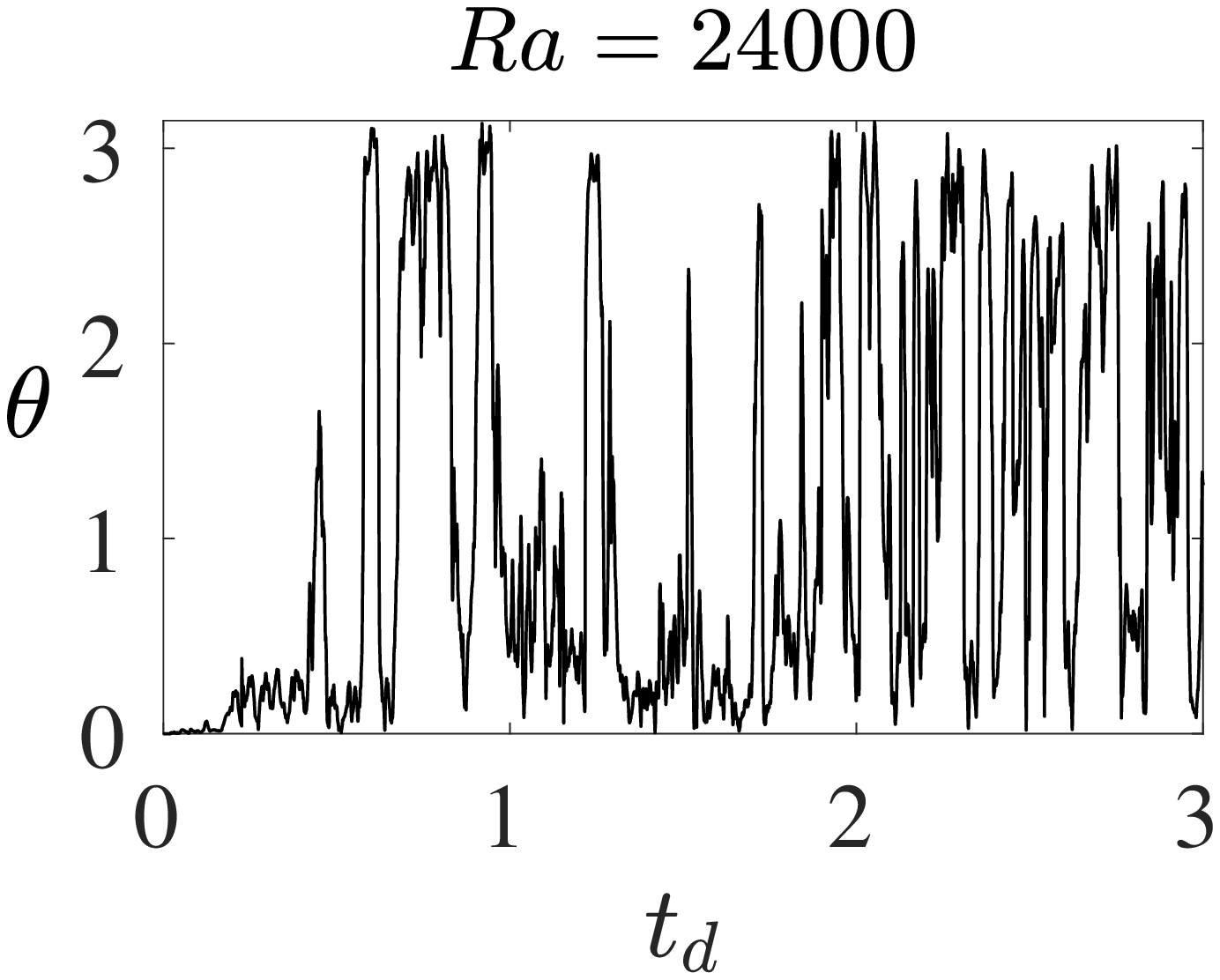}\\
	\hspace{-2 in}	(d)  \hspace{1.6 in} (e)  \hspace{1.6 in} (f) \\%
	\hspace{-0.26 in} 	\includegraphics[width=0.32\linewidth]
	{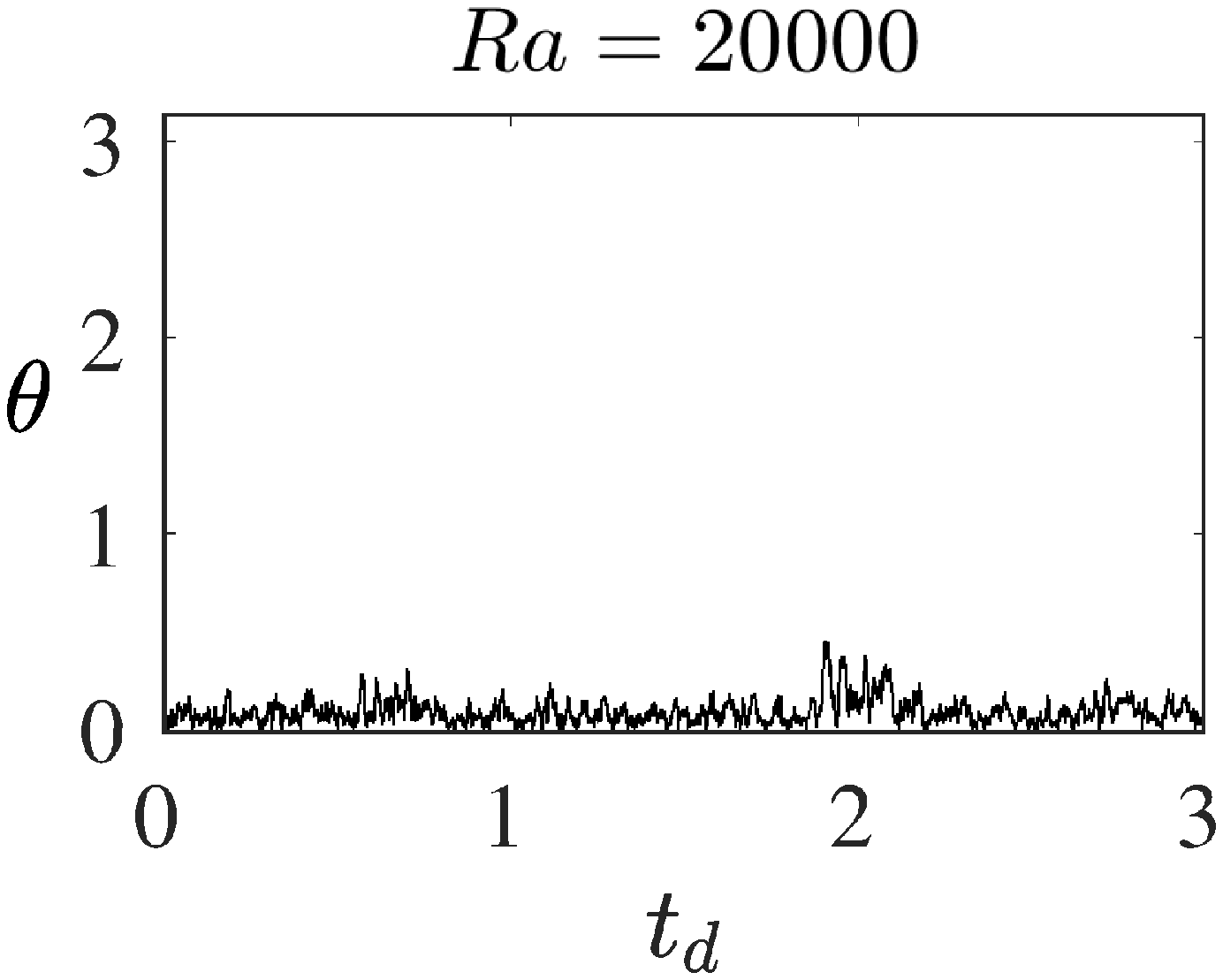}\hspace{-0.05 in} 
	\includegraphics[width=0.32\linewidth]{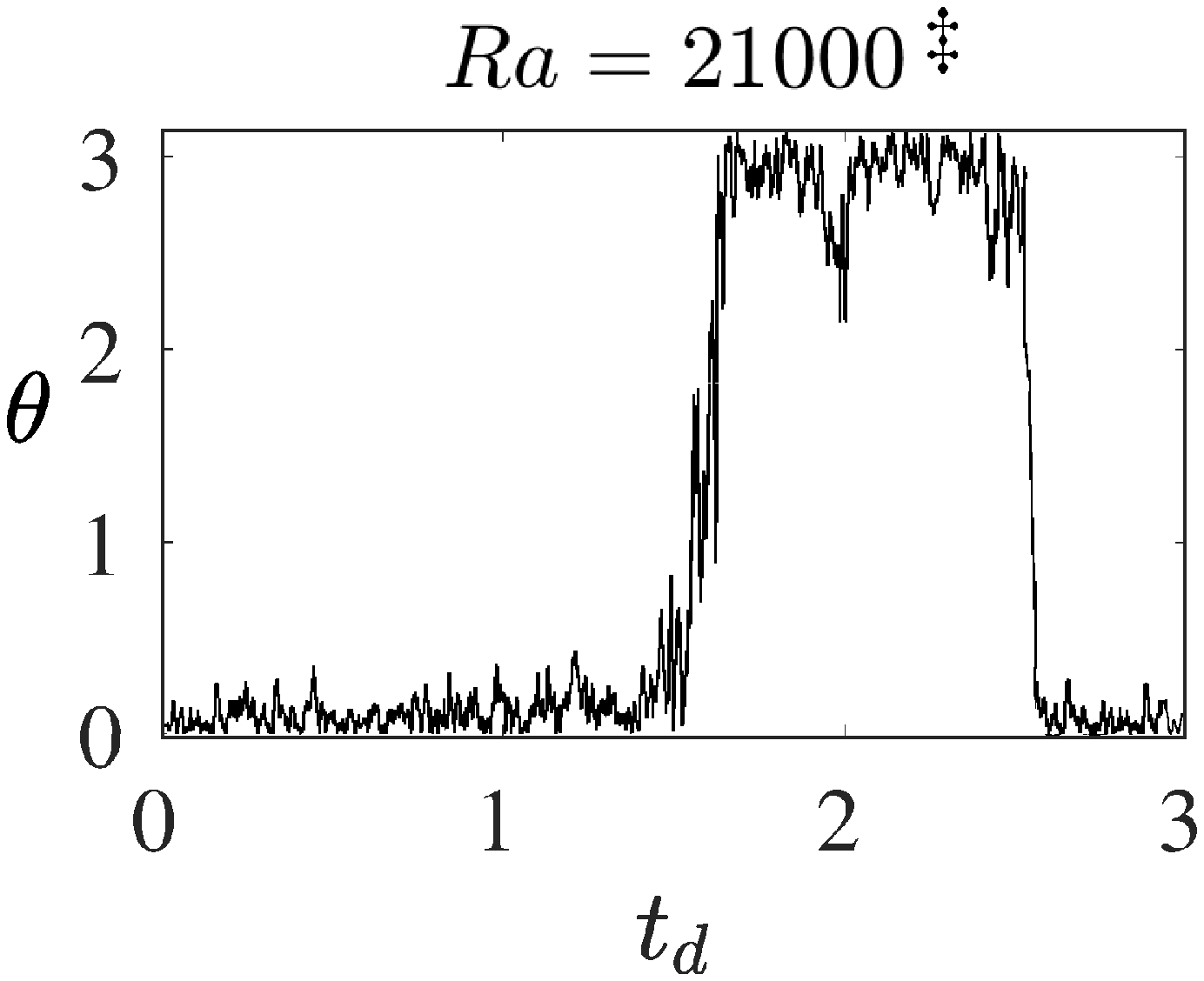}\hspace{-0.05 in} 
	\includegraphics[width=0.32\linewidth]{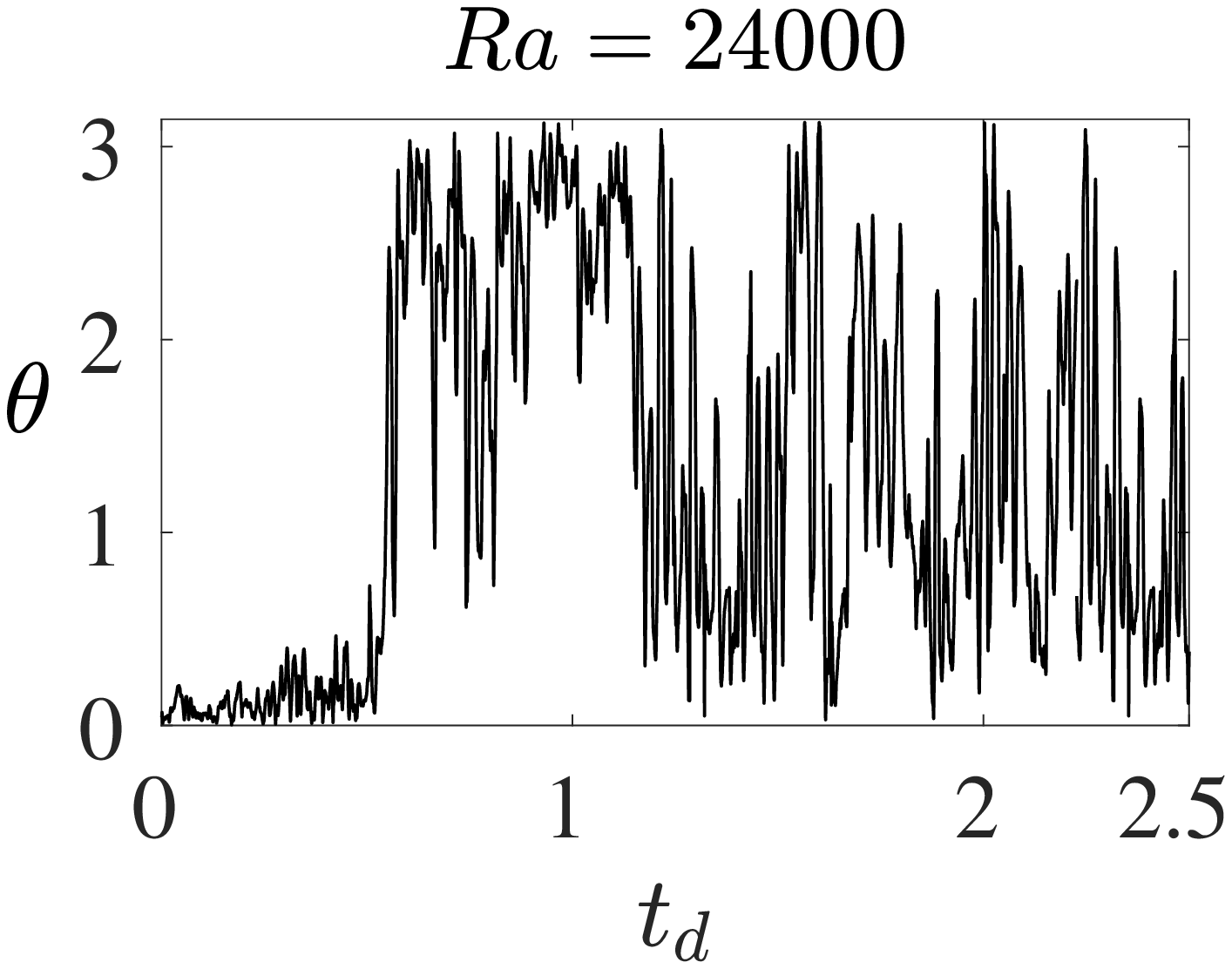}\\
	\caption{ { 
Evolution of dipole colatitude with magnetic diffusion time
for three closely spaced values of $Ra$ in strongly driven dynamo
simulations. 
The top panels (a)--(c) are for
 simulations  starting from a seed magnetic field while the bottom panels
(d)--(f) are for simulations starting from a strong magnetic field. 
For $Ra=21000$, the superscript $\dag$ denotes a seed field start
whereas the superscript $\ddag$ denotes a strong field start.
The other dynamo parameters are $E = 6 \times 10^{-5},Pm=Pr=5$. }
}
	\label{tilt}
\end{figure}

\begin{figure}
	\centering
	\hspace{-2.5 in}	(a)  \hspace{2.5 in} (b) \\
	\includegraphics[width=0.48\linewidth]{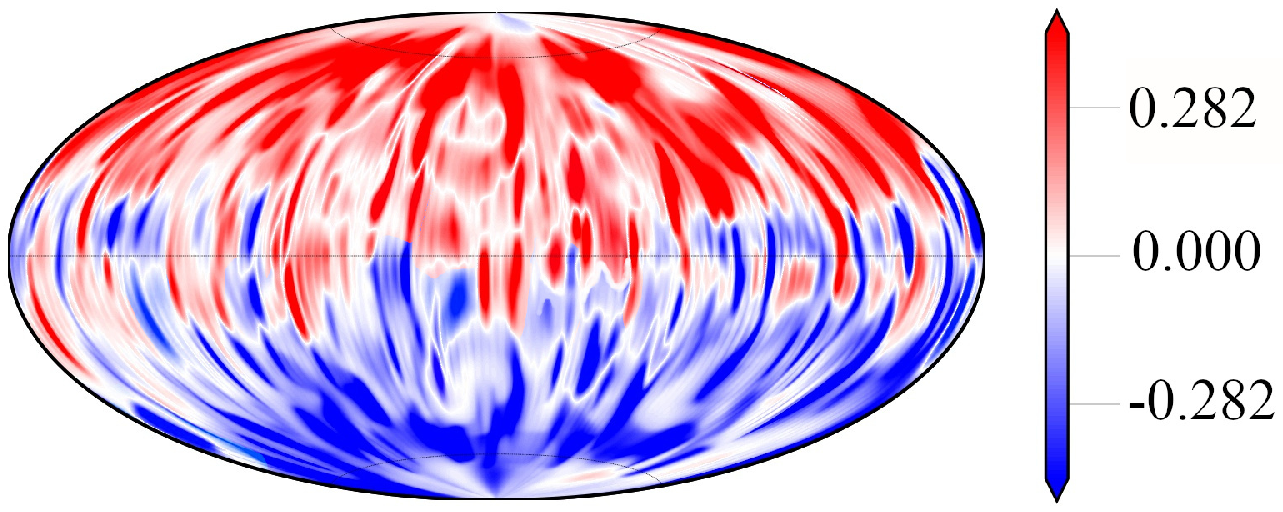}
	\includegraphics[width=0.48\linewidth]{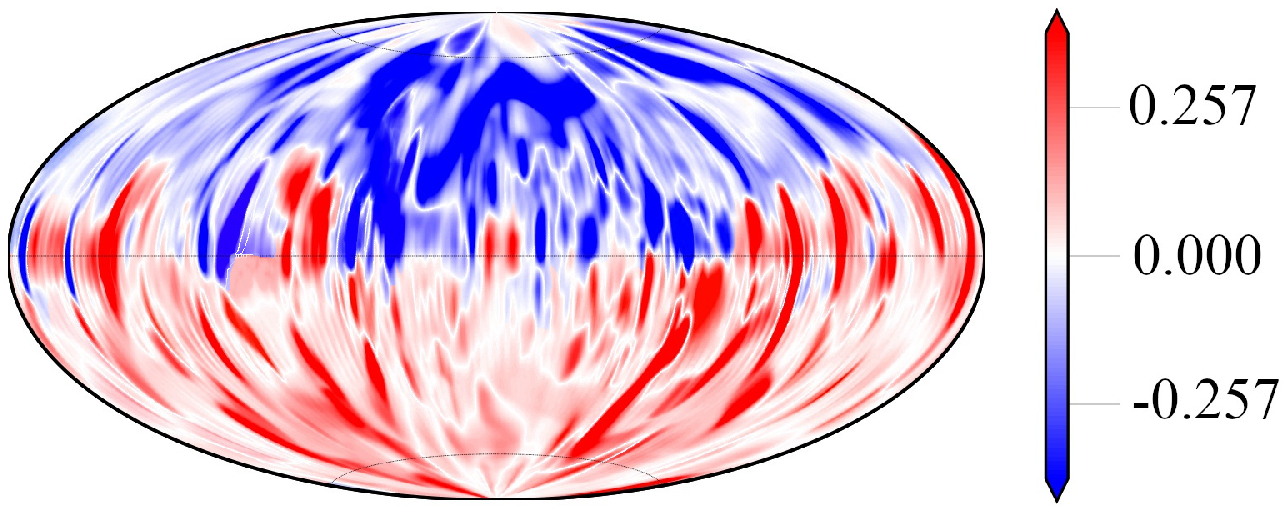}
	\caption{The contours of the radial magnetic field 
at the outer boundary for $Ra=21000$ at magnetic diffusion
		times (a) $t_d=1.1$ and (b) $t_d=2.3$ 
shown in figure \ref{tilt}(e).  The other
		dynamo parameters are $E = 6 \times 10^{-5}$,
$Pm=Pr=5$.}
	\label{br}
\end{figure}

Figure \ref{tilt} shows the magnetic colatitude
of the
dipole field, $\theta$ at the upper boundary obtained from spherical
harmonic Gauss coefficients, as follows:
\begin{equation}
\cos\theta = g_1^0/|\bm{m}|, \quad \bm{m}= (g_1^0, g_1^1, h_1^1),
\label{diplat}
\end{equation}
{ where $g_1^0$, $g_1^1$ and $h_1^1$ are  obtained from the 
 Schmidt-normalized 
expansion for the scalar potential of the field
\cite[][pp.142--143]{glatz2013}.
 For $E=6 \times 10^{-5}$ and $Pm=Pr=$5, the evolution
of $\theta$ is shown
for runs at three closely spaced values of $Ra$ in strongly
supercritical convection. For $Ra=20000$,  the run
that begins from a dipole seed field enters a 
chaotic multipolar state and
subsequently regains dipolarity (figure \ref{tilt}(a)), a behaviour
first noted by \cite{prf18}. 
For the same $Ra$, the run that begins
from the saturated (strong field of mean square
intensity $B^2 = O(1)$) state of $Ra=18000$ gives
a stable dipole field throughout (figure \ref{tilt}(d)). 
Both these runs saturate at
nearly identical dipolar 
states. For $Ra=21000$, the
run that begins from a seed field is multipolar 
throughout (figure \ref{tilt}(b)). At the same $Ra$,
the run that begins from a strong field
produces occasional polarity reversals
separated by well-defined periods of normal and reversed
polarity (figure \ref{tilt}(e)), as in earlier reversing
simulations \citep{glatz1999,kutzner2002} and
experiments \citep{monchaux2009}. The radial magnetic
fields at the outer boundary before and after a reversal are
shown in figure \ref{br}.
For still higher forcing ($Ra=24000$), both seed field and strong field runs give
multipolar solutions (figure \ref{tilt}(c) \& (f)). Therefore,
polarity reversals happen in a narrow range of $Ra$
that lies between dipolar and multipolar regimes, possibly due
to variations in outer boundary heat flux. }

\begin{figure}
\centering
\hspace{-2.5 in}	(a)  \hspace{2.5 in} (b) \\
 \includegraphics[width=0.48\linewidth]{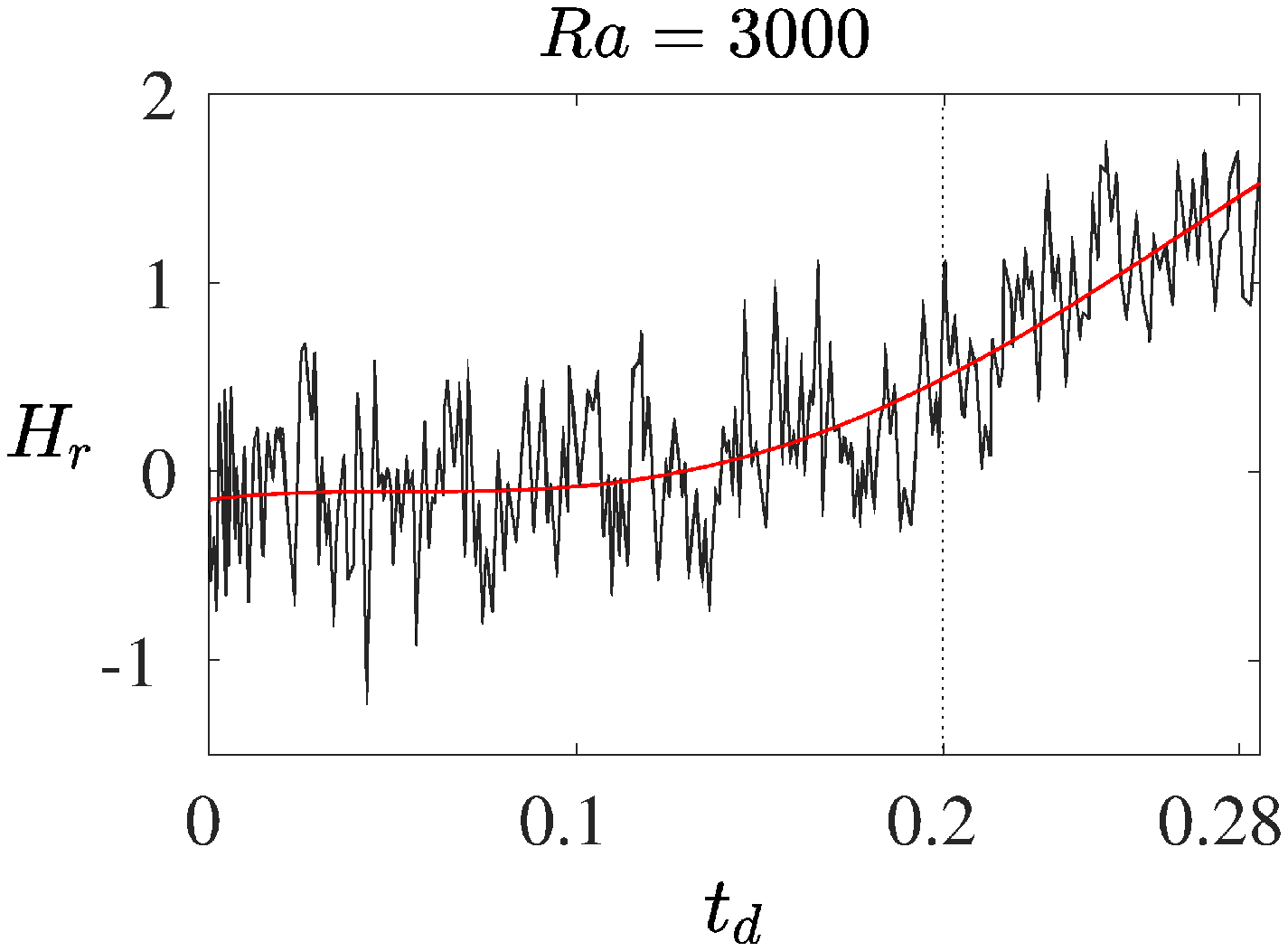}
\includegraphics[width=0.48\linewidth]{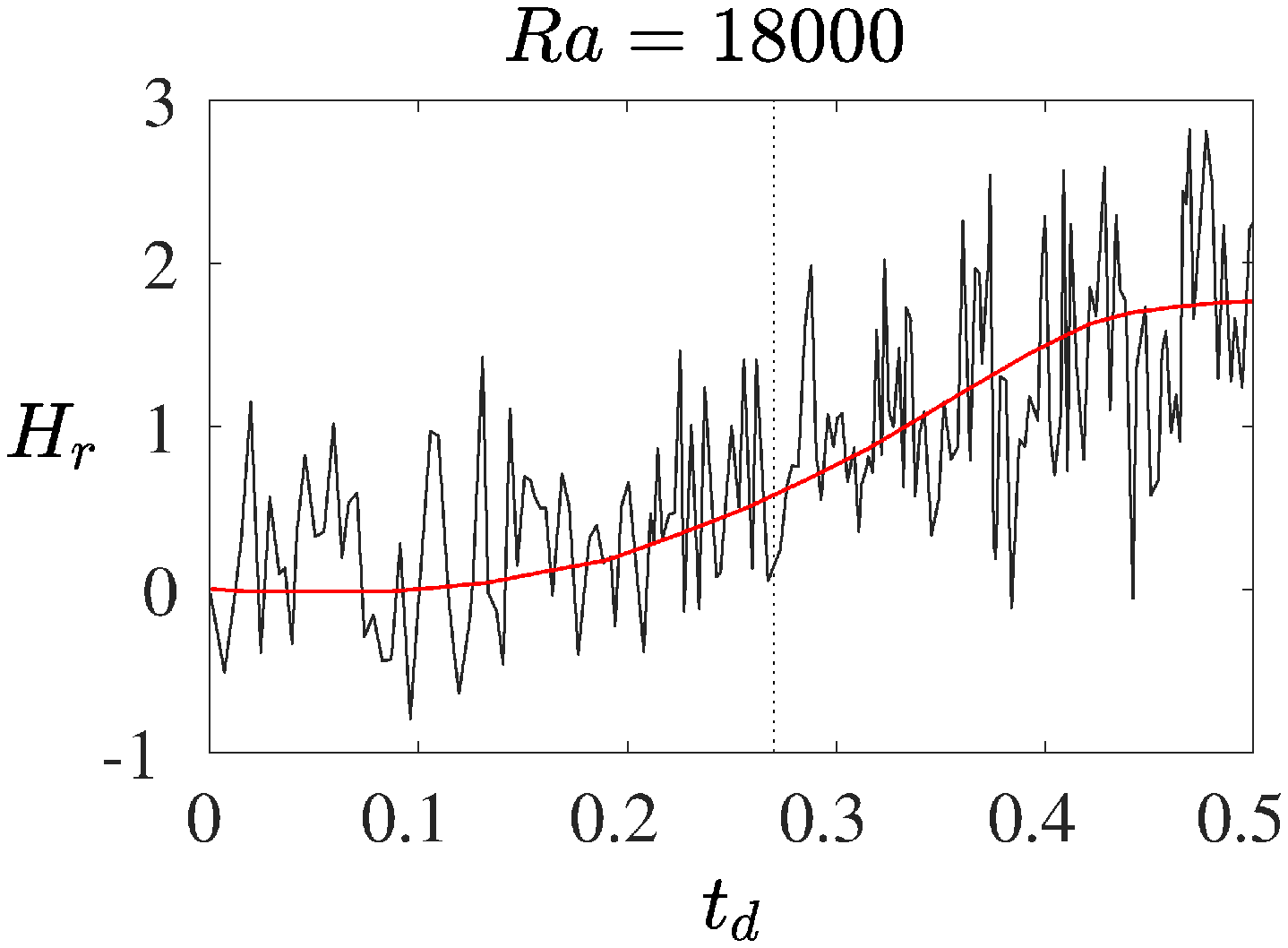}\\
\hspace{-2.5 in}	(c)  \hspace{2.5 in} (d) \\
 \includegraphics[width=0.48\linewidth]{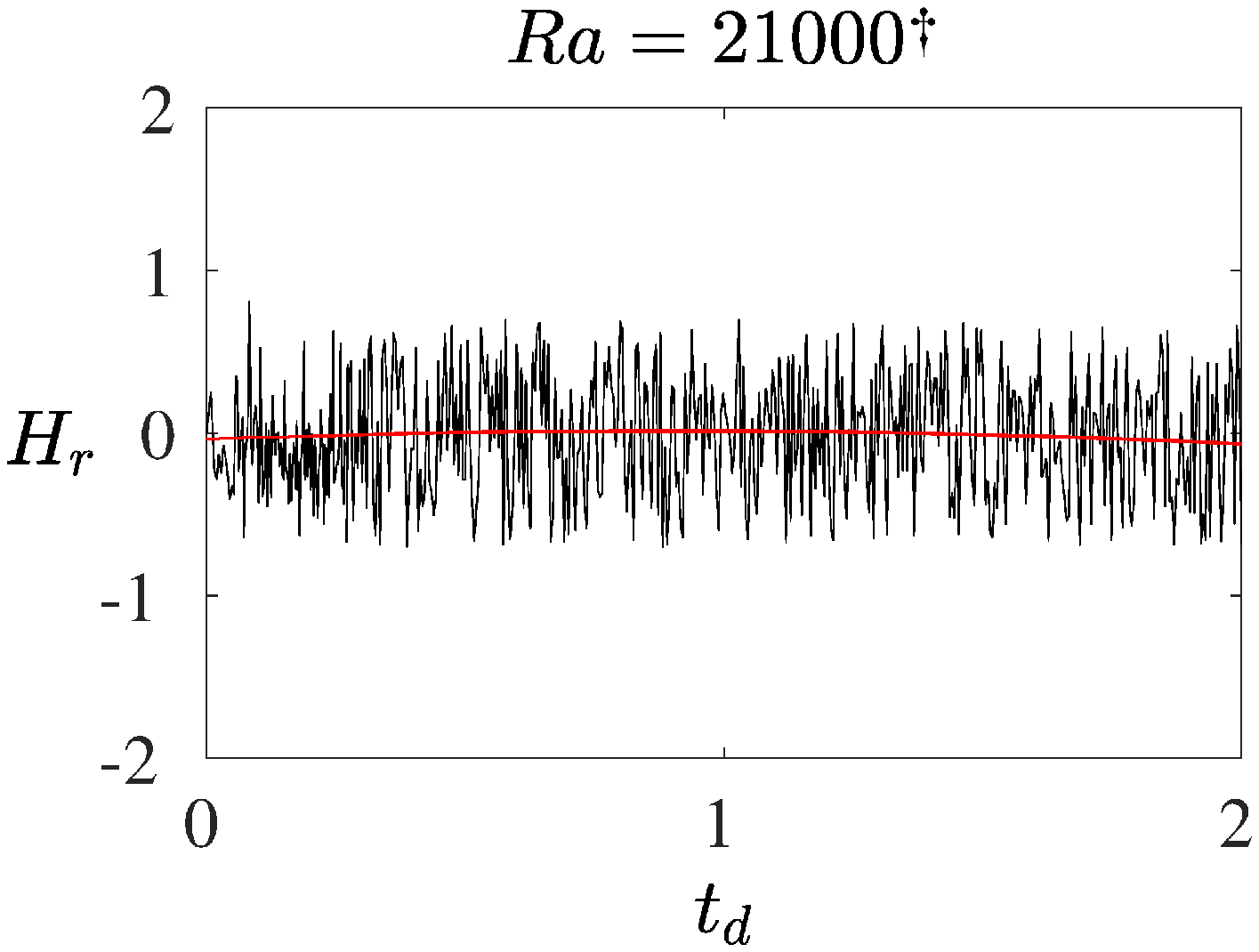}
\includegraphics[width=0.48\linewidth]{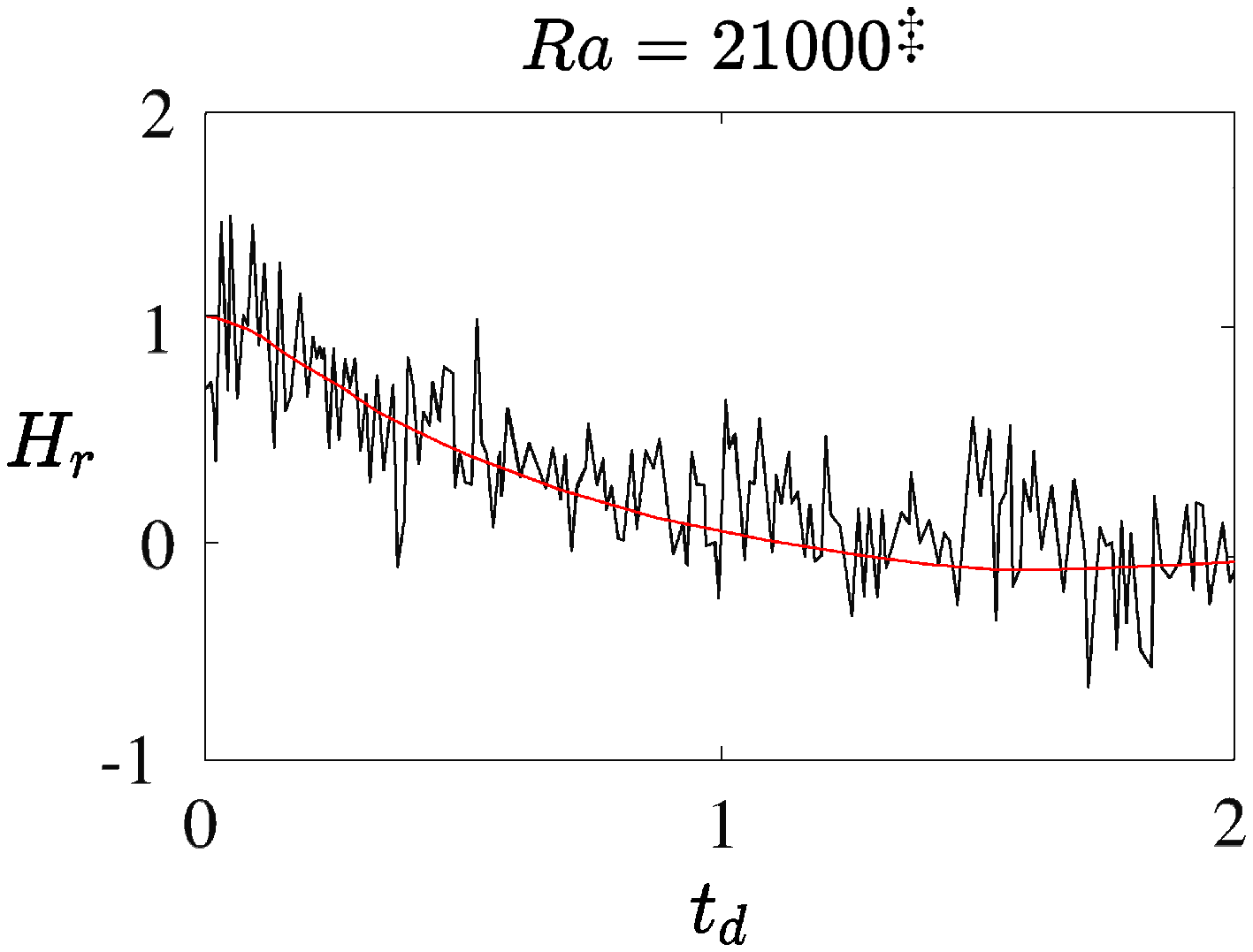}\\
	\caption{ { Relative helicity, $H_r$, as defined in \eqref{rhel},
in dynamo simulations for
 (a) $Ra=3000$ (seed field start),
(b) $Ra=18000$ (seed field start),
(c) $Ra=21000^\dag$ (seed field start),
(d) $Ra=21000^\ddag$ (strong field start).}
The superposed red lines are polynomial fits showing the trend
of the evolution. The dotted vertical lines in (a) and (b)
mark the approximate times of formation of the axial dipole
from a multipolar field at the outer boundary in the run.
The other dynamo parameters are $E = 6 \times 10^{-5}$, $Pm=Pr=5$.}
	\label{relhel}
\end{figure}
The range of spherical harmonic
degrees $l \leq l_E$ is of particular interest since 
kinetic helicity is known to be generated in the nonlinear dynamo
in this range of energy-containing scales (see \S \ref{intro}). 
A relative helicity is defined that measures the
augmentation of lower-hemisphere 
helicity in the nonlinear dynamo (magnetic)
run relative to that in the equivalent nonmagnetic simulation.
\begin{equation}
H_r = \dfrac{h_{m} - h_{nm}}{h_{nm}},
\label{rhel}
\end{equation}
for $l \leq l_E$. Here, the subscripts $m$ and $nm$ denote
the magnetic (dynamo) and nonmagnetic values respectively.
{ The variation of $H_r$ with time in dynamos at
$E=6 \times 10^{-5}$ and progressively increasing $Ra$
is given in figure \ref{relhel}. 
The runs that begin from a seed field and produce an axial dipole
show an approximately two-fold increase in helicity
(figures \ref{relhel}(a) \& (b)). For $Ra=21000$, the run that begins
 from a seed field and produces a multipolar solution
shows no noticeable
increase in helicity over the nonmagnetic 
state (figure \ref{relhel}(c)).
At the same $Ra$, the run that begins from a 
strong field and produces polarity reversals
undergoes a decrease in helicity from its initial value
to the nonmagnetic value (figure \ref{relhel}(d)). 
We emphasize that the effect of
the magnetic field on helicity generation can be experienced only
in the energy-containing range of harmonic degrees
 $l \leq l_E$. 
If the entire spectrum is considered
\citep{ranjan2020etal}, the dipolar dynamo helicity
would be lower than the nonmagnetic helicity.
 
The results in figure \ref{tilt} and figure \ref{relhel}  prompt us
to examine the nature of wave motions in dipolar
and reversing dynamos. An important aim
of this study is to show through the analysis of MAC waves that
 the formation of the dipole 
from a multipolar state and the onset of polarity
reversals lie in similar regimes. Such an
argument is essential to place a constraint 
on the Rayleigh number that admits polarity reversals
(see \S 3.3).} 

\subsection{The suppression of slow MAC waves in 
polarity-reversing dynamos}
\label{smac}
 
Isolated density disturbances in a rotating stratified fluid layer excite 
MAC waves whose frequencies depend on the 
fundamental frequencies $\omega_M$, $\omega_A$ and $\omega_C$,
representing Alfv\'en waves, internal gravity waves and linear inertial
waves respectively. In unstable stratification that drives planetary
core convection, $\omega_A^2 <0$, where $|\omega_A|$
 is simply a measure of the strength of buoyancy. 
 Since the dimensional frequencies $\omega_M^2$, $-\omega_A^2$ and
$\omega_C^2$ in the dynamo are given by \citep{aditya2022}
\begin{equation} 
\omega_{M}^2 =\frac{(\bm{B} \cdot \bm{k})^2}{\mu \rho},\quad	
-\omega_{A}^2= g\alpha\beta\left(\frac{ k_{z}^2+ k_{\phi}^2}{k^2}\right),\quad
\omega_{C}^2 = \frac{4(\bm{\varOmega} \cdot \bm{k})^2}{ k^2},\quad  
\label{om1}
\end{equation}
and scaling the frequencies by $\eta/L^2$,
we obtain in dimensionless units,
\begin{equation}
\omega_M^2 = \frac{Pm}{E} (\bm{B} \cdot {\bm k})^2, \quad 
-\omega_A^2 = \frac{Pm^2 Ra}{Pr \,E} \, \left(\frac{{k_z}^2
+ k_{\phi}^2}{k^2}\right), \quad
\omega_C^2 = \frac{Pm^2}{E^2} \frac{k_z^2}{k^2},
\label{om2}
\end{equation}	
where $k_s$, $k_{\phi}$ and $k_z$ are the radial, azimuthal and
axial wavenumbers in cylindrical
coordinates $(s,\phi,z)$,   {$k_\phi=m/s$, where $m$ is the
spherical harmonic degree, and $k^2=k_s^2+k_\phi^2+k_z^2$}. Here,
$\omega_A$ is evaluated on the
equatorial plane where the buoyancy force is maximum.
The magnetic (Alfv\'en) wave frequency 
$\omega_M$ is based on the three components of the measured magnetic
field at the peak-field location.  The wavenumber $k_\phi$ is evaluated
at $s=1$, approximately mid-radius of the spherical shell.

In the limit of zero magnetic diffusion ($\omega_\eta=0$), the characteristic
equation \eqref{maineqn} reduces to
\begin{equation}
\lambda^4-\lambda^2(\omega_A^2+\omega_C^2+2\omega_M^2)+\omega_A^2\omega_M^2+\omega_M^4=0,
\label{char2nd}
\end{equation}
the roots of which are \citep{jfm21},
\begin{eqnarray}
\lambda_{1,2} = \pm\frac{1}{\sqrt{2}} \sqrt{\omega_A^2+\omega_C^2+2\omega_M^2+\sqrt{\omega_A^4+2\omega_A^2\omega_C^2+4\omega_M^2\omega_C^2+\omega_C^4}},\label{ndr1}\\
\lambda_{3,4} = \pm\frac{1}{\sqrt{2}} \sqrt{\omega_A^2+\omega_C^2+2\omega_M^2-\sqrt{\omega_A^4+2\omega_A^2\omega_C^2+4\omega_M^2\omega_C^2+\omega_C^4}}.
\label{ndr2}
\end{eqnarray}
For the inequality $|\omega_C| > |\omega_M| > |\omega_A|$,
$\lambda_{1,2}$ and $\lambda_{3,4}$ give the frequencies of the 
fast ($\omega_f$) and slow ($\omega_s$) MAC waves respectively.
While the fast waves
are linear inertial waves weakly modified by the magnetic
field and buoyancy, the slow waves are magnetostrophic.
\begin{figure}
	\centering
\hspace{-2.5 in}	(a)  \hspace{2.5 in} (b) \\
\includegraphics[width=0.45\linewidth]{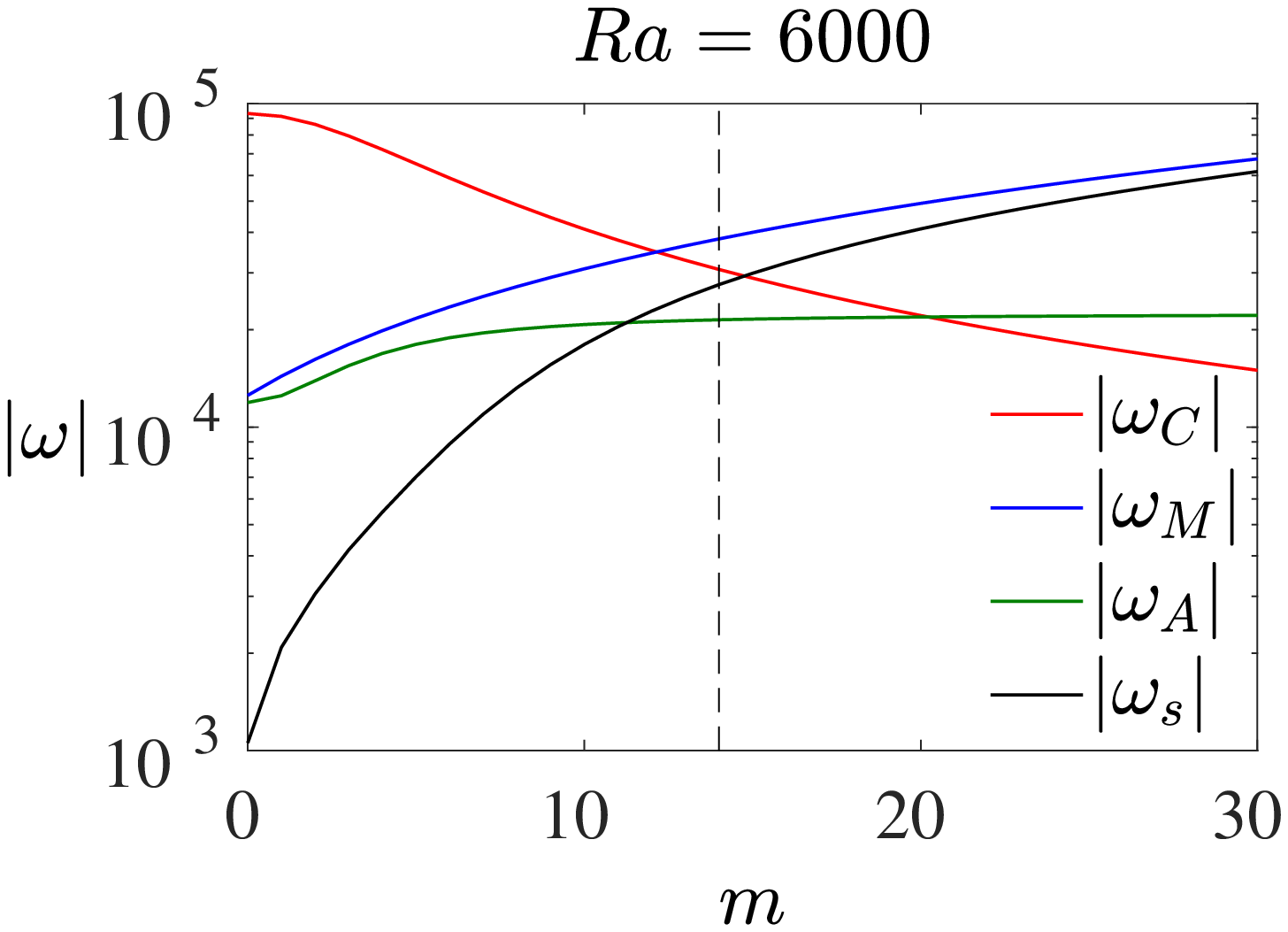}
\includegraphics[width=0.45\linewidth]{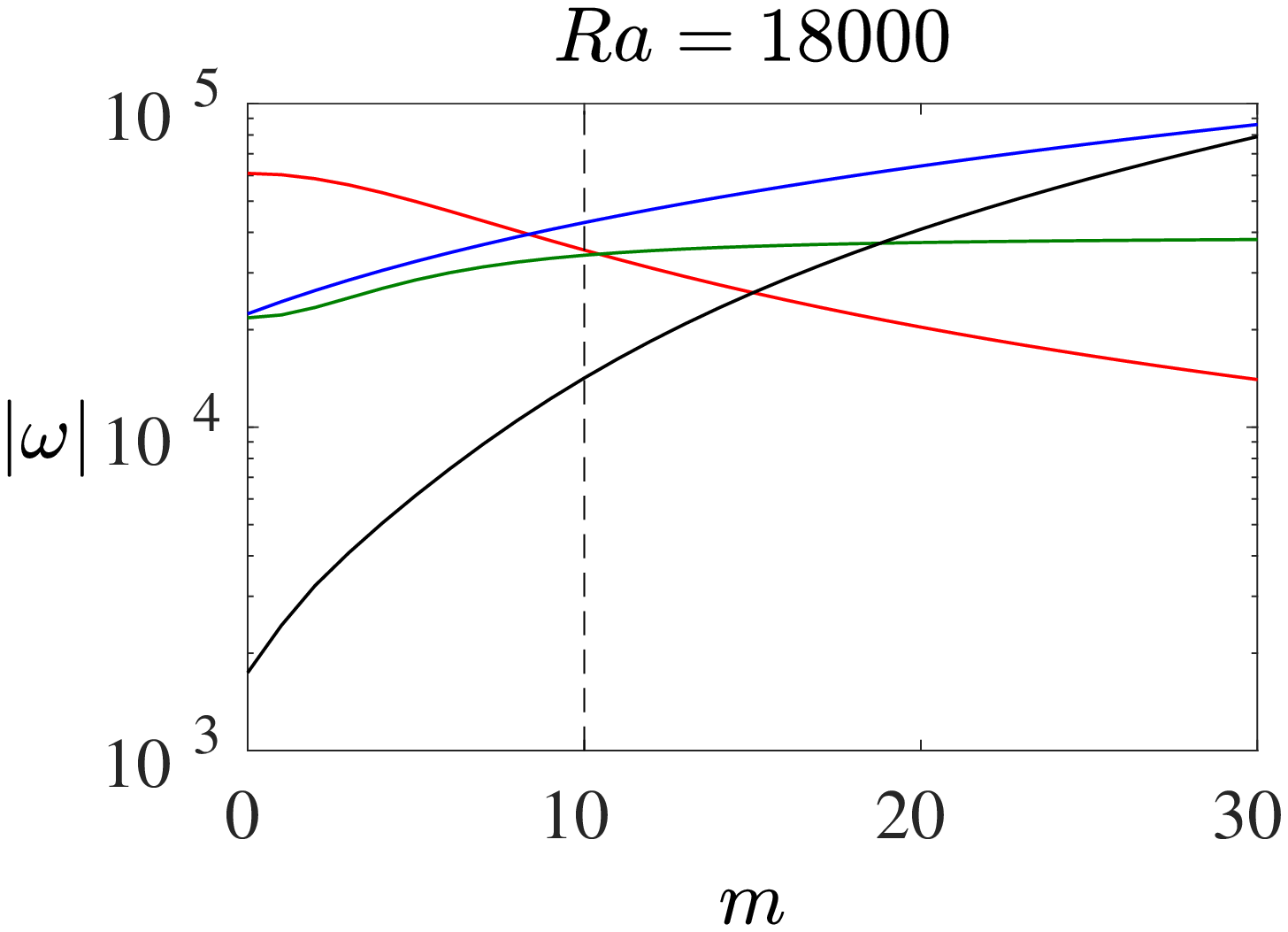}\\
\includegraphics[width=0.45\linewidth]{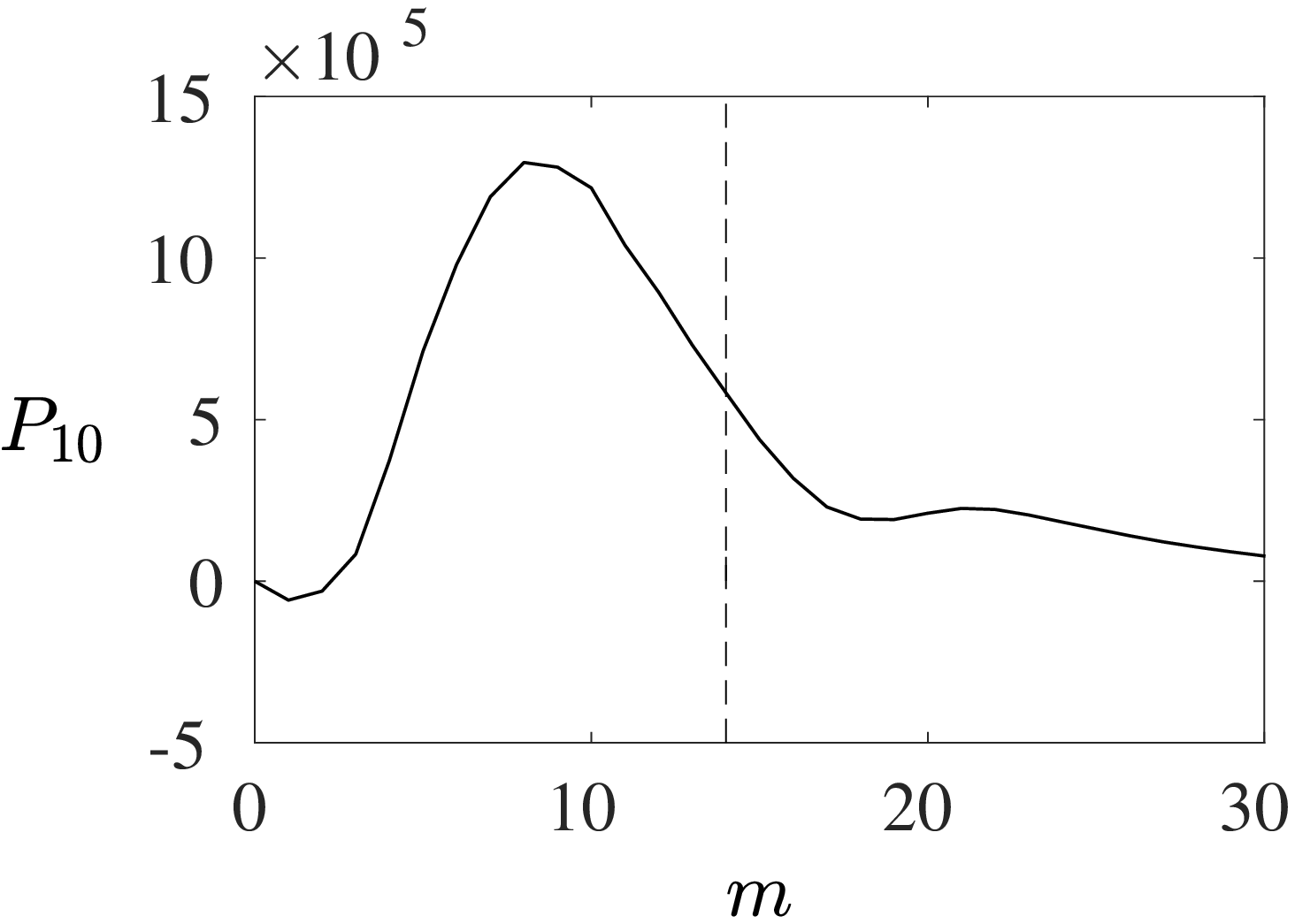}
\includegraphics[width=0.45\linewidth]{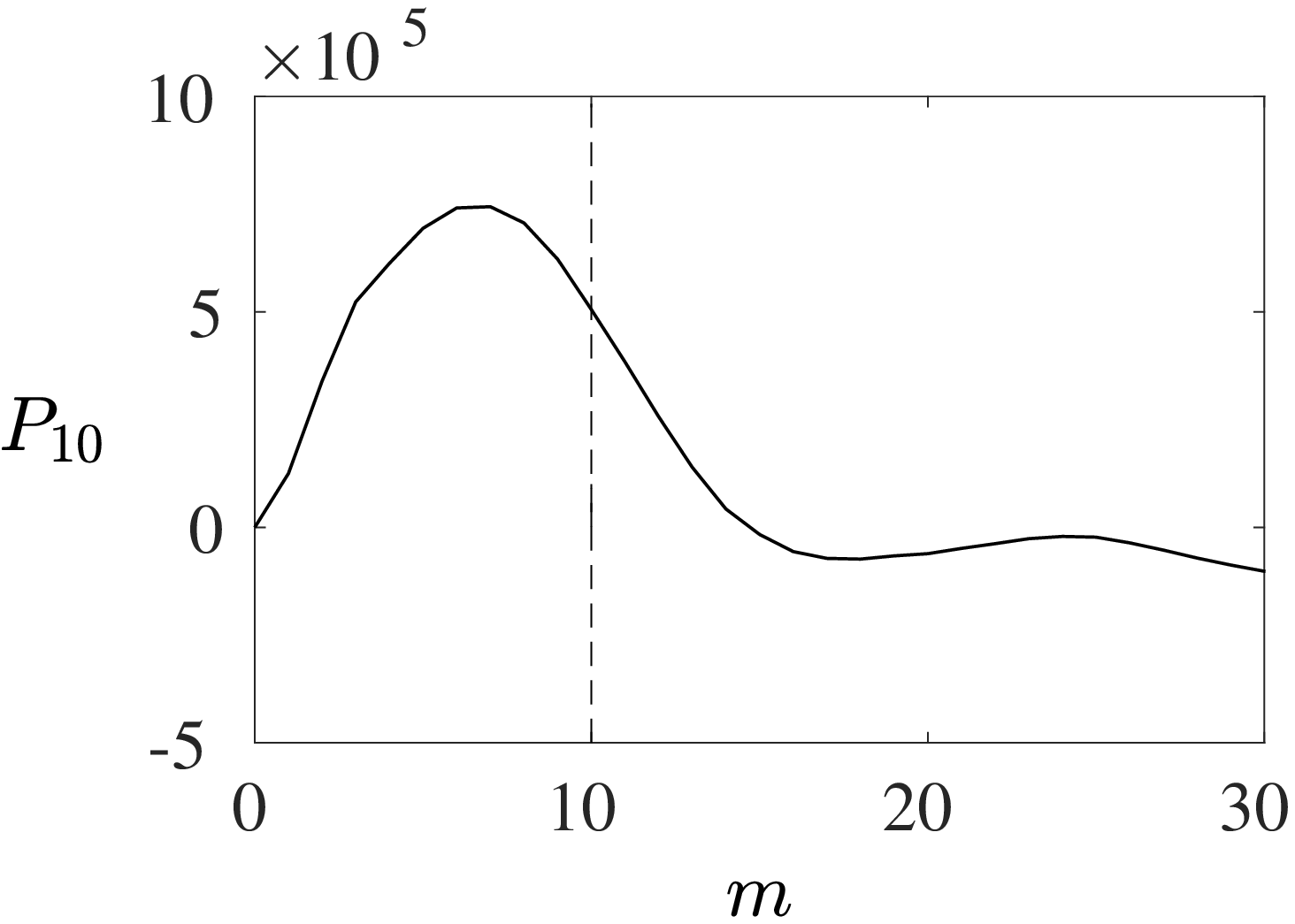}\\	
\hspace{-2.5 in}	(c)  \hspace{2.5 in} (d) \\
\includegraphics[width=0.45\linewidth]{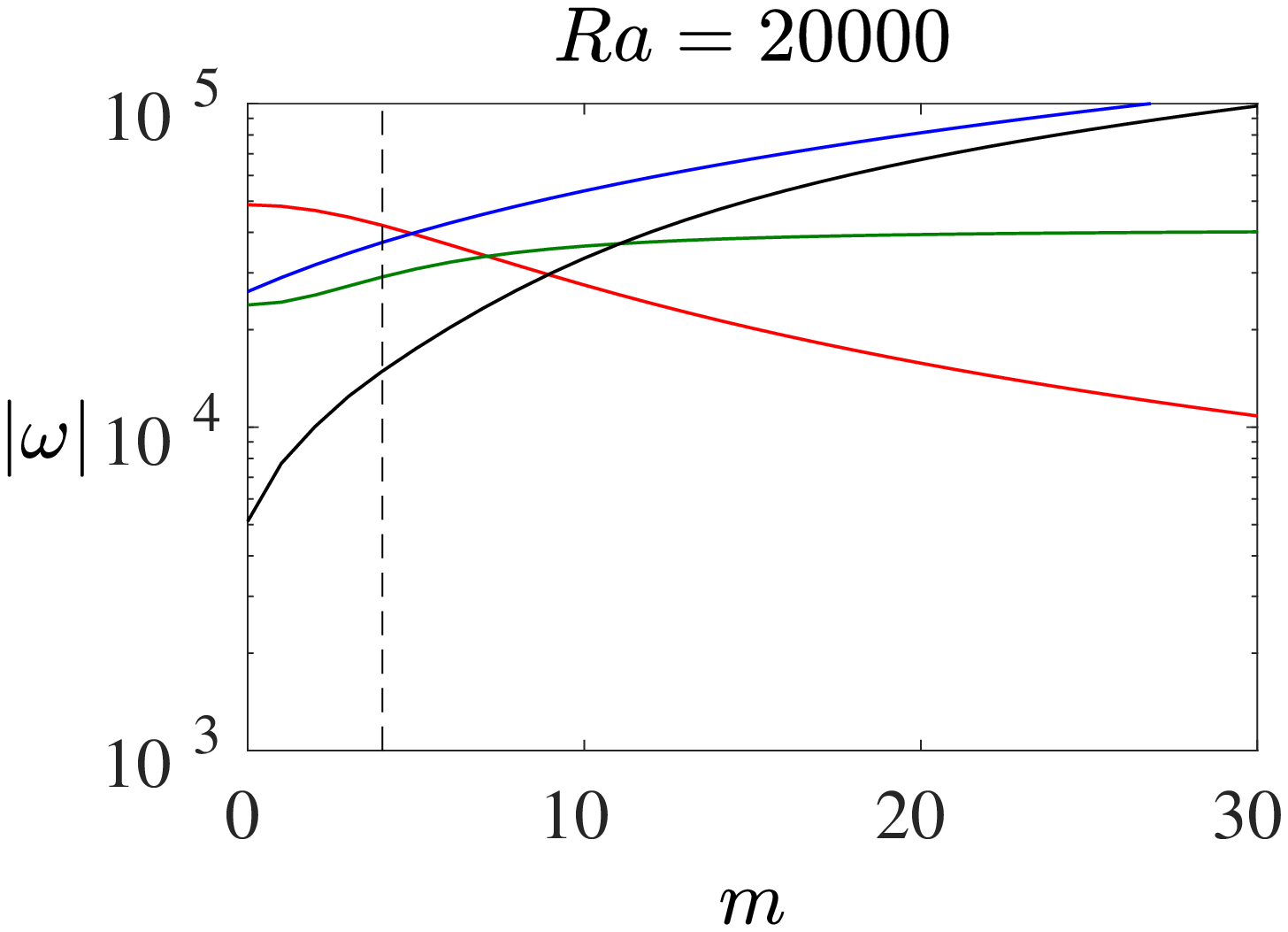}
\includegraphics[width=0.45\linewidth]{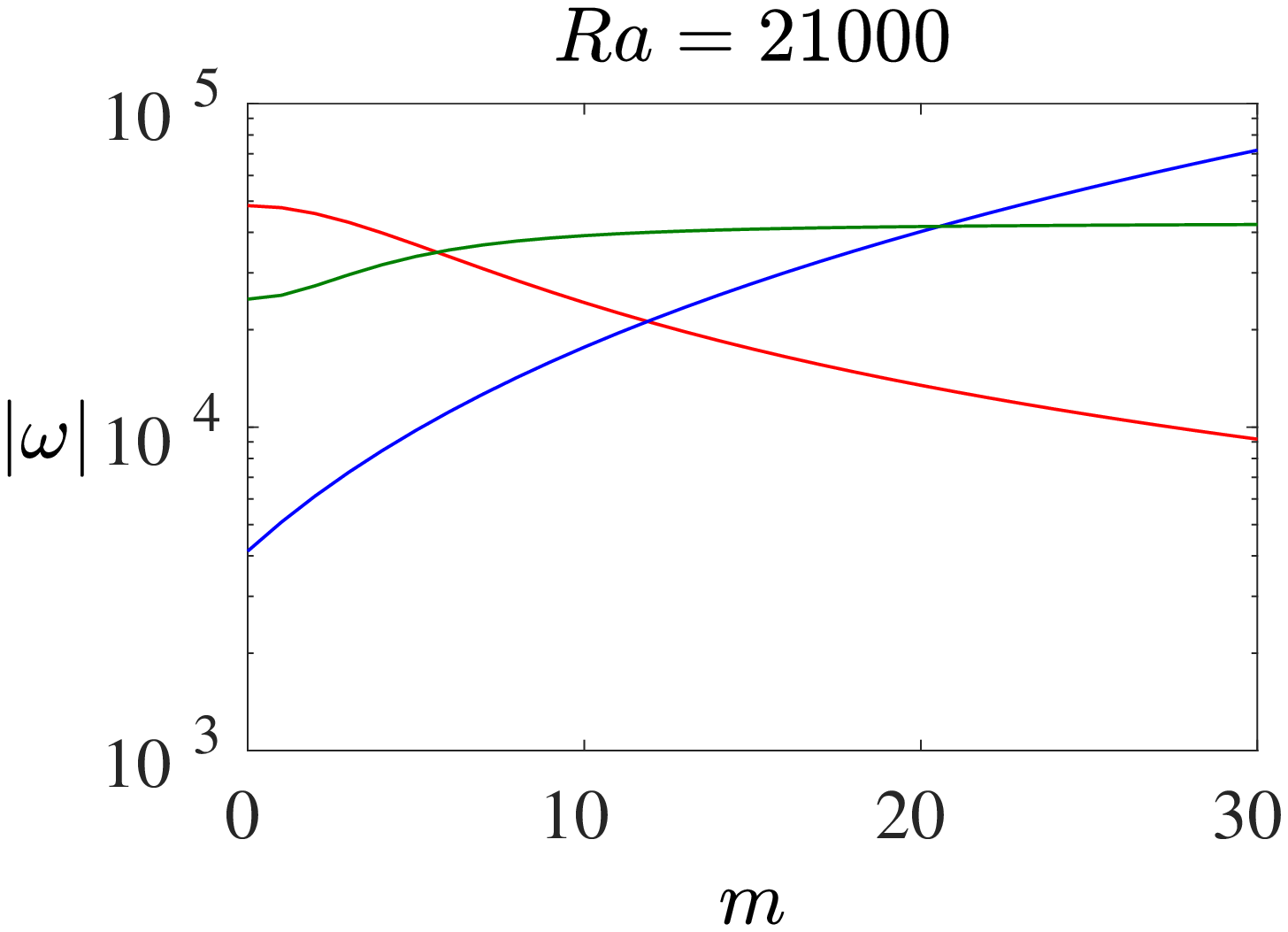}\\
\includegraphics[width=0.45\linewidth]{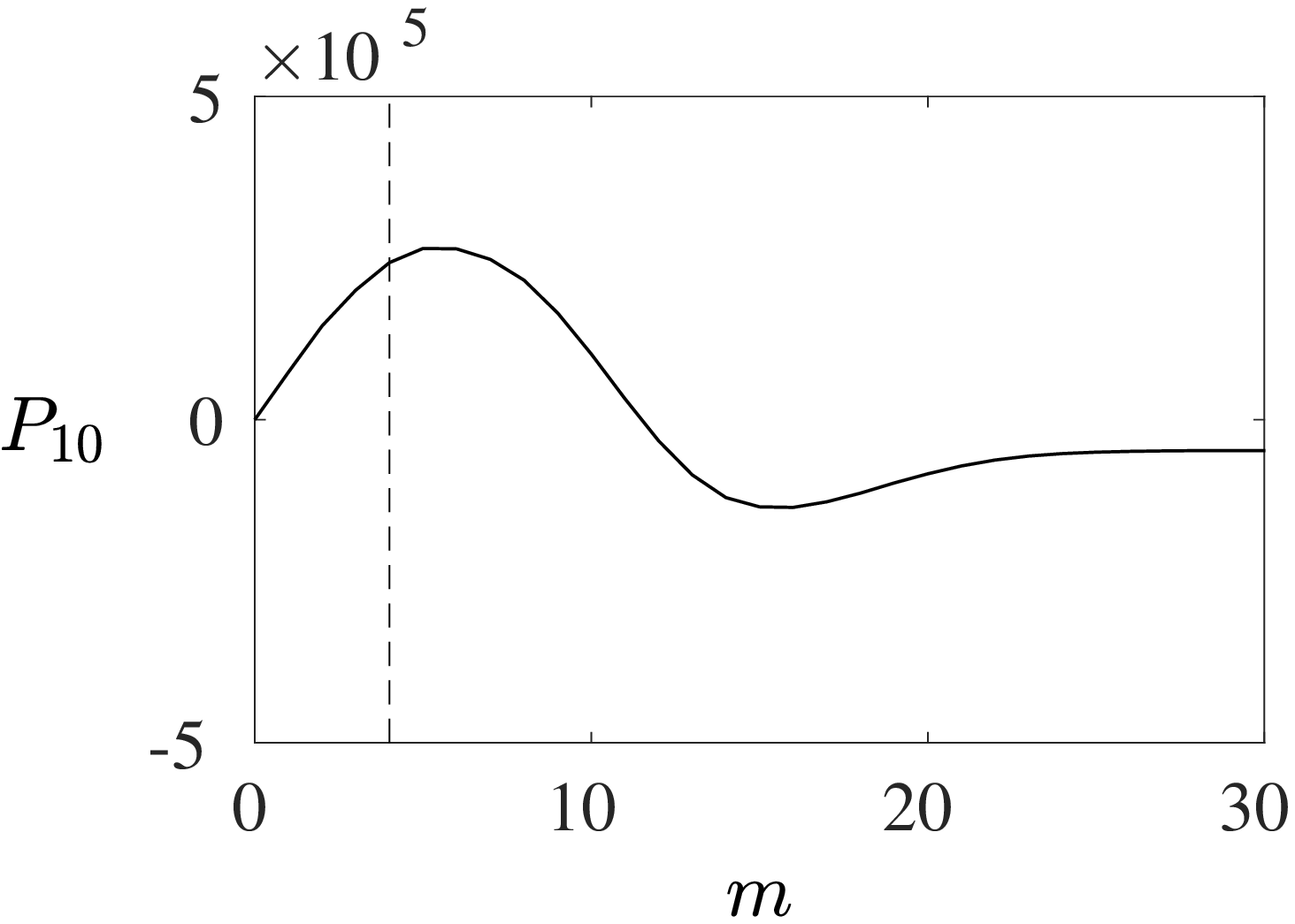}
\includegraphics[width=0.45\linewidth]{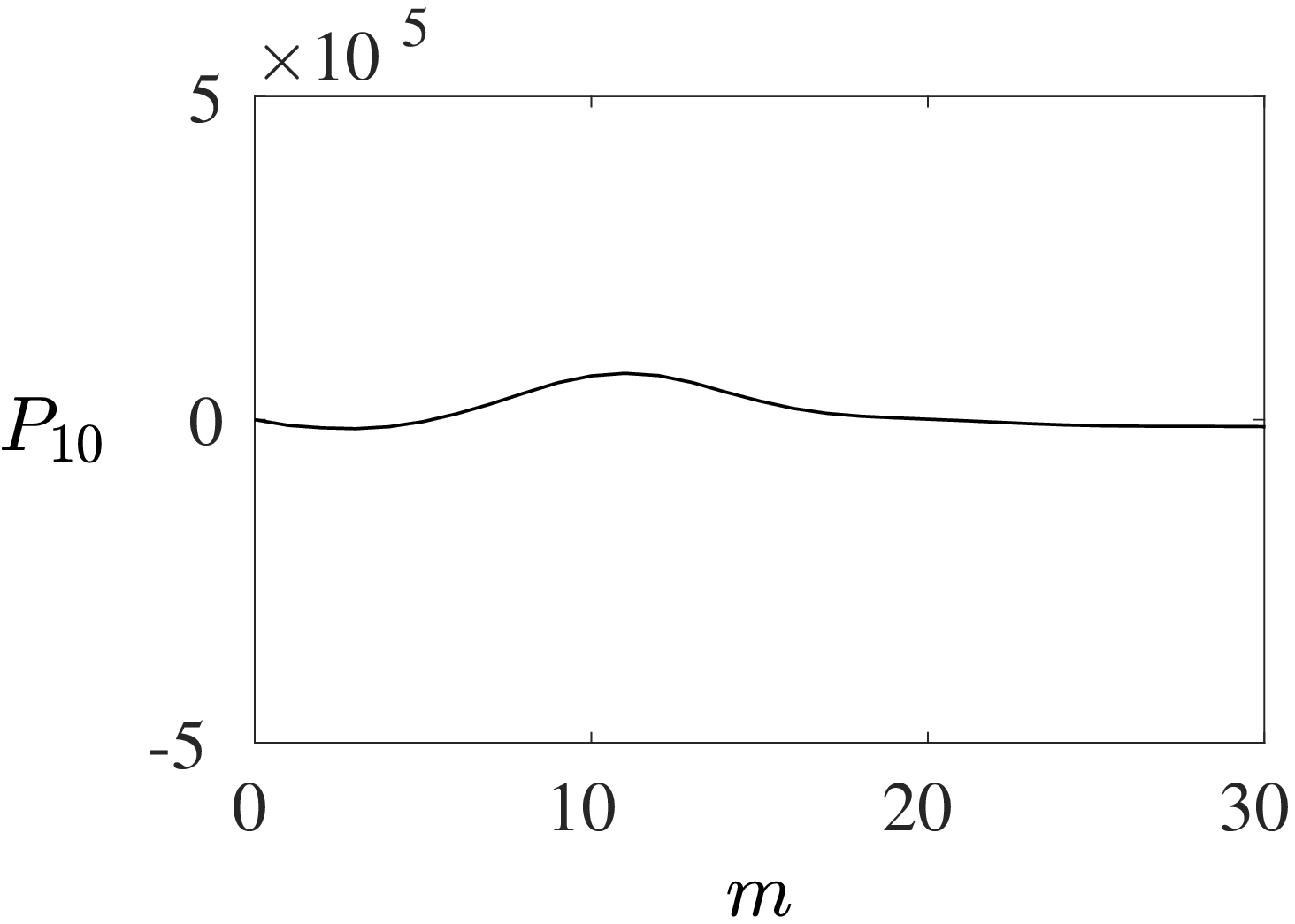}\\	
\caption{The upper panels of (a)--(d) show the absolute values of 
wave frequencies 
			 plotted for the saturated state of 
	the dynamo.
		    The dashed vertical 
			lines show the upper 
boundary of the range of wavenumbers $m$
for which the the helicity of the dynamo run is greater than
that of the equivalent nonmagnetic run. 
The lower panels of (a)--(d) show the spectral distribution of the power 
		    supplied to the axial dipole, defined in
\eqref{ps10}. The modified Rayleigh number $Ra$ 
of the model is given above
each panel.
			The other dynamo parameters are 
			$E=6 \times 10^{-5}$, and $Pm=Pr=5$.}
\label{freqdipole}
\end{figure}

In figure \ref{freqdipole}, the
magnitudes of the fundamental frequencies are shown
as a function of the spherical harmonic {{order}} $m$ in the
saturated state of the dynamo run at
$E=6 \times 10^{-5}$, and $Pr=Pm=5$.
The frequencies are computed from \eqref{om2} using
the mean values of the $s$ and $z$ wavenumbers.
For
example, real space integration over $(s,\phi)$
gives the kinetic energy as a function of $z$, the Fourier
transform of which gives the one-dimensional spectrum
$\hat{u}^2 (k_z)$. Subsequently, we obtain
\begin{equation}
\bar{k}_z = \frac{\Sigma k_z \, \hat{u}^2 (k_z)}
{\Sigma \hat{u}^2 (k_z)}.
\label{kzmean}
\end{equation}
A similar approach gives $\bar{k}_s$. The computed
frequencies in figure \ref{freqdipole}(a)--(c),
shown for dynamos with $Ra=$ 6000--18000,
satisfy the inequality $|\omega_C| > |\omega_M| >
|\omega_A|$ in a range of the spherical harmonic
{{order}} $m$. The dashed
vertical lines show the value of $m$
below which
the helicity in the nonlinear
dynamo is greater than that in the nonmagnetic run at the same parameters. 
 The frequency root $\lambda_3$ in \eqref{ndr2}, shown
by the black lines, represents the slow MAC waves
of frequency $\omega_s$ when the inequality 
$|\omega_C| > |\omega_M| > |\omega_A|$ holds true.
Evidently, the scales of helicity generation 
in the nonlinear dynamo overlaps with the scales
where the slow MAC waves are generated.  The range of $m$
over which the above frequency inequality
 holds narrows down with $Ra$,
and for the polarity-reversing dynamo with 
$Ra=21000$, this inequality does not exist at 
any $m$ (figure \ref{freqdipole}(d)). 

Figures \ref{freqdipole} (a)--(d) also show
the spectral distribution of the power supplied to the 
poloidal part of the axial dipole field $B^P_{\mathrm{10}}$, 
 given by \citep[e.g.][]{buffett2002energetics}
\begin{equation}
P_{\mathrm{10}}= \int_{V} \bm {B}^{P}_{\mathrm{10}} \cdot
[\mathbf \nabla \times(\bm {u} \times \bm{ B})_{m}] \, dV,
\label{ps10}
\end{equation}
 where $\bm{u}$ and $\bm{B}$ share the same value of
$m$ \citep{bullard1954}. The axial dipole is predominantly
generated in the wavenumbers where the MAC waves are
 generated,  except in the dynamo close to
 reversals (figure \ref{freqdipole} c) where the 
 peak of $P_{10}$ lies outside the MAC wave
 window. In the reversing dynamo without
the MAC wave window, the power supplied to the
dipole is small. 

\begin{figure}
	\centering
	\hspace{-2 in}	(a)  \hspace{1.6 in} (b)  \hspace{1.6 in} (c) \\%
	\hspace{-0.26 in} 	
\includegraphics[width=0.35\linewidth]{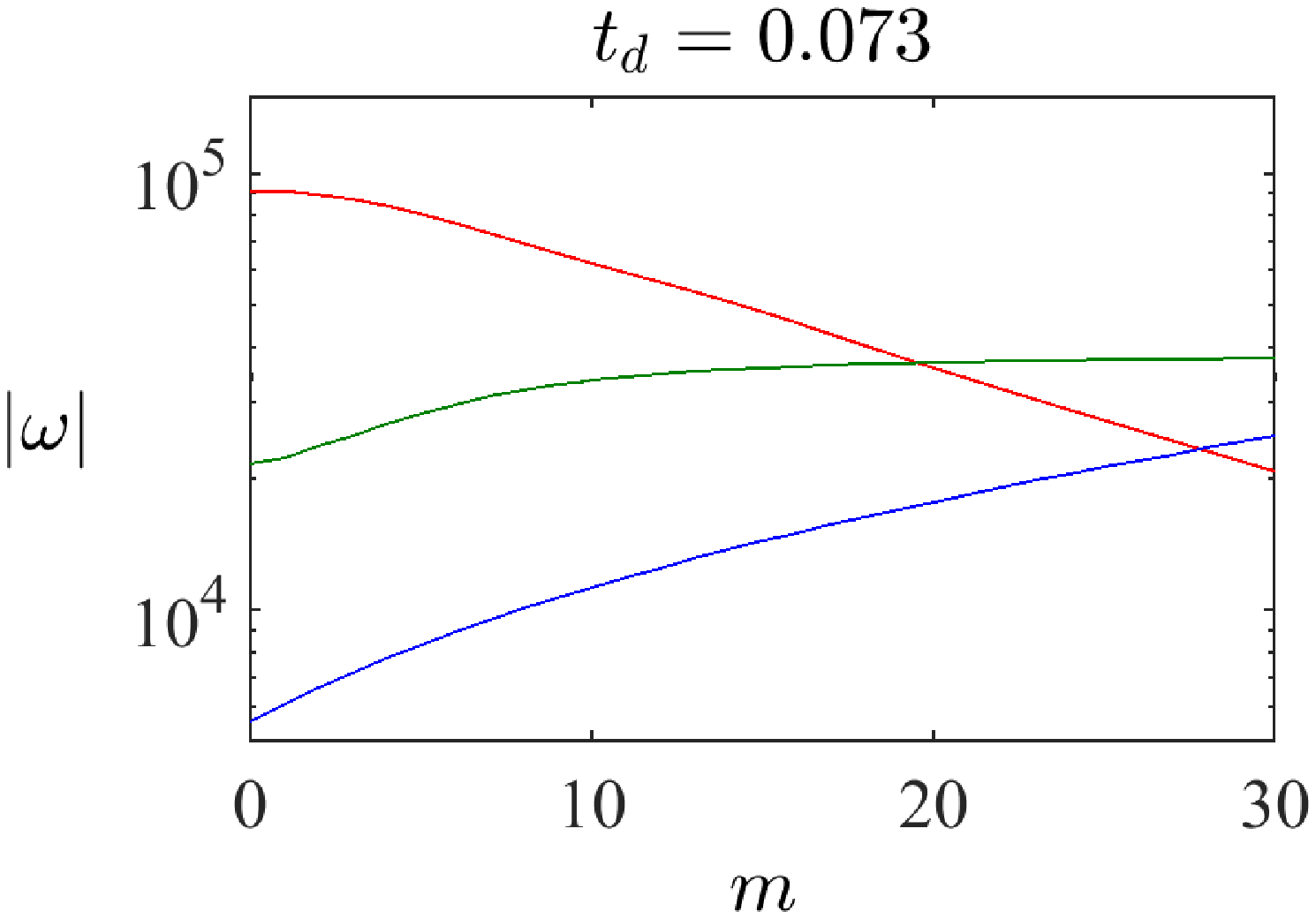}\hspace{-0.05 in} 
	\includegraphics[width=0.35\linewidth]{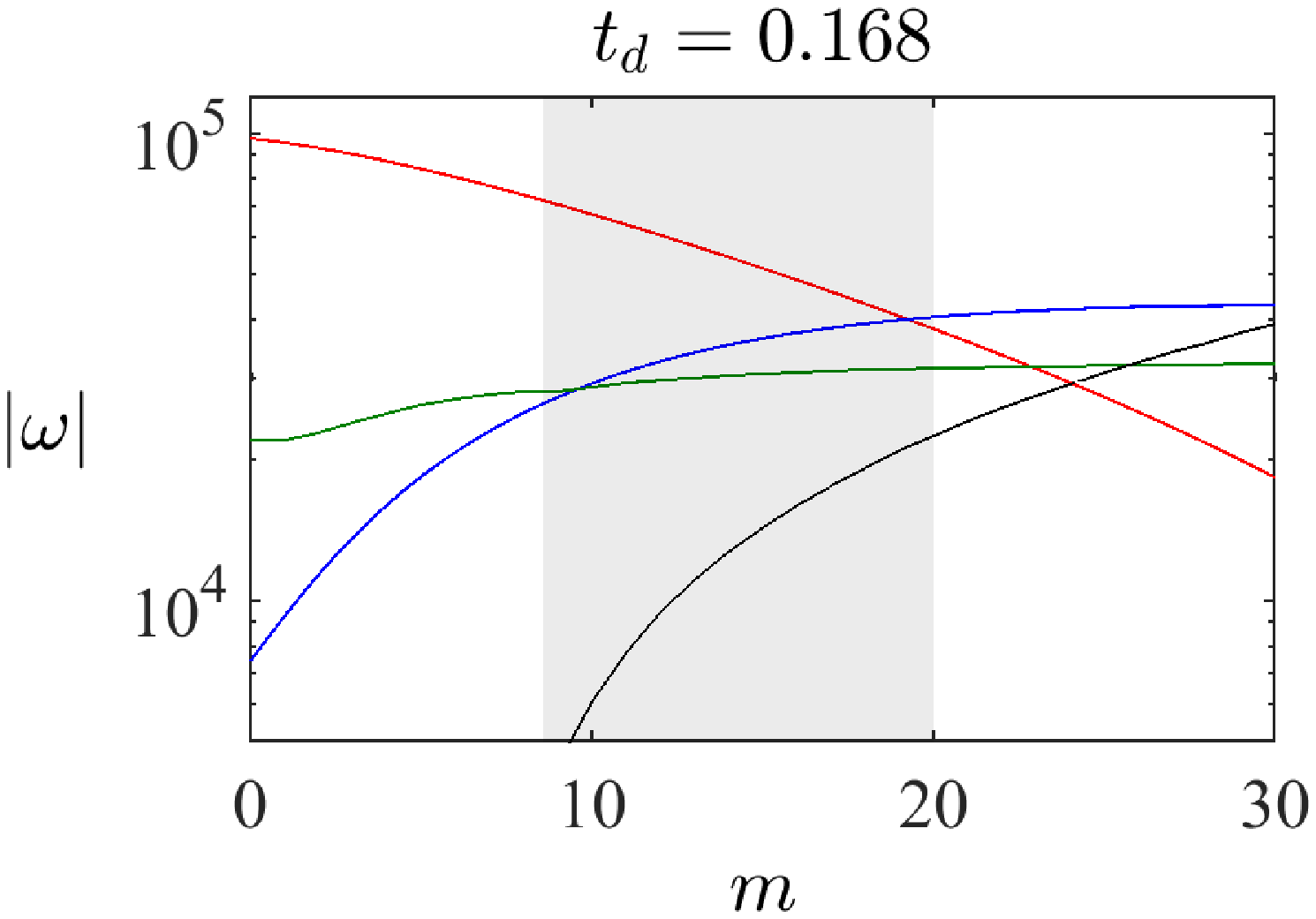}\hspace{-0.05 in} 
	\includegraphics[width=0.35\linewidth]{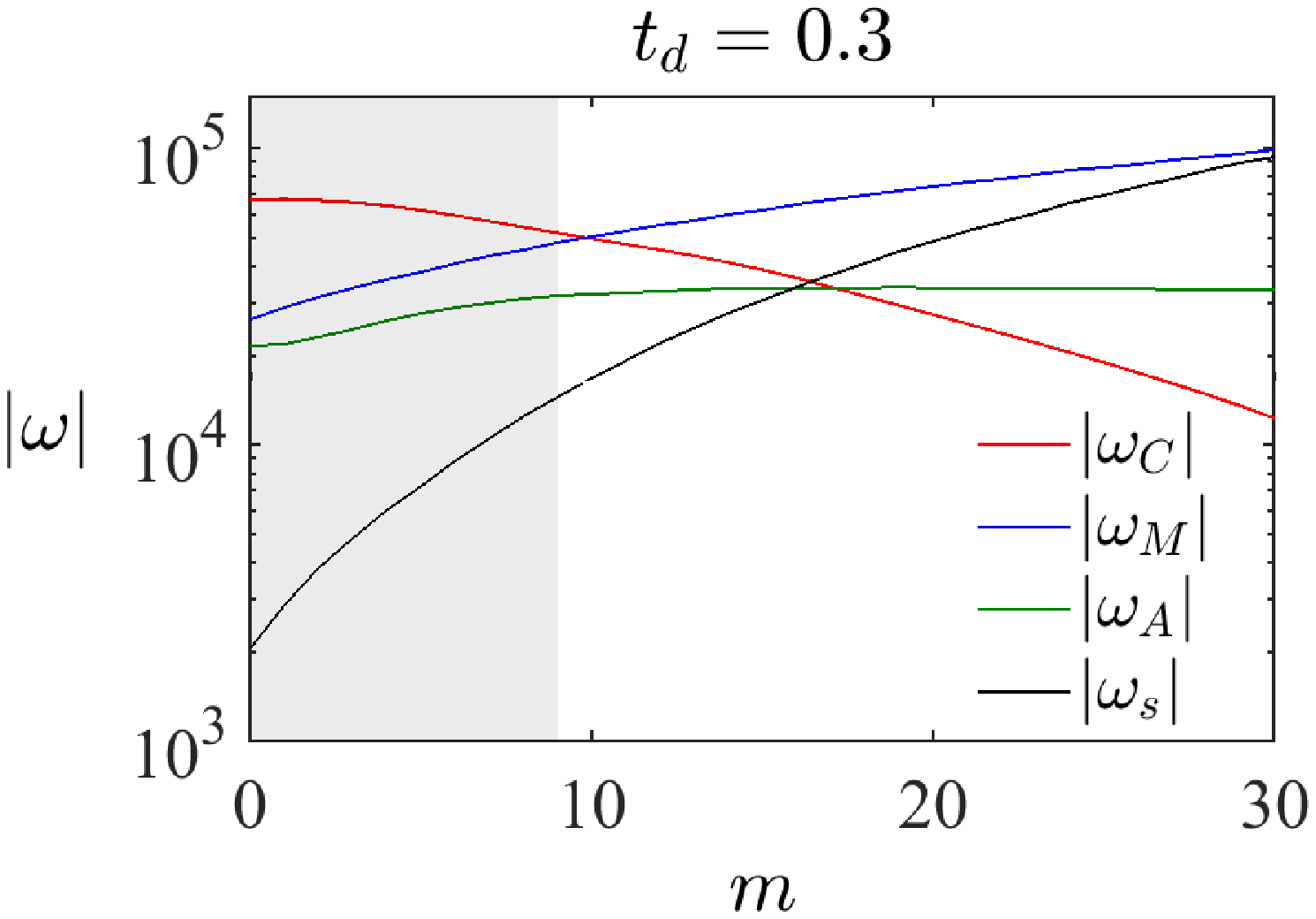}\\
	\hspace{-0.26 in} \includegraphics[width=0.35\linewidth]
	{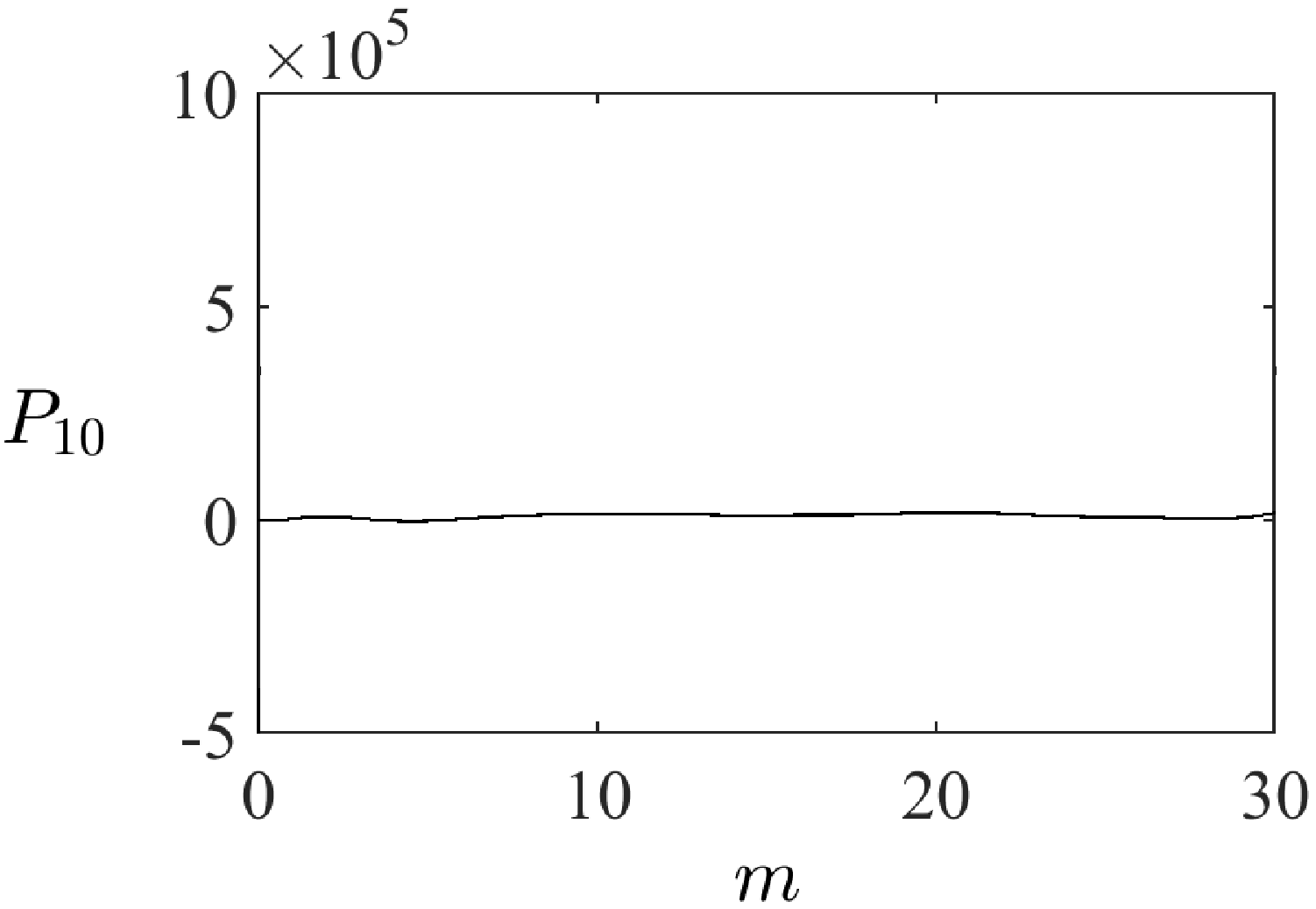}\hspace{-0.05 in} 
	\includegraphics[width=0.35\linewidth]{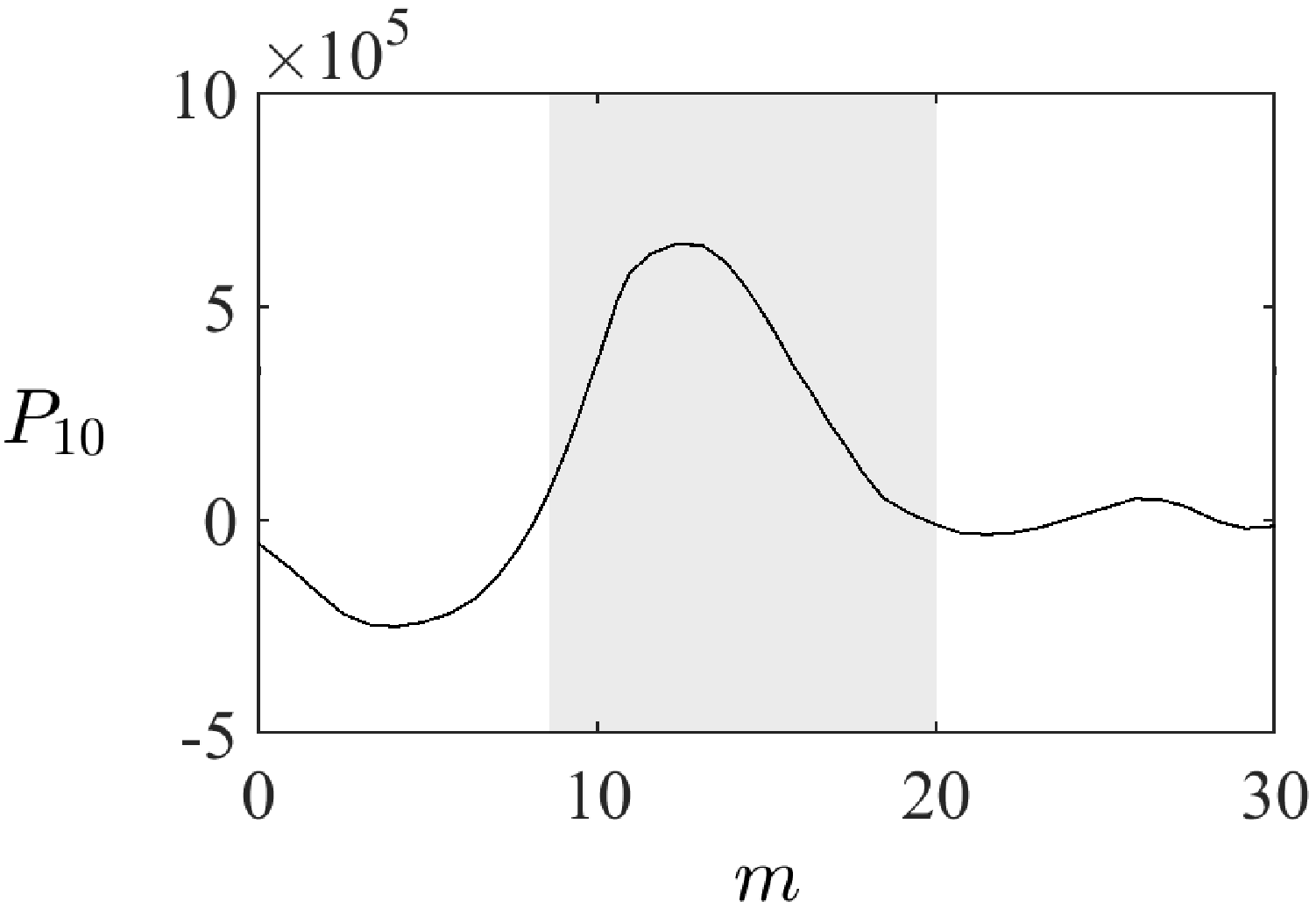}\hspace{-0.05 in} 
	\includegraphics[width=0.35\linewidth]{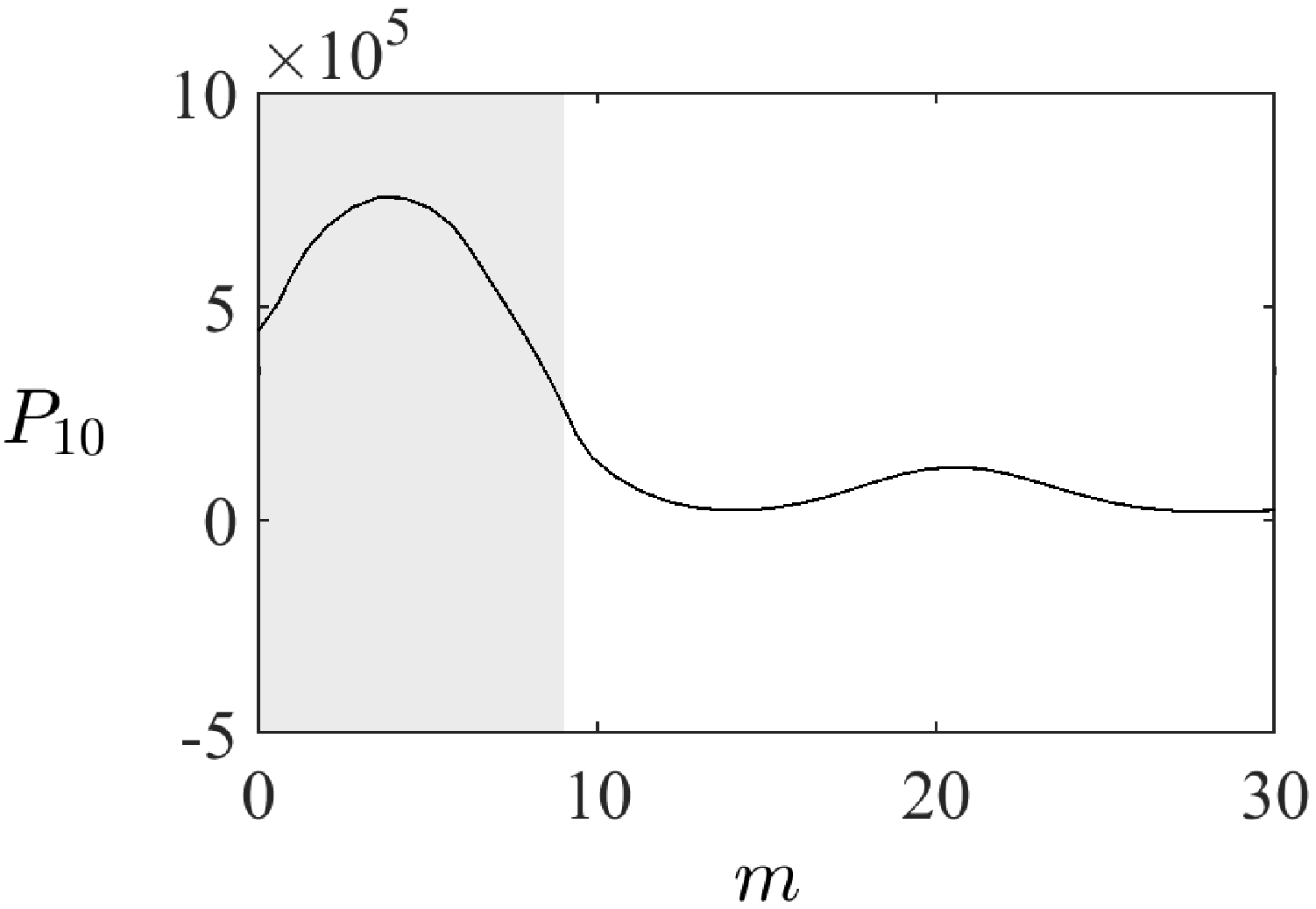}\\	
	\caption{{ The upper panels of (a)--(c) show the absolute values of 
		wave frequencies 
		plotted for three different times in the growth phase of a dynamo
starting from a seed field.
		The shaded grey area shows the range of 
scales where the helicity of the dynamo 
		run is greater than
		that of the equivalent nonmagnetic run. 
		The lower panels of (a)--(c) show the spectral distribution of the power 
		supplied to the axial dipole, defined in equation 
		\eqref{ps10}.
		The dynamo parameters are 
		$Ra= 18000, E=6 \times 10^{-5}$, and $Pm=Pr=5$.}
}
	\label{seed3k_freq_energy}
\end{figure}

In figure \ref{seed3k_freq_energy}, the upper panels show
the absolute values of the frequencies for a range of spherical harmonic 
order $m$ at three different times during the evolution of the
dynamo from a seed field. 
The lower panels show the $m$-distribution of the power 
	supplied to the axial dipole.
As the magnetic field increases from a seed, the
slow MAC waves of frequency $\omega_s$ are absent at early times
but are subsequently excited in the scales 
where the inequality 
$|\omega_{C}|>|\omega_{M}|>|\omega_{A}|$ 
is satisfied. Kinetic helicity is generated in these scales, where
a peak of the axial dipole power $P_{10}$ is also noted. 
 In the 
growth phase of the dynamo where the field is multipolar,
the dominant contribution to $P_{10}$ comes
from higher wavenumbers of $\bm{u}$ and $\bm{B}$ within
the MAC wave window (figure \ref{seed3k_freq_energy} b); 
here helicity is generated
via slow MAC waves at the peak field
locations (figures S1 and S2 in Supplementary
Material). Once the axial dipole is established, it can
hold itself up \citep{jfm11} by inducing the generation
of slow waves (figure \ref{seed3k_freq_energy} c).

\begin{figure}
	\centering
	\hspace{-2.5 in}	(a)  \hspace{2.5 in} (b) \\
	\includegraphics[width=0.48\linewidth]{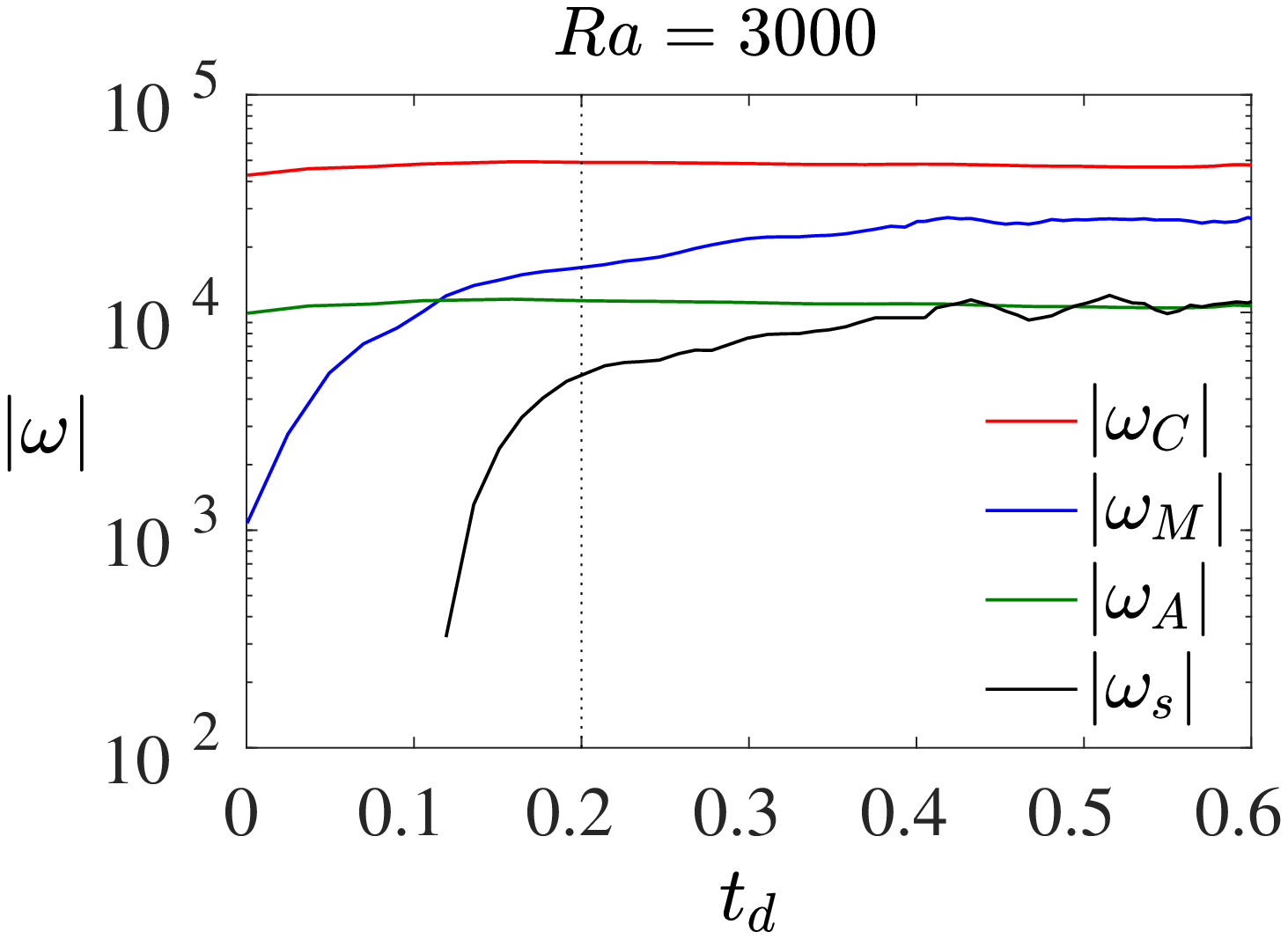}
	\includegraphics[width=0.48\linewidth]{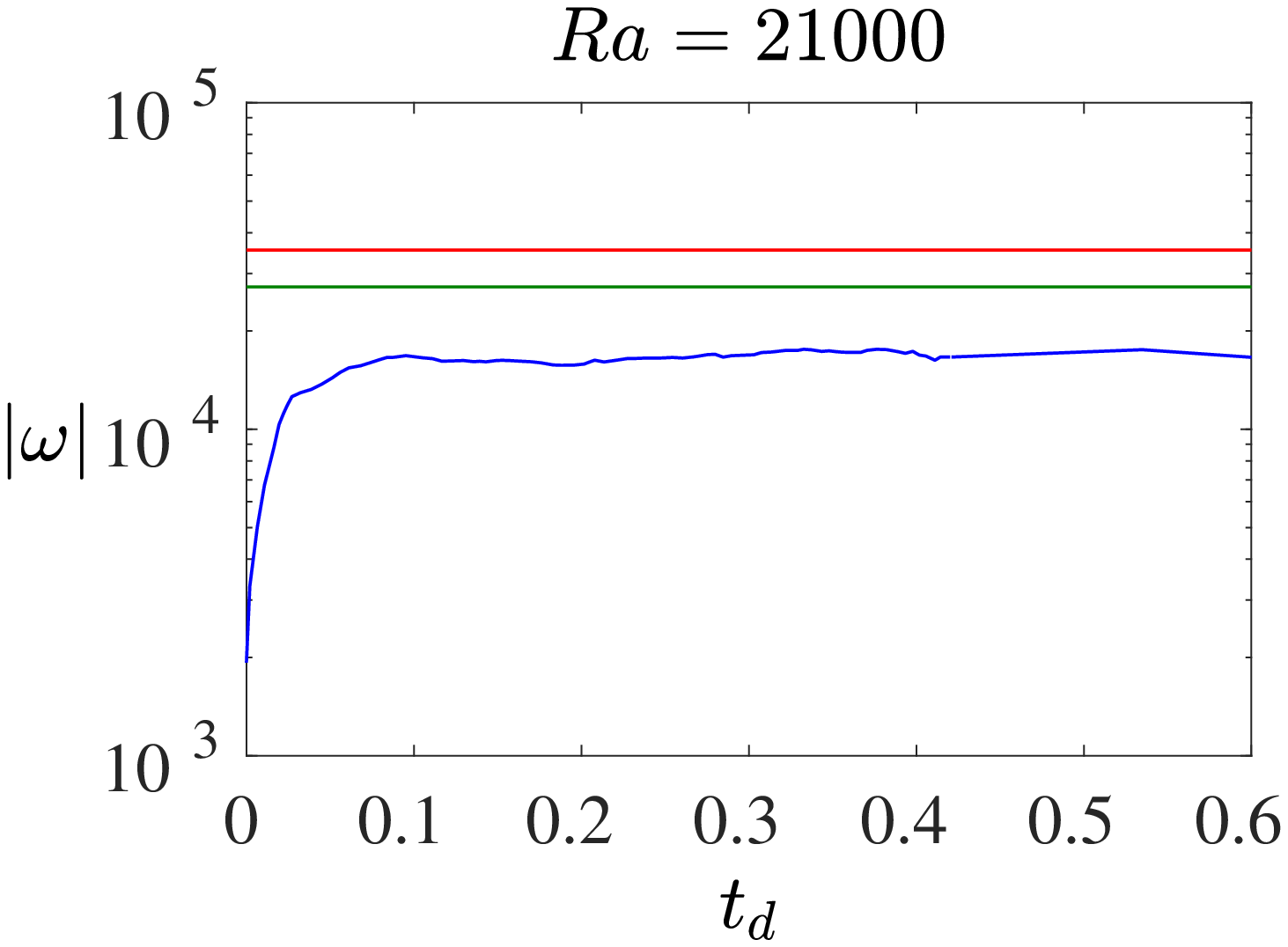}\\
	\hspace{-2.5 in}	(c)  \hspace{2.5 in} (d) \\
	\includegraphics[width=0.48\linewidth]{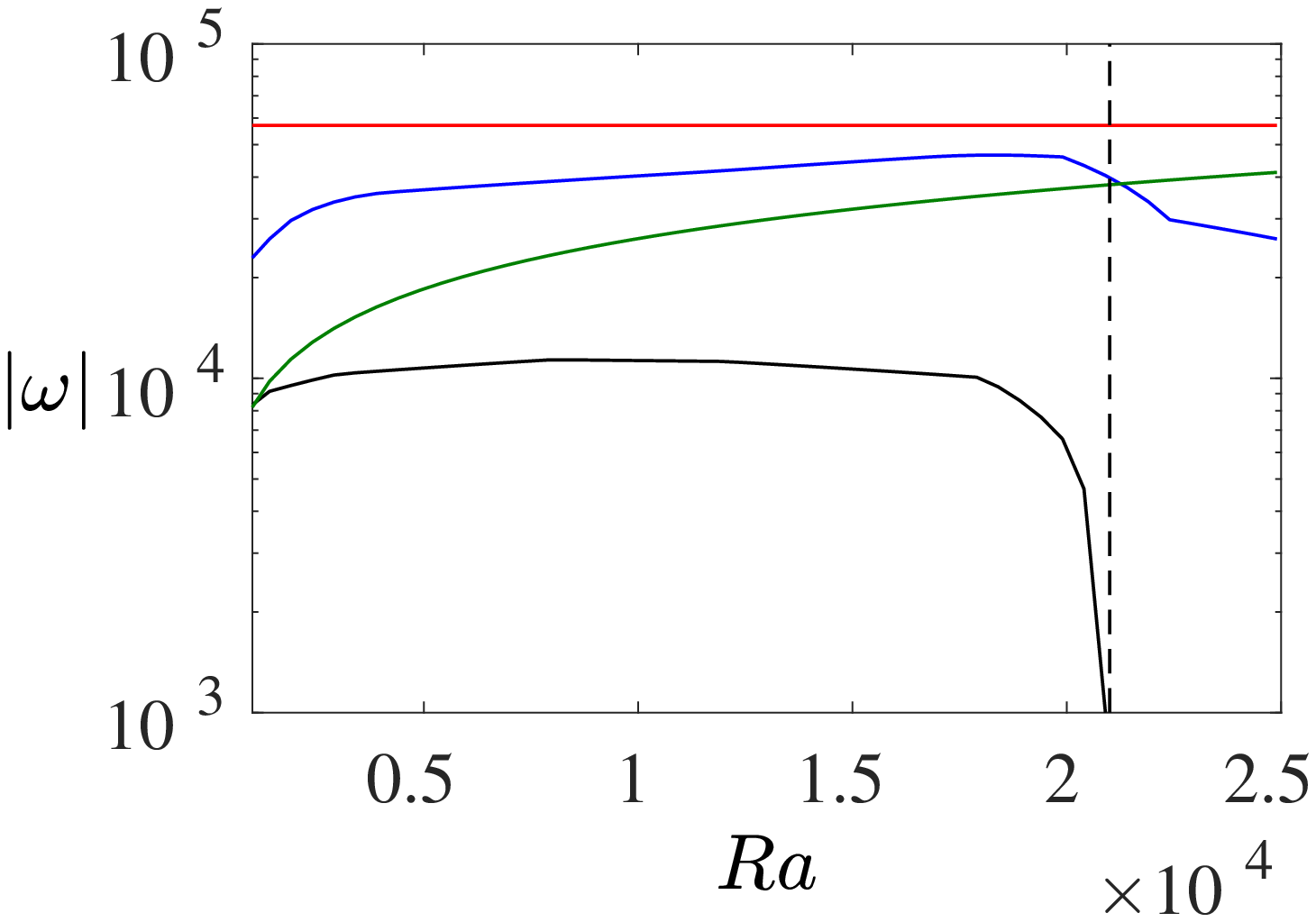}
	\includegraphics[width=0.48\linewidth]{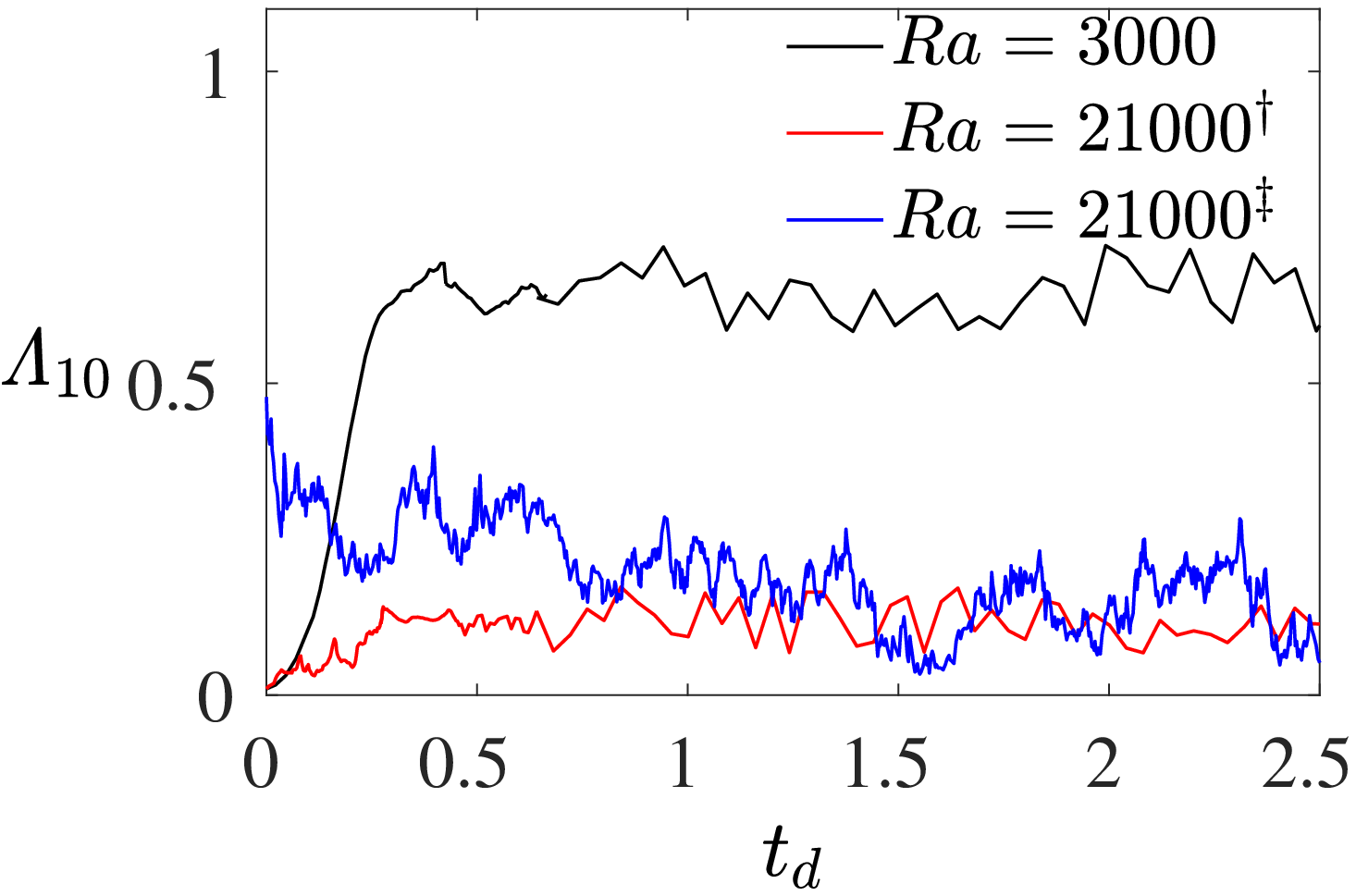}\\
	\caption{(a) \& (b) Absolute values of the 
measured frequencies
$\omega_M$, $\omega_A$, $\omega_C$ and $\omega_s$
 plotted  against time 
(measured in units of the magnetic diffusion time, $t_d$).
 These simulations study the evolution of 
the dynamo starting from a small seed
magnetic field. The modified 
Rayleigh number $Ra$ is shown above each panel.
The dotted vertical line in (a) marks the time of formation of the
axial dipole from a multipolar field.
(c) Frequencies in the saturated dynamo shown against $Ra$. 
The dashed vertical line here gives the value of $Ra$
at which the slow MAC wave frequency $\omega_s$ goes
to zero.
		(d) The Elsasser number of the axial dipole field
component, based on its root mean square value,
 for $Ra=3000$, $Ra=21000^\dag$ (seed field start) and
 $Ra=21000^\ddag$ (strong field start). 
 The dynamo parameters are $E = 6 \times 10^{-5}$, $Pm=Pr=5$.
The colour codes in (a) are also used in (b) and (c).}
	\label{compare_ra}
\end{figure}

In figure \ref{compare_ra}, the 
dynamo frequencies are computed from \eqref{om2} using
the mean values of the $s$, $\phi$  and $z$ wavenumbers
in the saturated state of the dynamo
in the range $l \le l_E$, which
represents the energy-containing scales.  The
mean spherical harmonic order $\bar{m}$ is evaluated
through a weighted average as in \eqref{kzmean}, but  
over the range of $m$ within $l \le l_E$. As the field increases from
a small seed value in the dipolar dynamo run at
$Ra=3000$ (figure \ref{compare_ra}(a)), slow MAC waves 
of frequency $\omega_s$ are 
first excited when $|\omega_M| > |\omega_A|$.
 The formation of the axial dipole
from a chaotic field, marked by the dotted vertical line,
follows slow wave excitation. 
In the run at $Ra=21000$ that
begins from a seed field and produces a multipolar
solution,  $|\omega_M|$ remains lower than
$|\omega_A|$ throughout (figure \ref{compare_ra}(b)), so 
the slow waves are never excited.
From the variation of the frequencies with increasing strength
of forcing, given in figure  \ref{compare_ra}(c), we note
that  $|\omega_M|$ falls below $|\omega_A|$ at 
$Ra \approx 21000$, which indicates that polarity
reversals would indeed onset in the regime 
$|\omega_M| \approx |\omega_A|$ when slow MAC waves
disappear. The volume-averaged mean square value of the 
axial dipole field is much smaller in 
the multipolar regime
of $Ra=21000$ than in the stable dipole regime of $Ra=3000$
(figure \ref{compare_ra}(d)), which suggests that the slow
MAC waves have an important role in the formation of
the axial dipole. { In the run
at $Ra=21000$ that begins from a strong field and 
produces polarity reversals, the dipole intensity
decreases and eventually reaches the same order
of magnitude as that in the multipolar run. This
is consistent with the suppression of the slow
waves at this $Ra$, and reflected in the decrease
in kinetic helicity to the nonmagnetic value (figure \ref{relhel}(d)). }
\begin{table}
	\centering
	\begin{tabular}{ llllllllllllllll } 
		$Ra$& $Ra/Ra_c$ & $N_r$ & $l_{max}$ & $Rm$ & $Ro_\ell$ & $l_E$ 
		& $l_C$ & $\bar{m}$ &
		$\bar{k}_s$ & $\bar{k}_z $ & {$E_k$} & {$E_m$} & 
{$f_{dip}$} & $Ra_\ell$ & $\varLambda$ \\
		&&&&&&&&&&&$\times 10^5$&$\times 10^5$&&&\\
		\multicolumn{16}{c}{ $E = 3 \times 10^{-4},Pm=Pr=20$}\\
		400  & 18.2  &72 &64 & 56  & 0.002 & 11 & 8   & 3.52 &2.92&2.87&0.22&2.44&0.84& 1274    &24 \\
		800  & 36.4  &72 &64 & 75  & 0.003 & 11 & 9   & 3.91 &3.65&2.81&0.41&3.73&0.77& 2066   &48 \\
		1600 & 72.7  &72 &72 & 106 & 0.005 & 13 & 10  & 4.12 &3.32&2.43&0.78&5.57&0.72& 3721   &64 \\
		2000 & 90.9  &72 &72 & 119 & 0.005 & 14 & 11  & 4.34 &3.01&2.97&1.01&4.81&0.68& 4192   &87 \\
		2400 & 109.1 &72  &96 & 130 & 0.006 & 14 & 11  & 4.26 &3.14&2.66&1.22&4.69&0.69& 5221   &114\\
		3000 & 136.4 &72  &96 & 148 & 0.007 & 15 & 11  & 4.31 &3.52&2.24&1.56&3.80&0.68& 6376   &128\\
		4000 & 181.8 &72  &96 & 171 & 0.009 & 15 & 11  & 4.19 &3.25&2.63&2.12&3.36&0.61& 8995   &160\\
		4500 & 204.6 &132 &128 & 192 & 0.009 & 15 & 11 & 4.34 &2.87&2.54&2.47&3.43&0.60& 9432  &167\\
		4750 & 215.9 &132 &128 & 199 & 0.010 & 15 & 11 & 4.29 &3.43&2.46&2.58&3.03&0.59& 10189  &179\\
		4875 & 221.6 &132 &132 & 203 & 0.010 & 15 & 11 & 4.16 &3.26&2.19&2.61&2.81&0.57& 11121  &185\\
		4950 & 225.0 &132 &132 & 105 & 0.011 & 15 & 11 & 4.12 &3.11&2.42&2.73&2.37&0.55& 11513  &195\\
		5000$^{\star}$ & 227.3 &132 &132 & 210 & 0.011 & 15 & 11 & 4.02 &3.35&2.31&2.99& 0.07
 &0.31& 12215   &210 \\
		\multicolumn{16}{c}{$E=6\times10^{-5},Pm=Pr=5$}\\
		300   & 10.3  &88  &96  & 67   & 0.002 & 10 & 9  & 3.89 &3.81&3.36&0.29&2.42&0.96& 783   &26\\
		400   & 13.8  &88  &96  & 74   & 0.002 & 11 & 9  & 3.78 &4.21&3.25&0.33&5.51&0.94& 1105  &29\\
		1000  & 34.5  &128 &120 & 98   & 0.004 & 16 & 12 & 4.69 &3.86&3.31&0.67&8.37&0.81& 1795  &39\\
		3000  & 103.5 &160 &160 & 169  & 0.009 & 20 & 15 & 5.23 &4.75&4.29&2.06&18.72&0.76& 4330 &96\\
		6000  & 206.9 &160 &160 & 243  & 0.014 & 22 & 16 & 5.92 &5.51&4.13&4.47&19.43&0.72& 6759 &126\\
		8000  & 275.9 &160 &180 & 288  & 0.020 & 23 & 17 & 6.14 &5.27&3.62&6.14&17.54&0.71& 8377 &149\\
		12000 & 413.8 &160 &180 & 365  & 0.024 & 24 & 17 & 7.13 &4.68&3.34&9.86&17.29&0.70& 9319 &160\\
		14000 & 482.8 &160 &180 & 402  & 0.026 & 25 & 18 & 7.54 &5.11&2.97&12.41&16.32&0.62& 9722 &185\\
		18000 & 620.7 &160 &180 & 456  & 0.032 & 25 & 19 & 7.78 &4.84&3.30&15.22&16.91&0.61& 11740 &200\\
		20000 & 689.7 &160 &180 & 505  & 0.035 & 25 & 19 & 8.12 &4.64&3.46&17.17&15.06&0.59& 11975 &215\\
		21000$^{\star}$ & 724.1 &160 &180 & 549  & 0.039 & 25 & 19 & 8.05 &4.22&3.38&20.54&11.47&0.30&  12793 &227\\
		\multicolumn{16}{c}{$E = 1.2 \times 10^{-5},Pm=Pr=1$}\\
		300   & 10.3  &90  &96  &78   & 0.004 & 15 & 15  & 4.87  &3.87&2.86&0.45 &1.22 &0.94&   499  &17\\
		700   & 24.1  &90  &96  &102  & 0.005 & 19 & 17  & 5.02  &4.25&3.02&0.64 &5.33 &0.90&   1097 &18\\
		1000  & 34.5  &132 &144 &112  & 0.006 & 21 & 20  & 5.12  &4.58&2.89&0.89 &17.81&0.82&   1506 &36\\
		2500  & 86.2  &168 &160 &174  & 0.011 & 26 & 20  & 6.54  &4.04&2.82&2.22 &27.75&0.81&   2308 &38\\
		4000  & 137.9 &180 &168 &224  & 0.017 & 28 & 20  & 7.94  &4.14&3.42&3.41 &31.01&0.82&   2505 &41\\
		10000 & 344.8 &192 &180 &384  & 0.033 & 33 & 24  & 9.13  &4.62&3.07&10.63&32.21&0.83&   4736 &67\\
		15000 & 517.2 &192 &180 &500  & 0.045 & 34 & 24  & 9.87  &4.87&2.97&18.48&31.86&0.82&   6079 &95\\
		20000 & 689.7 &192 &180 &573  & 0.052 & 35 & 25  & 9.93  &4.43&3.12&23.84&33.57&0.78&   8007 &139\\
		25000 & 862.1 &192 &180 &655  & 0.061 & 36 & 25  & 10.01 &4.48&2.78&31.34&36.44&0.76&   9850 &176\\
		27000 & 931.0 &192 &180 &698  & 0.065 & 36 & 25  & 9.96  &4.69&2.87&35.56&32.03&0.75&   10745&190\\
		28000$^{\star}$ & 965.5 &192 &180 &775  & 0.073 & 36 & 25  & 10.05 &4.81&2.91&39.74&19.67&0.25&  10944& 196\\
	\end{tabular}
	\caption{Summary of the main input and output parameters
		in the dynamo simulations considered in this study. Here,
		$Ra$ is the modified Rayleigh number, $Ra_c$ is the modified
		critical Rayleigh number for onset of nonmagnetic
		convection, $N_r$ is the number of radial grid points, $l_{max}$
		is the maximum spherical harmonic degree, $Rm$ is the magnetic
		Reynolds number, $Ro_\ell$ is the local Rossby number, $l_C$
		and $l_E$ are the mean spherical harmonic degrees of convection
		and energy injection respectively
		(defined in \eqref{elldef}), $\bar{m}$ is the 
		mean spherical harmonic order in the range $l \leq l_E$, 
		$\bar{k}_s$ and $\bar{k}_z$
		are the mean $s$ and $z$ wavenumbers in the range $l \leq l_E$, 
		$E_k$ and $E_m$ are the time-averaged total kinetic 
and magnetic energies defined in \eqref{energies}, $f_{dip}$ is
the relative dipole field strength, $Ra_\ell$ is the local Rayleigh number defined in \eqref{raell}
 and $\varLambda$
		is the peak Elsasser number obtained from the square of the measured 
		peak field at the earliest time of excitation of slow MAC waves
		in the dynamo run starting
		from a small seed field. $^*$The last run in each Ekman number series
		is a polarity-reversing dynamo, 
		for which $\varLambda$ is the square of
		the measured peak field when slow MAC waves cease to exist in
		the run starting from the saturated state of the penultimate run
		in that series (see also \S \ref{selfs}).}
	\label{parameters}
\end{table}

\begin{figure}
	\centering
\hspace{-1.4 in}(a)  \hspace{1.4 in} (b) \hspace{1.6 in} (c)  \\
\includegraphics[width=0.32\linewidth]{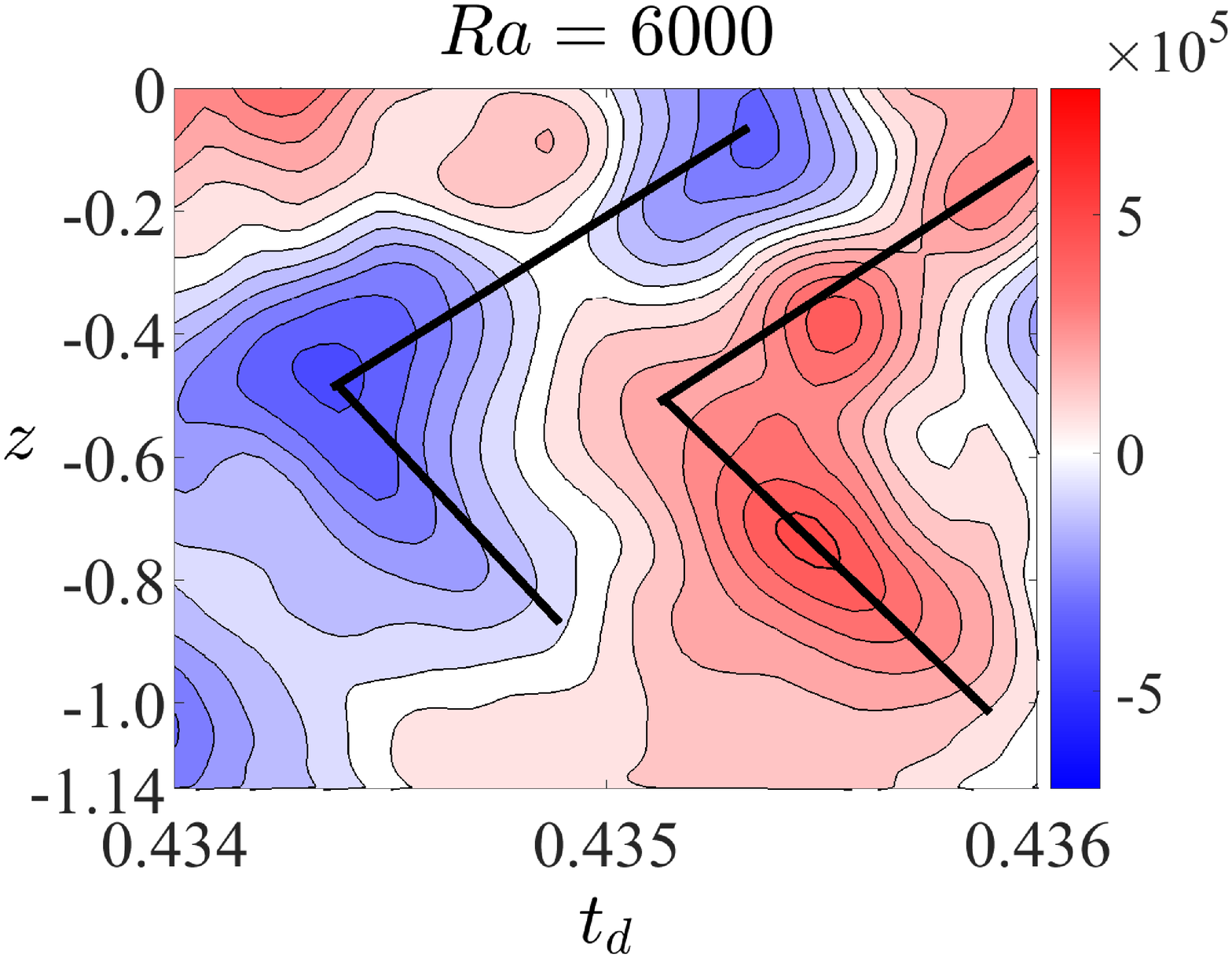}
\includegraphics[width=0.32\linewidth]{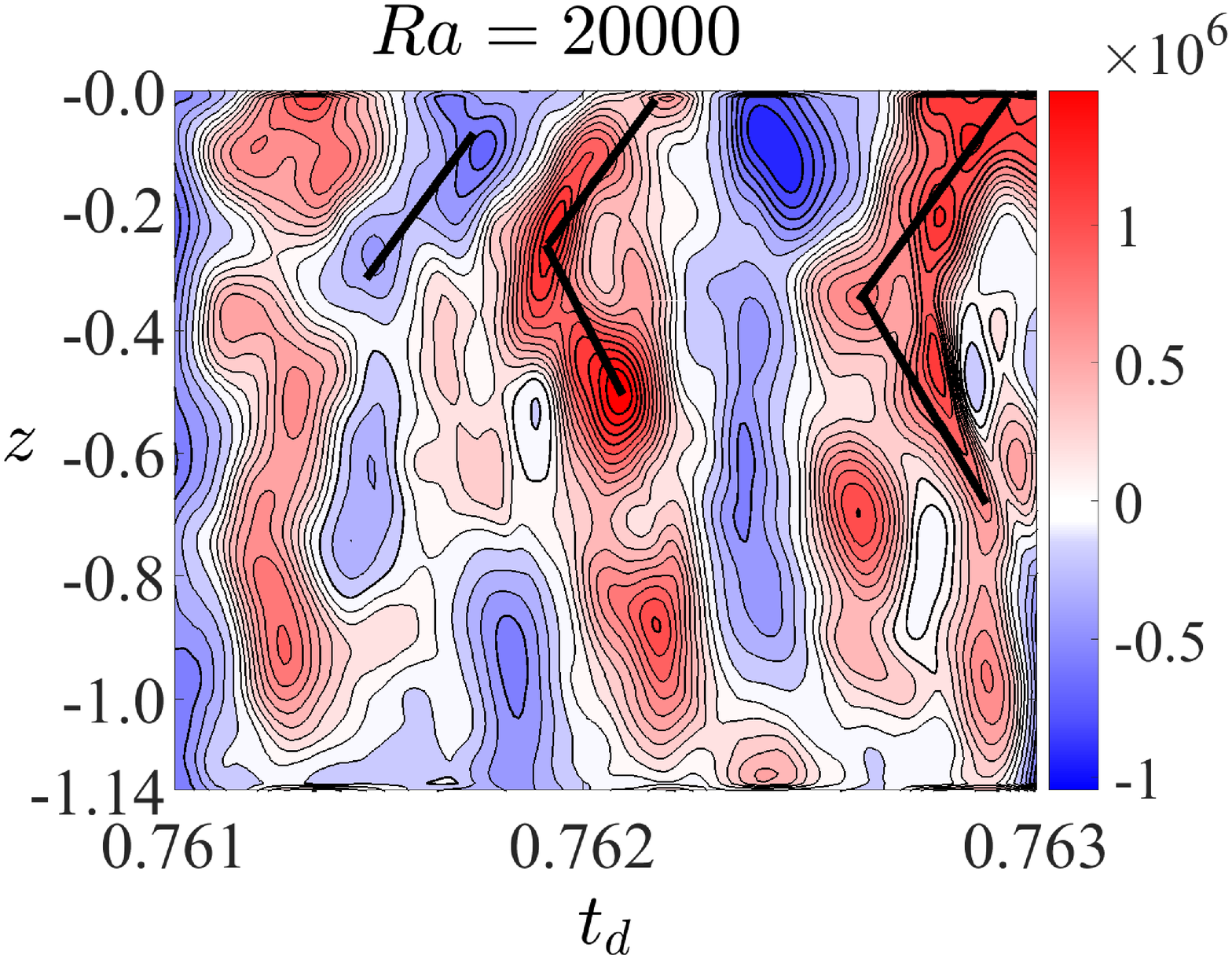}
\includegraphics[width=0.32\linewidth]{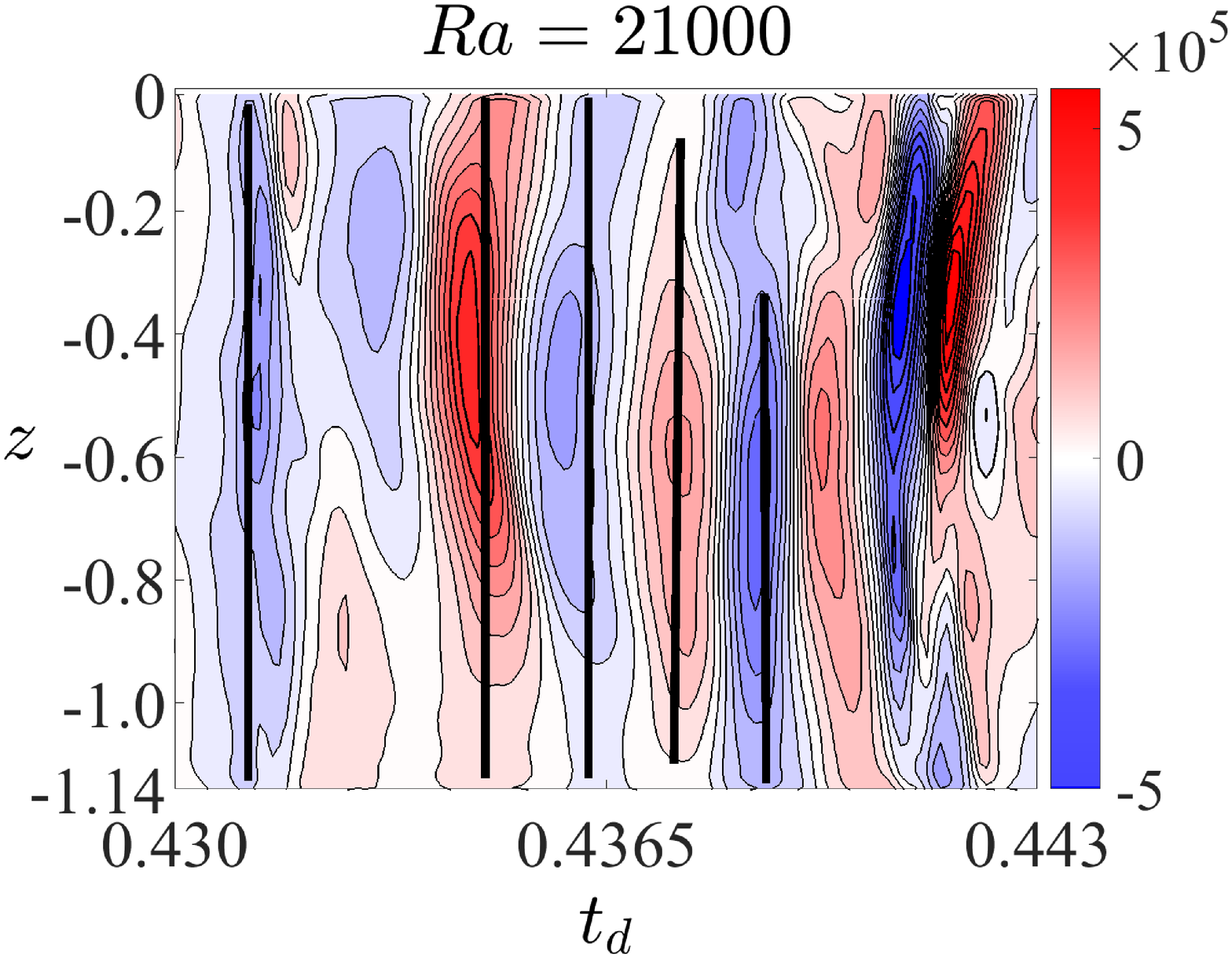}\\
\caption{Contour plots of $\partial{u}_z/\partial t$ at
cylindrical radius $s=1$  for $l\leq l_E$ and small intervals
of time in the saturated state of three dynamo simulations. 
The parallel black lines indicate the predominant
direction of travel of the waves and their slope gives the
group velocity $U_{g,z}$. The Rayleigh number $Ra$ of the simulation is given above each
panel. The other dynamo parameters are $E=6 \times 10^{-5}$, 
$Pr=Pm=5$. The estimated group 
velocity of the fast and slow MAC waves ($U_f$ and
$U_s$ respectively) and $U_{g,z}$ are given in table \ref{tablecg}. }
\label{grvel}
\end{figure}

Figure \ref{grvel} shows the measurement of wave motion
in the saturated state of dynamos at $E=6 \times 10^{-5}$ and $Pr=Pm=5$
and three values of $Ra$ spanning the dipole-dominated
regime and reversals. Contours of $\dot{u}_z$ at
cylindrical radius $s=1$ are plotted over small
time windows in which the ambient magnetic field
and wavenumbers are approximately constant.
These contours show the propagation paths
of the fluctuating $z$ velocity.
In line with the discussion so far, the measurement
of axial motions is limited to the energy-containing scales
$l \leq l_E$, with no restriction on the wavenumber.
The measured axial group velocity of the waves, $U_{g,z}$ -- obtained
from the slope of the black lines in figure \ref{grvel} -- 
is compared with the estimated fast ($U_f$) and slow ($U_s$)
group velocities obtained by taking the derivatives of the
respective frequencies in \eqref{ndr1} and \eqref{ndr2}
with respect to $k_z$ (table \ref{tablecg}). The theoretical
frequencies $\omega_f$ and $\omega_s$
are estimated using the three components of the magnetic field at the
peak-field location and the mean values of
$k_s$, $k_z$ and $m$ over the range of
energy-containing scales, $l \leq l_E$. 
In the dipole-dominated run at $Ra=6000$, 
the slow and fast waves coexist, but the intensity of slow wave
motions measured at the peak-field locations is at least
as high as that of the fast wave motions.
At $Ra=20000$, the increasing occurrence of the fast
waves alongside the slow waves is noted from the nearly vertical
propagation paths (figure \ref{grvel}(b)). The measured
slow wave group velocity at $Ra=20000$ is greater than that at
$Ra=6000$, which reflects the larger self-generated
field at the higher Rayleigh number. In the reversing
dynamo at $Ra=21000$, slow waves are totally absent while the fast
waves are abundant (figure \ref{grvel}(c)). 

\begin{figure}
	\centering
\hspace{-2.5 in}(a)  \hspace{2.5 in} (b) \\
\includegraphics[width=0.48\linewidth]{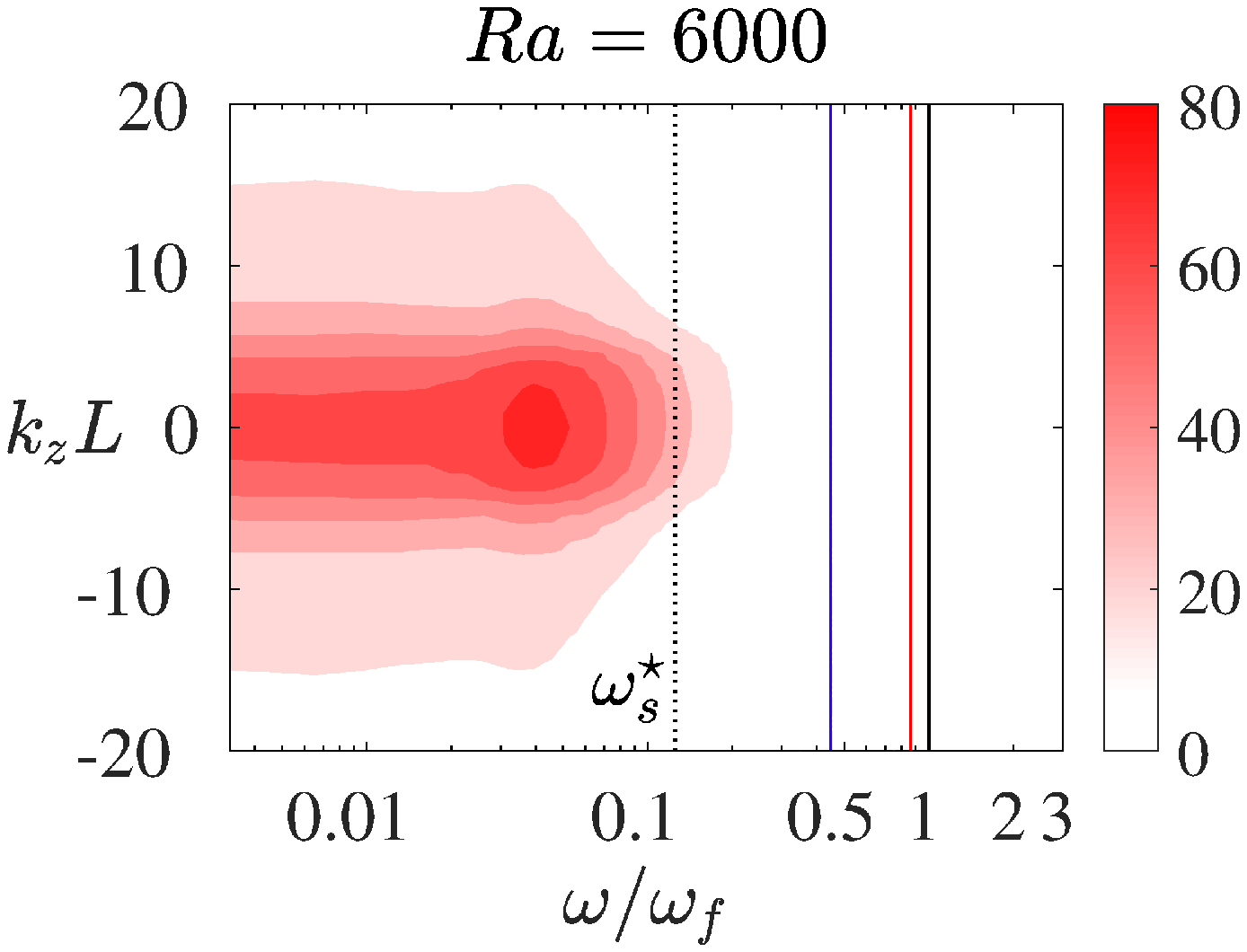}
	\includegraphics[width=0.48\linewidth]{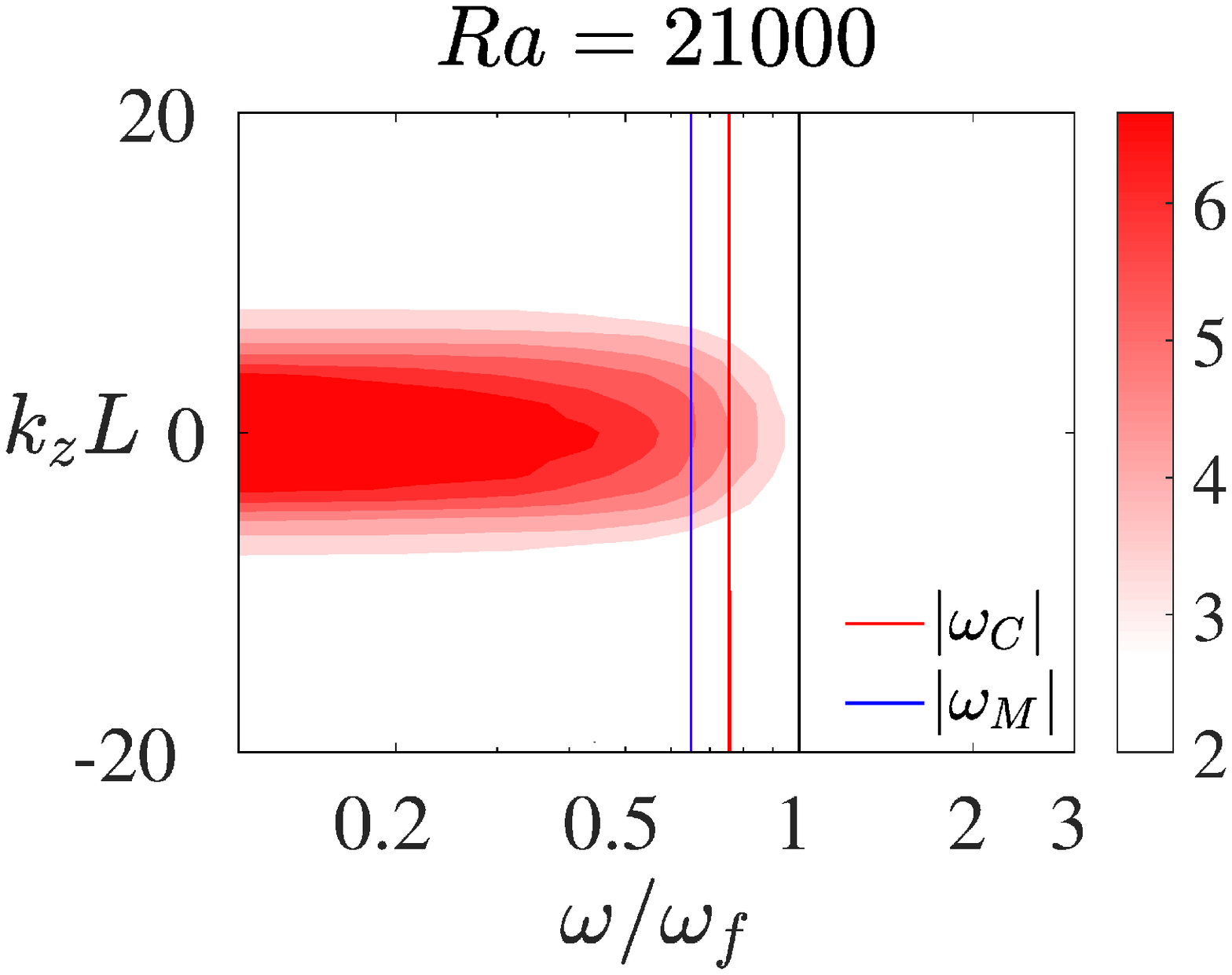}\\
	\caption{(a) FFT spectrum of $\partial u_z/\partial t$ at cylindrical radius 
		$s = 1$ for the scales $l \le l_E$. The spectra
are computed at discrete $\phi$ points and then averaged azimuthally. 
The solid vertical line in (a) and (b) 
		correspond to $\omega/\omega_f=1$ and the dashed vertical line in (a)
corresponds to $\omega_s^\star=\omega_s/\omega_f$, where $\omega_f$ and $\omega_s$ are the
estimated fast and slow MAC wave frequencies.
The coloured vertical lines 
give $\omega_C$ and $\omega_M$, both normalized by
$\omega_f$.
The Rayleigh number in the simulation is given above each panel. The other
dynamo parameters are $E = 6 \times 10^{-5}$ and $Pm=Pr=$5.  }
	\label{freqcg}
\end{figure}

The dominance of the slow MAC waves in 
the dipole-dominated dynamo and the fast MAC waves in the reversing
dynamo is further evident in figure \ref{freqcg}, where the fast
Fourier transform (FFT) of $\dot{u}_z$ is shown. The flow largely
consists of waves of frequency $\omega \sim \omega_s$ 
in the dipolar dynamo (figure \ref{freqcg}(a)),
whereas in the reversing dynamo, waves of much higher
frequency $\omega \sim \omega_f$ are dominant (figure \ref{freqcg}(b)).
{ The coloured vertical lines in figures \ref{freqcg}(a) and (b)
give the estimated magnitudes of $\omega_C$ and $\omega_M$ normalized by
$\omega_f$, where $\omega_M$ is based on the peak magnetic field.}

\begin{table}   
	\centering
	\begin{tabular}{lllllllllllll}
		& $E$& $Ra$& Fig.&$\omega_n^2$&
		$\omega_C^2$&$\omega_M^2$     &$-\omega_A^2$&$\omega_f$&$\omega_s$& $U_f$& $U_s$& $U_{g,z}$\\
		
		&&& No.&$(\times10^{10})$& $(\times10^{8})$& $(\times10^{8})$& $(\times10^{8})$&$(\times10^{4})$&$(\times10^{4})$&&&\\
		
	    	1&$6 \times 10^{-5}$ &6000  &\ref{grvel}(a) &1.6&22.27&5.98&3.67&5.48&0.67& 7454 &368& 391\\
		
	    	2&$6 \times 10^{-5}$ &20000 & \ref{grvel}(b)  &1.44&18.94&17.72&13.45&6.24&1.39& 7029 &2124&  1146\\
		
	    	3&$6 \times 10^{-5}$ &21000 &\ref{grvel}(c) &0.53&17.84&13.93&14.05&5.63&0& 7214 & -- & 10587 \\	
	\end{tabular}
\caption{Summary of the data for MAC wave measurement in the dynamo models.
 The sampling frequency
$\omega_n$ is chosen to ensure that the fast MAC waves are not missed 
in the measurement of group velocity. The values of $\omega_M^2$, $-\omega_A^2$ and
$\omega_C^2$ are calculated from \eqref{om2} using the mean values of
$m$, $k_s$ and $k_z$ over the range of
 energy-containing scales, $l \leq l_E$. The measured group
velocity in the $z$ direction ($U_{g,z}$) is 
compared with the estimated fast ($U_f$) 
or slow ($U_s$) MAC wave velocity.}
\label{tablecg}
\end{table}

\subsection{Self-similarity of the dipole--multipole transition}
\label{selfs}

\begin{figure}
		\centering
	\hspace{-2.6 in}	(a)  \hspace{2.3 in} (b) \\
	\includegraphics[width=0.48\linewidth]{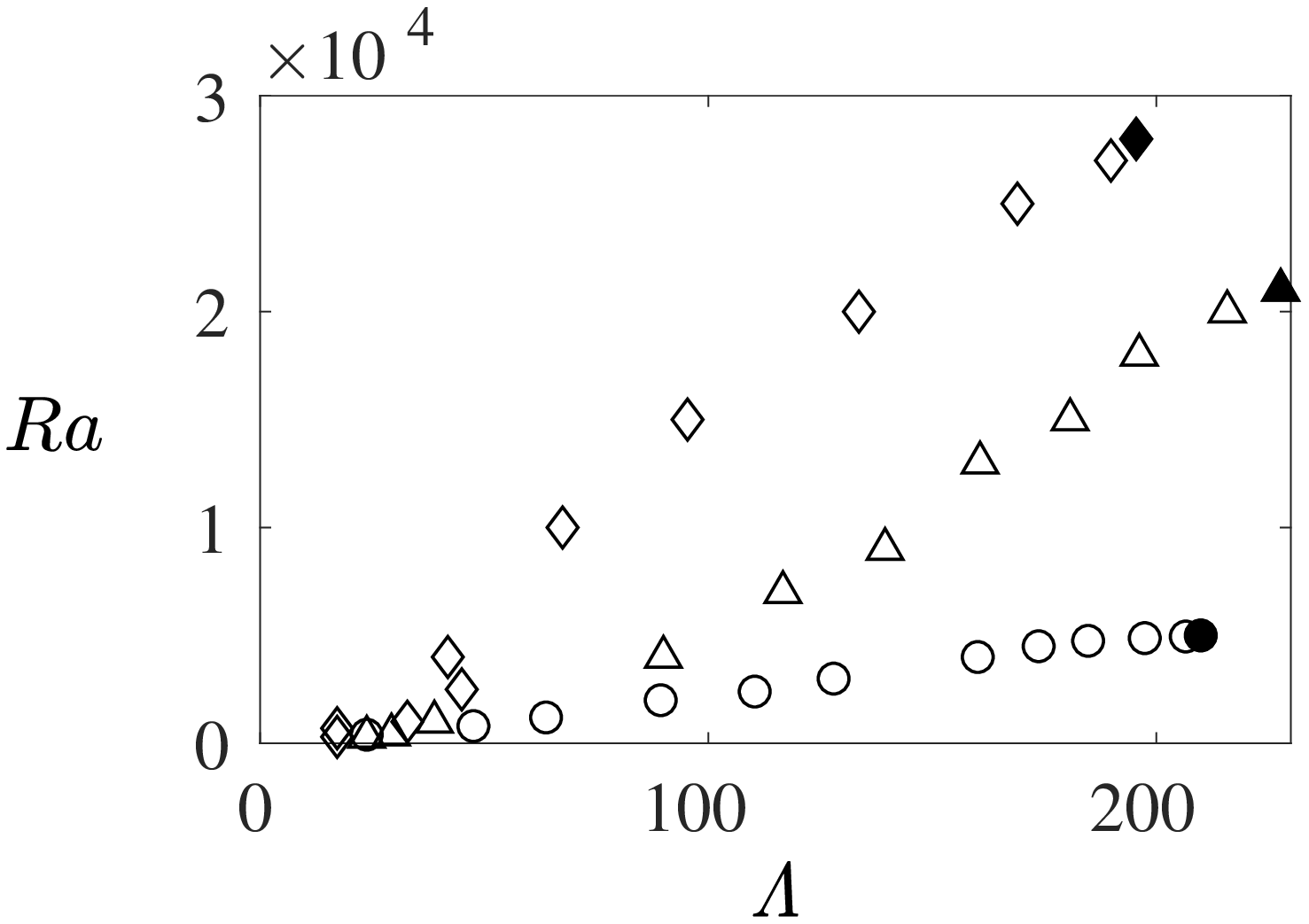}
	\includegraphics[width=0.48\linewidth]{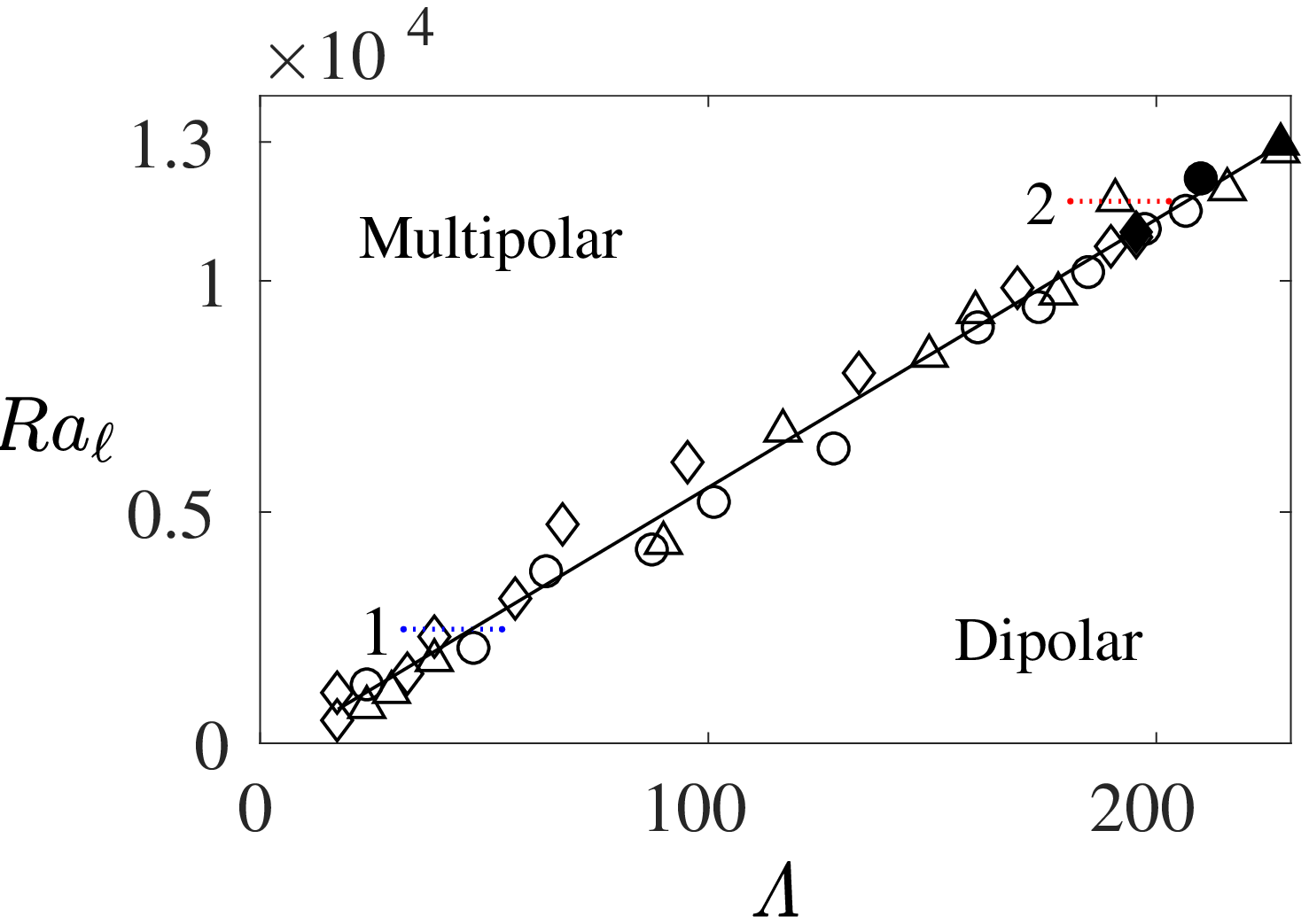}
	\caption{(a) Variation of the modified Rayleigh
number $Ra$  with the peak Elsasser number $\varLambda$
 (square of the
peak magnetic field) at both excitation and
suppression of slow MAC waves. 
The hollow symbols represent the states
where slow waves are first 
excited as the dynamos evolve from a small seed
field; the filled symbols represent the states
where slow waves are suppressed in the
polarity-reversing dynamos.
The parameters of the three dynamo series and their symbolic
representations are as follows:
$E=3\times10^{-4},Pm=Pr=20$ (circles), 
	$E=6\times10^{-5},Pm=Pr=5$ (triangles),  
	$E=1.2\times10^{-5},Pm=Pr=1$ (diamonds). 
(b) Variation of the local Rayleigh number $Ra_\ell$, defined
in \eqref{raell}, with $\varLambda$. The values of $Ra$, $Ra_\ell$
and $\varLambda$ in the plots are given in table \ref{parameters}.
 The sections 1 and 2
marked on the self-similar line are analysed further
in figures \ref{freqcg2} and \ref{brcmb} below.}
	\label{trans2}
	
\end{figure}

In the presence of small but finite magnetic diffusion,
the slow MAC waves are known to disappear in an unstably
stratified medium for $|\omega_A/\omega_M| \approx 1$. 
The same condition must hold for
the appearance of slow waves in a medium where the magnetic
field progressively increases from a small value.
 In simulations starting
from a small seed field, the earliest time of excitation of
the slow MAC waves is noted from group velocity measurements
at closely spaced times during the growth phase of the dynamo.
The peak Elsasser number $\varLambda=B_{peak}^2$ 
at this time is obtained from the
three components of the field at the peak-field location,
and presented in the last column of table \ref{parameters}.
In each of the three dynamo series considered in this study,
the last run is a polarity-reversing dynamo, for which
$\varLambda$ is the measured 
peak Elsasser number when slow
MAC waves cease to exist in the run starting from the
saturated (strong field) 
state of the penultimate run in that
series. 
Figure \ref{trans2}(a) shows that the variation of $Ra$ with
$\varLambda$ in the three dynamo series 
is nearly linear. The Rayleigh number corresponding to
reversals (at which slow MAC waves disappear) also
lies on this line, indicating that the appearance and
disappearance of MAC waves are in { similar} regimes.

Following the analysis in figure \ref{trans1}, where the Rayleigh
number based on the length scale of the buoyant perturbation was studied,
we define a local Rayleigh number in the dynamo,
\begin{equation}
Ra_\ell = \dfrac{g \alpha \beta}{2 \varOmega \eta}\left(\dfrac{2 \pi }{\bar{m}}\right)^2,
\label{raell}
\end{equation}
where $\bar{m}$ is the mean spherical harmonic order evaluated
over $m$ within the energy-containing scales $l \leq l_E$.
The behaviour of $Ra_\ell$, which is defined for the scales where
the MAC waves are excited by buoyancy, 
is approximately self-similar
(figure \ref{trans2}(b)).
The values of $Ra_\ell$ at the
onset of polarity reversals, where the slow MAC waves disappear,
also lie on the same self-similar branch. While the conventional
Rayleigh numbers $Ra$ at the onset of reversals in the three
dynamo series lie far apart (see the
filled symbols in figure \ref{trans2}(a)), 
the respective local Rayleigh numbers $Ra_\ell$ are remarkably
close, and $\sim 10^4$ (see the 
filled symbols in figure \ref{trans2}(b)
and table \ref{parameters}). The magnetic
Ekman number based on $\bar{m}$ in the three dynamo series
takes values of
$E_\eta \sim 10^{-5}$ for a wide range of $Ra$, which indicates
an energy-containing length scale $\sim$ 10 km for Earth 
(see \S 1).

\begin{figure}
	\centering
		\centering
	\hspace{-2.5 in}	(a)  \hspace{2.5 in} (b) \\
	\includegraphics[width=0.48\linewidth]{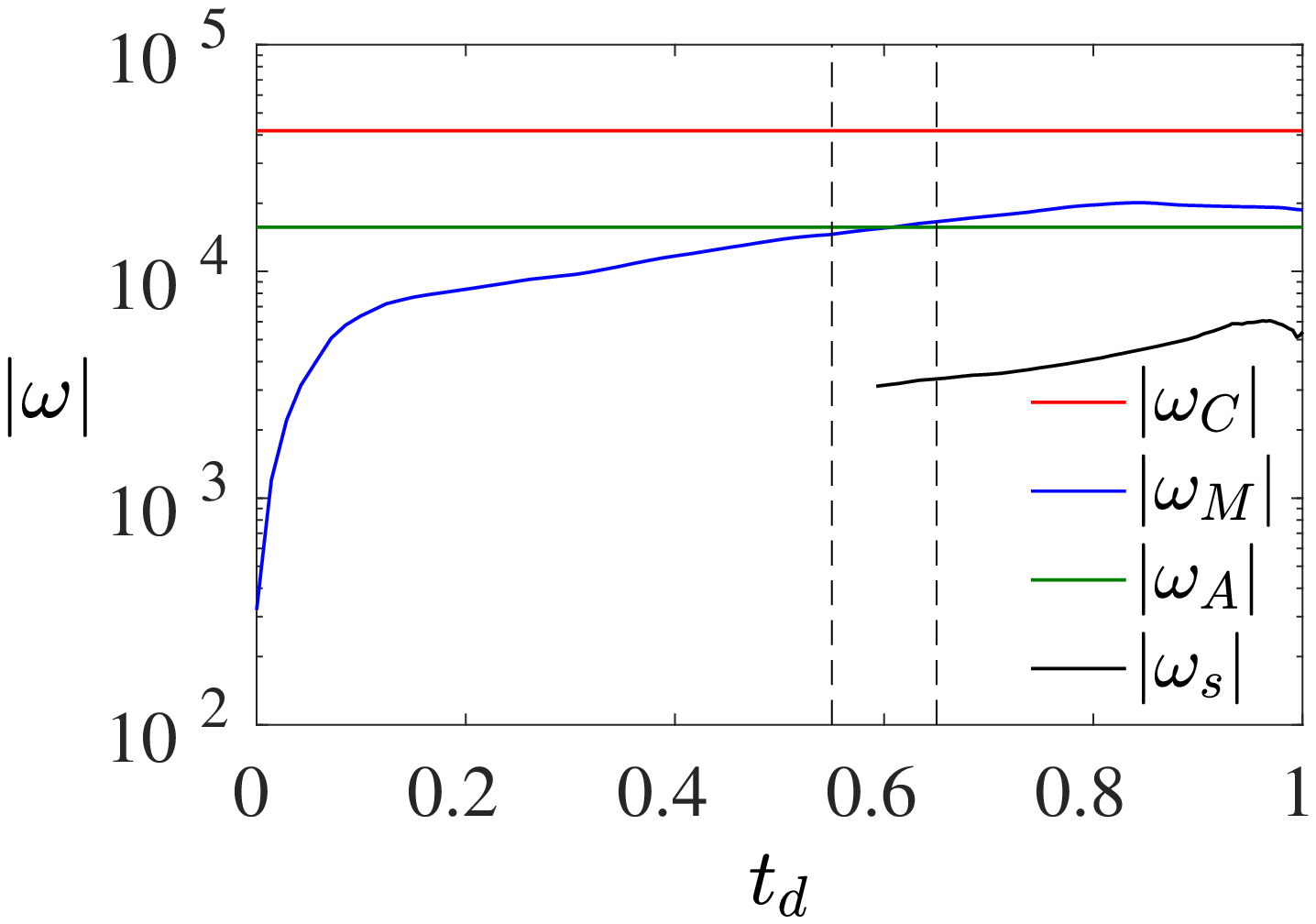}
	\includegraphics[width=0.48\linewidth]{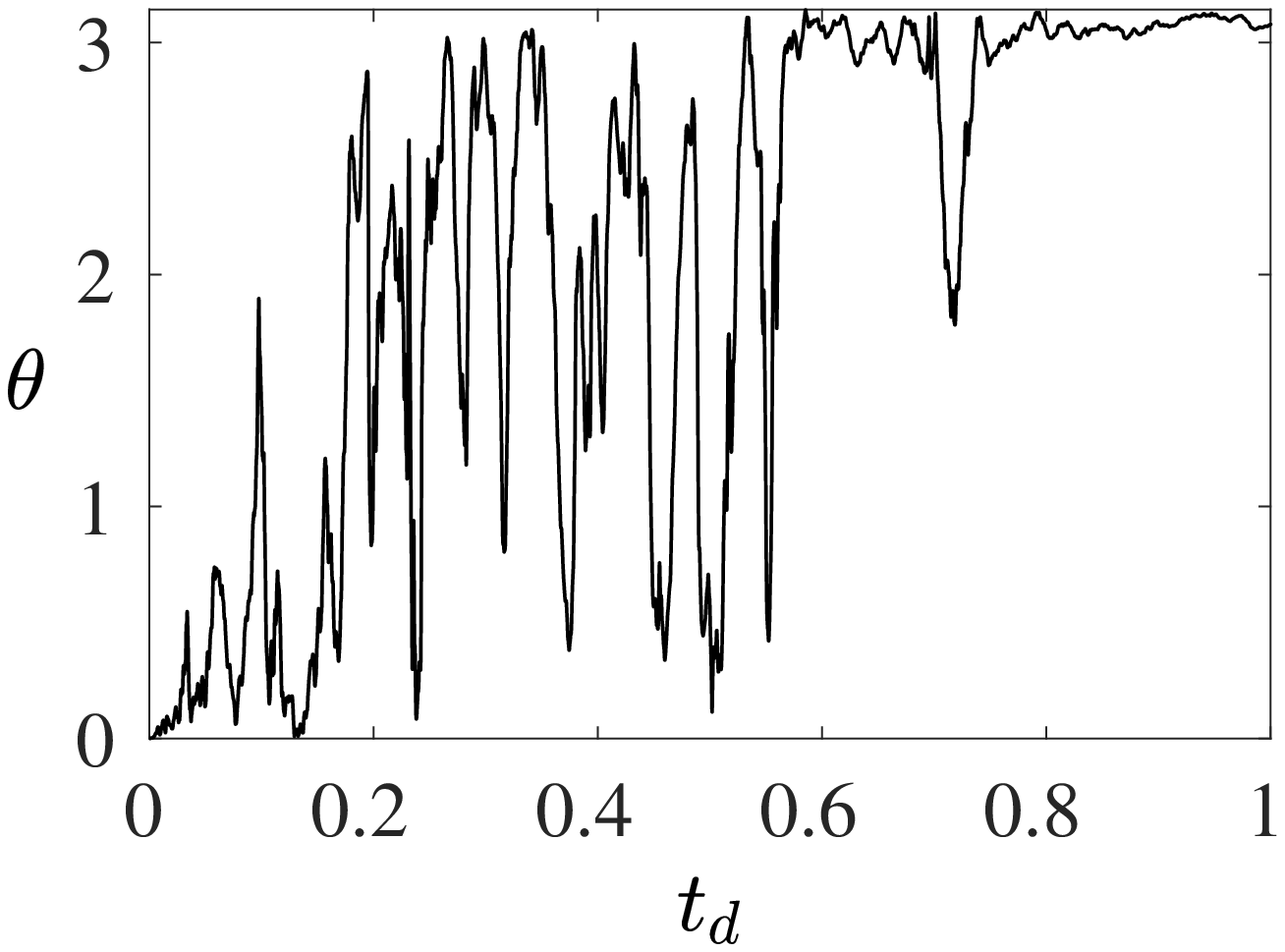}\\
	\hspace{-2.5 in}(c)  \hspace{2.5 in} (d) \\
	\includegraphics[width=0.48\linewidth]{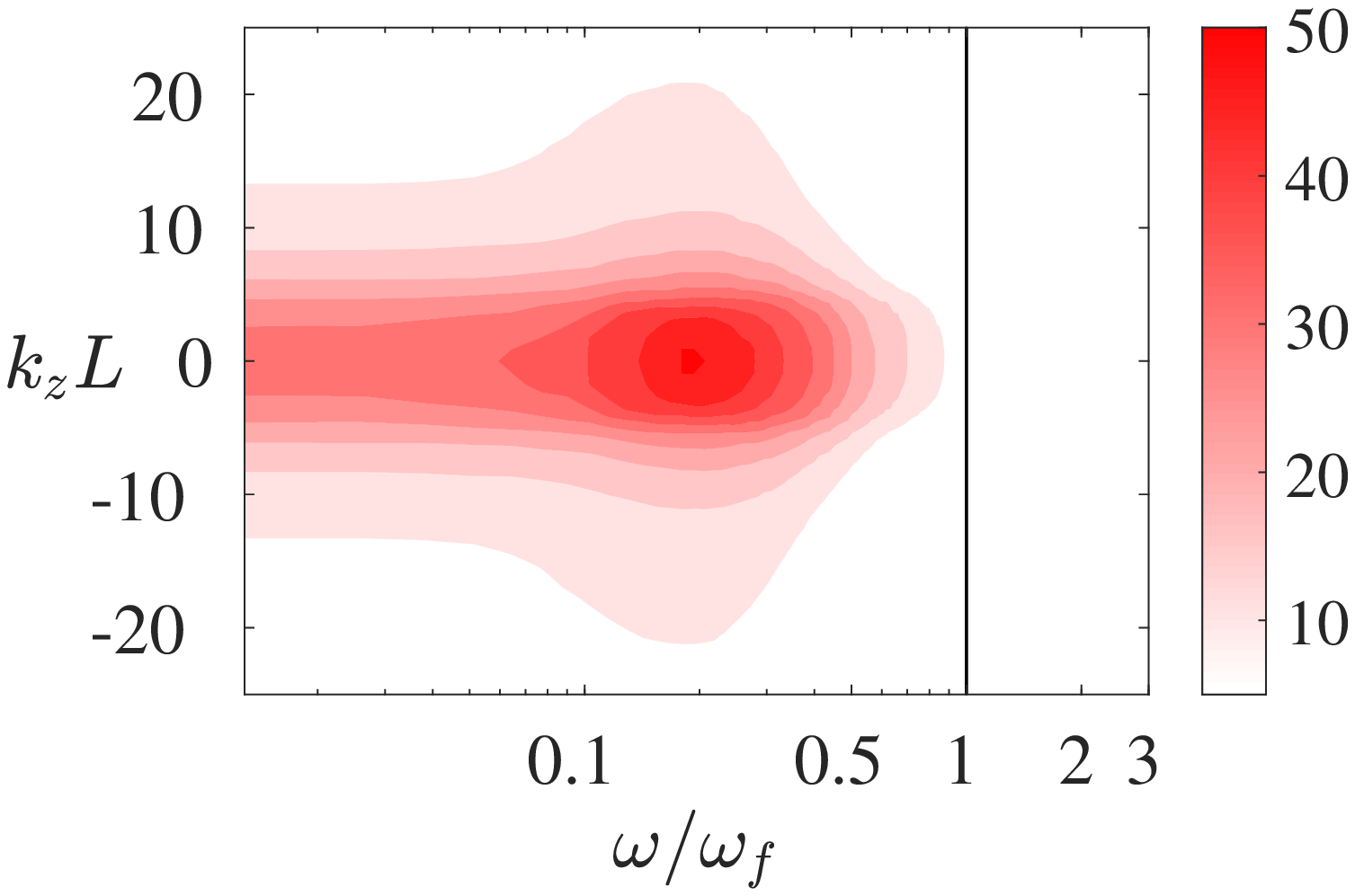} 
	\includegraphics[width=0.48\linewidth]{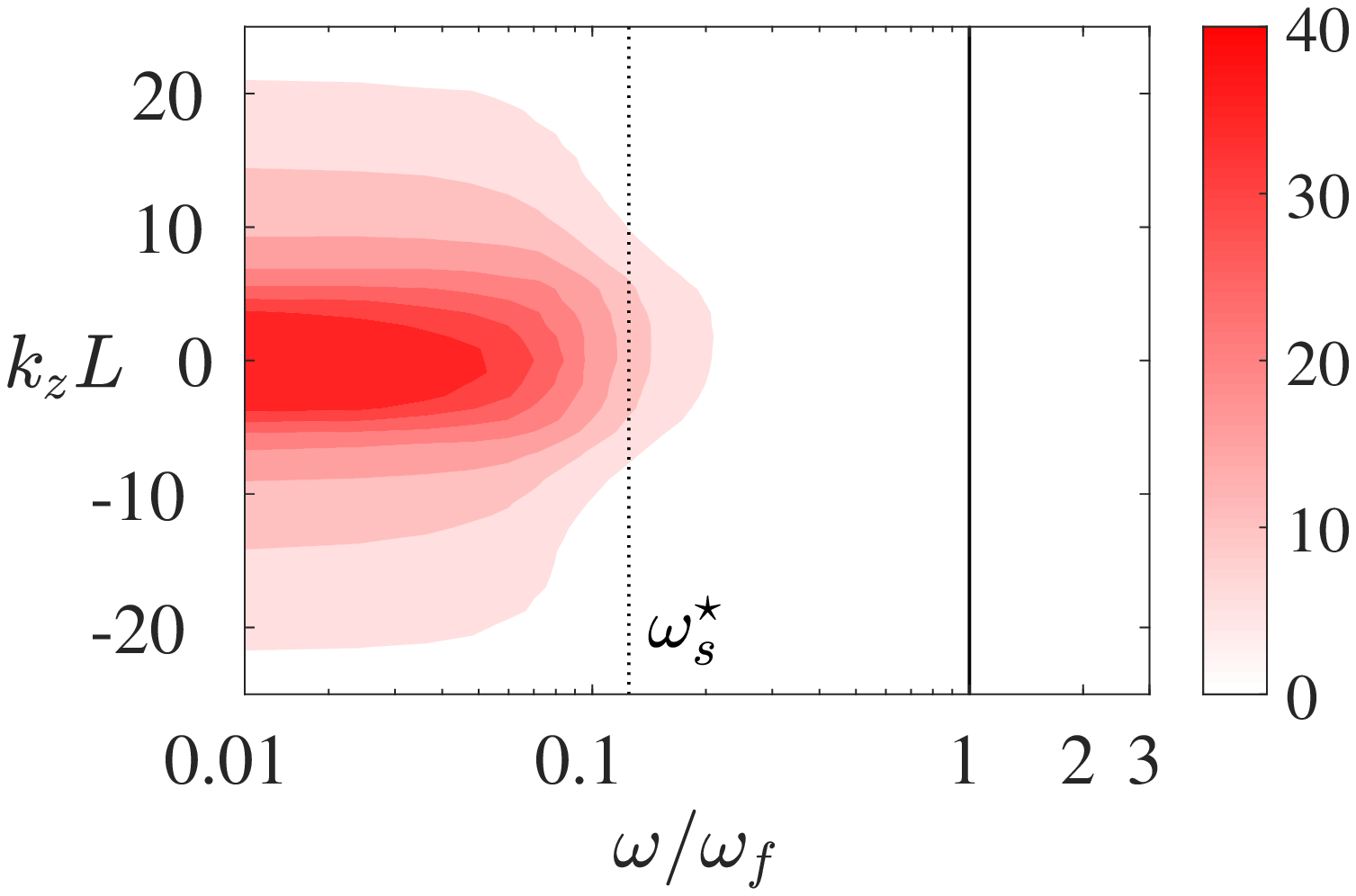}\\ \vspace{-4mm}
\caption{(a) Evolution the dynamo frequencies in a simulation
beginning from a small seed magnetic field at
$E = 1.2 \times 10^{-5}$, $Pm=Pr=1$, $Ra=4000$. The two dashed
vertical lines at $t_d=0.55 $ and $t_d=0.66$
represent the end points of section 1 marked
in figure \ref{trans2}(b) with peak Elsasser
numbers $\varLambda =$ 32 and 56 respectively.
(b) Evolution of the dipole colatitude in the above
simulation.{ (c) \& (d) FFT spectra of 
$\partial u_z/\partial t$ at cylindrical 
	radius $s = 1$ for the scales $l \le l_E$
at $\varLambda =$ 32 and 56 respectively.
The spectra are computed at discrete $\phi$ points and then
averaged azimuthally. In (d), $\omega_s^* = \omega_s/\omega_f$,
where $\omega_f$ and $\omega_s$ are the estimated fast and
slow MAC wave frequencies.}}
	\label{freqcg2}
\end{figure}

\begin{figure}
	\centering
	\hspace{-2.5 in}	(a)  \hspace{2.5 in} (b) \\
	\includegraphics[width=0.48\linewidth]{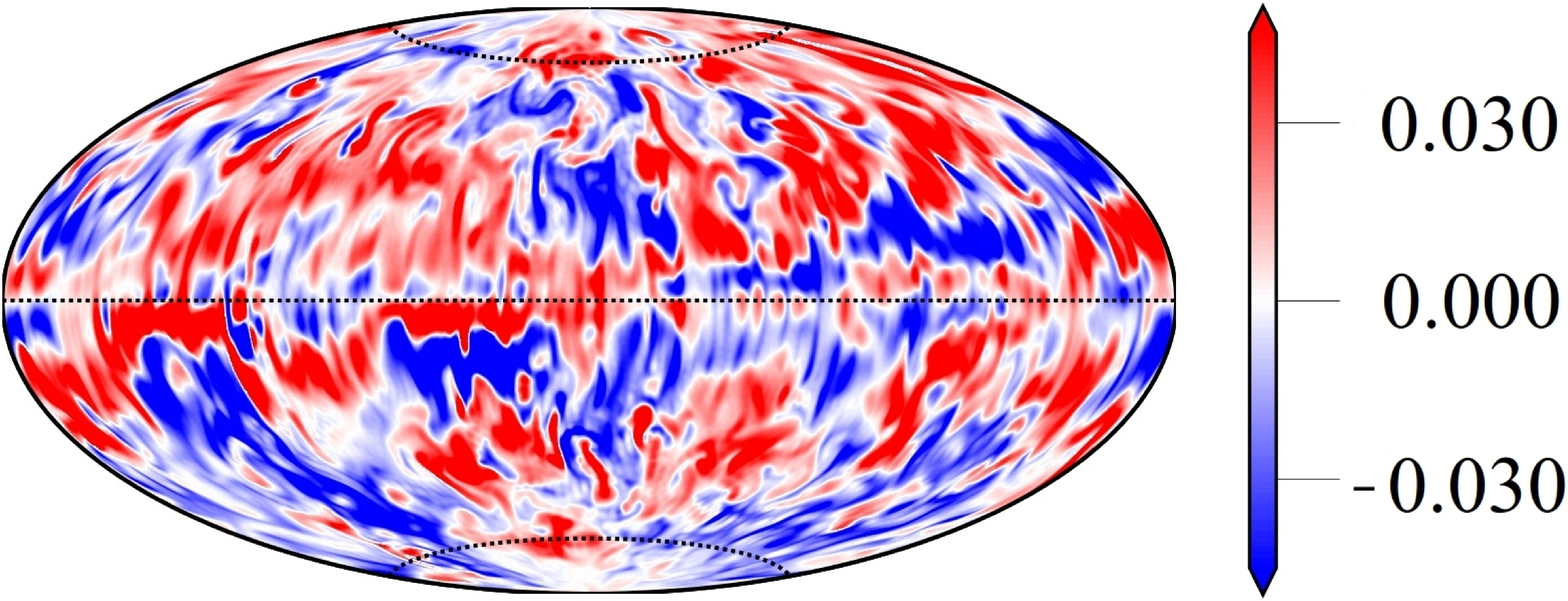}
	\includegraphics[width=0.48\linewidth]{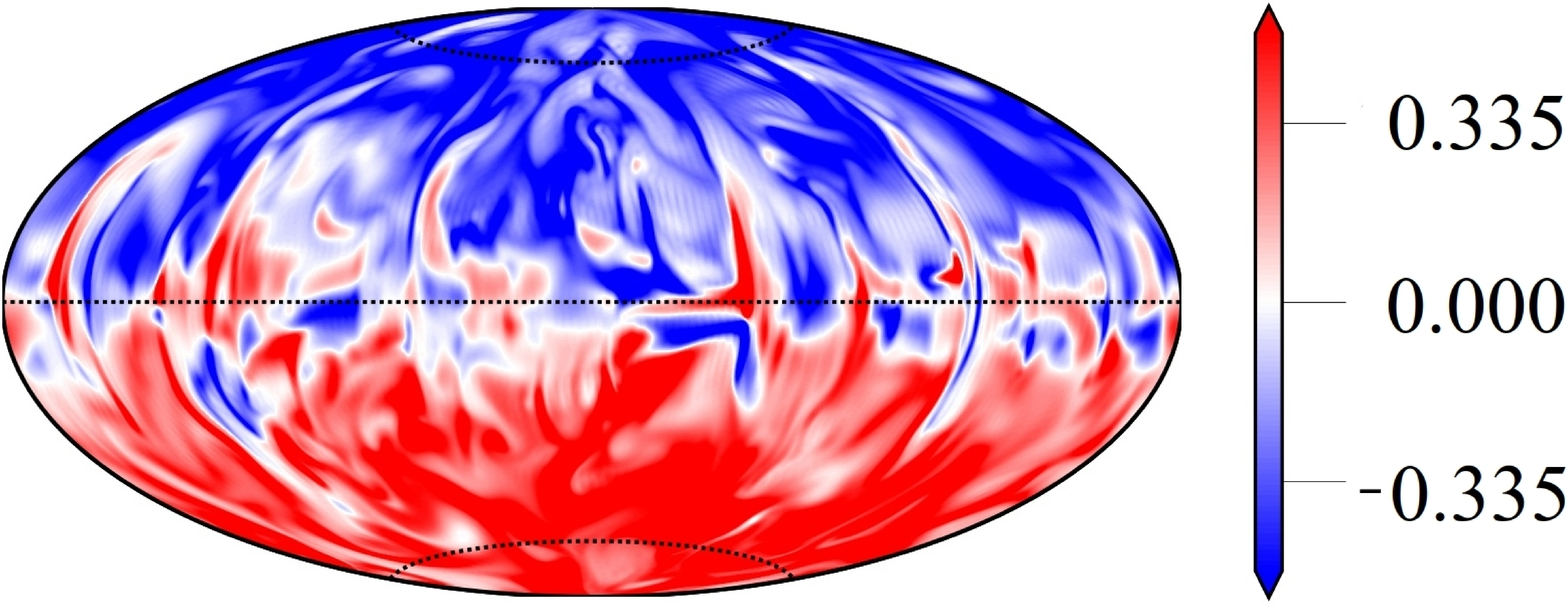}\\
	\hspace{-2.5 in}	(c)  \hspace{2.5 in} (d) \\
	\includegraphics[width=0.48\linewidth]{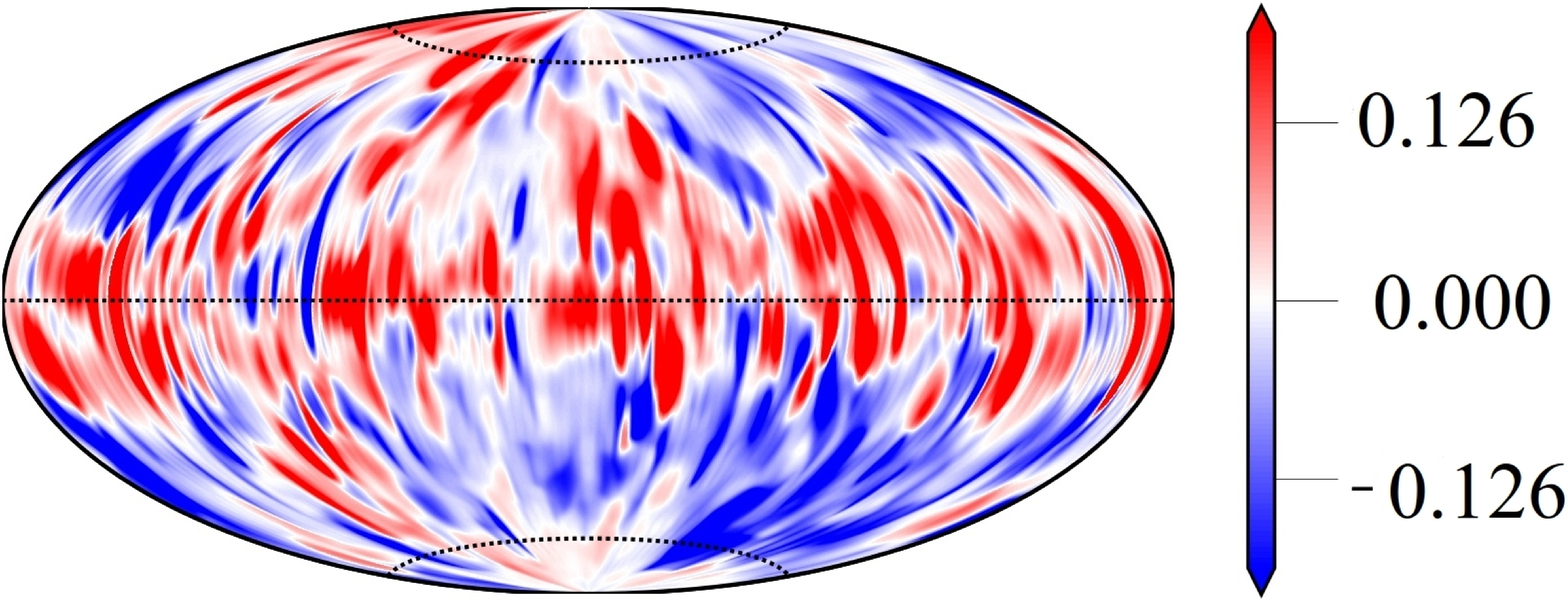}
	\includegraphics[width=0.48\linewidth]{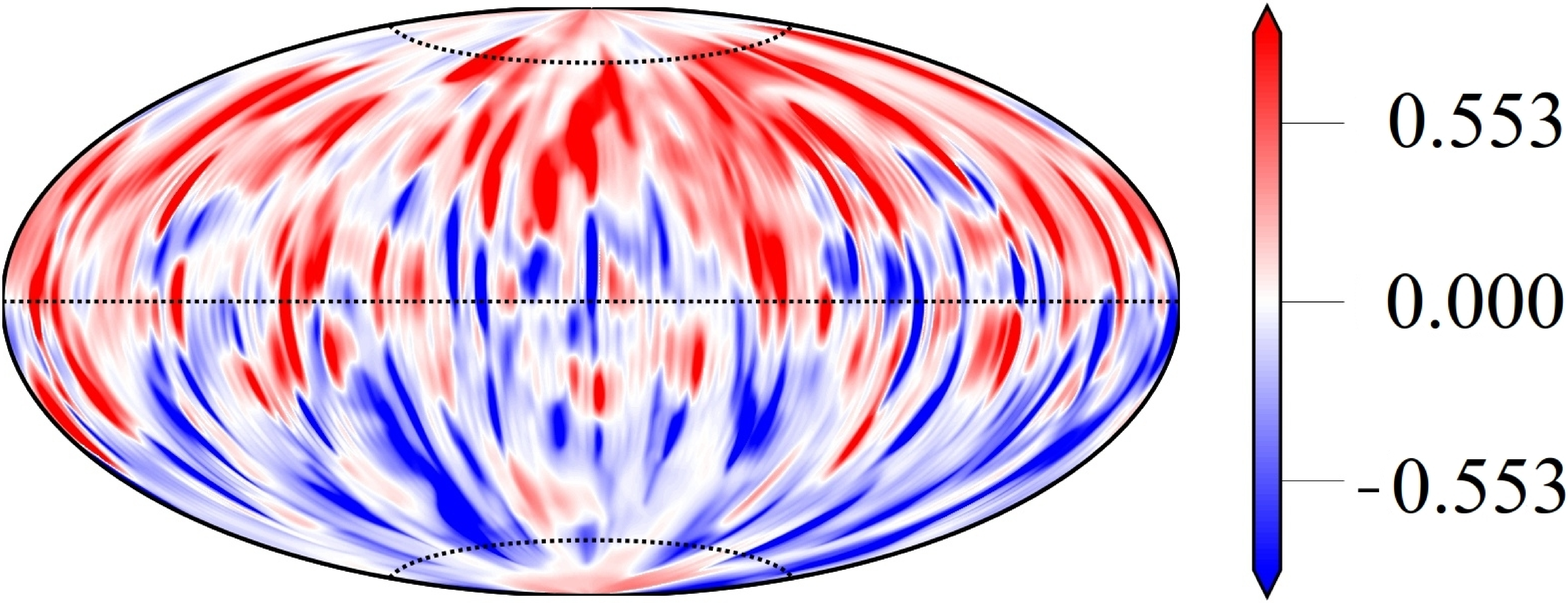}\\
	\caption{Shaded contours of the radial magnetic field at the
outer boundary in two dynamo simulations.
(a) and (b): $Ra_\ell=2505$, corresponding to
section 1 in figure \ref{trans2}(b), at $\varLambda=$ 
32 and 56 respectively.
The dynamo parameters are 
$E=1.2\times10^{-5}$, $Ra=4000$, $Pm=Pr=1$. 
(c) and (d): $Ra_\ell=11740$, corresponding to
section 2 in figure \ref{trans2}(b), at $\varLambda=$ 
187 and 216 respectively.
The dynamo parameters are 
$E=6\times10^{-5}$, $Ra=18000$, $Pm=Pr=5$. }
	\label{brcmb}
\end{figure}

The fact that the linear 
variation of $Ra_\ell$ with $\varLambda$
demarcates the boundary between dipolar and multipolar
states is evident by traversing the sections 1 and 2 marked on
the $Ra_\ell$ line from left to right (figure \ref{trans2}(b)).
In practice, this is done by following the evolution
of the dynamo from a small seed field. Figure \ref{freqcg2}(a)
shows the section 1 within dashed vertical lines, where
$|\omega_M|$ crosses $|\omega_A|$. The variation of the
dipole colatitude $\theta$, shown in figure \ref{freqcg2}(b),
indicates a multipolar field until this crossing, and a stable
dipole thereafter.  While the
flow is predominantly made up of fast MAC waves of
frequency $\omega \sim \omega_f$ before the transition,
the slow waves of frequency $\omega \sim \omega_s$
are dominant after the transition (figures \ref{freqcg2}(c) 
\& (d)).  The multipole--dipole
transitions are further evident in the contour plots of the 
radial magnetic field at the outer boundary,
given in figure \ref{brcmb} for sections 1 and 2 
marked in figure \ref{trans2}(b).

\section{Concluding remarks}
 \label{concl}

The present study investigates the dipole--multipole transition
in rotating dynamos through the analysis of MHD
wave motions. The limit of small Rossby number, based
not only on the planetary core depth but also on the length scale
of core convection, is considered. 
In this inertia-free limit, the dynamo 
polarity
depends on the relative magnitudes of $\omega_M$ and $\omega_A$,
which in turn depend on the peak intensity of the 
self-generated field and
the strength of buoyant forcing in the unstably stratified fluid
layer. The linear magnetoconvection model shows that
the slow MAC (magnetostrophic) waves have greater
helicity than the fast MAC waves for 
$|\omega_C| > |\omega_M|
> |\omega_A| > |\omega_\eta|$, the regime
thought to be relevant to dipole-dominated dynamos. Although
buoyancy-induced helicity 
can be spatially segregated
about the equator 
by linear inertial waves in the absence of the 
magnetic field \citep{davidson2018spatial},
this purely hydrodynamic process is not likely
to generate the axial dipole in rotating dynamos, 
where the helicity of slow MAC waves is at least as
high as that of the fast waves of frequency
$\sim \omega_C$, which already exist in the
seed field state (figure S2 in Supplementary Material).
Moreover, the time scale in which the slow waves establish the dipole
field can be a significant 
fraction of the magnetic diffusion time (see figures
\ref{relhel}(a) \& (b)),
whereas the fast waves would have to form the dipole
on a much shorter time scale. Since the
slow MAC waves are produced by 
magnetostrophic balances at the peak field
locations, the analysis of wave motions with $\omega_M$
based on the peak field intensity is performed here
rather than a scale-dependent analysis of volume-averaged
forces \citep[e.g.][]{schwaiger2019}.
The present study shows that the condition
of vanishing slow waves defines
polarity transitions in rapidly rotating dynamos.
The variation of the local Rayleigh number $Ra_\ell$ with
the Elsasser number
$\varLambda$ at the transition 
is very similar in the linear 
magnetoconvection and dynamo models, which is 
remarkable given that 
the Alfv\'en frequency $\omega_M$ 
is determined by the imposed field
in the linear model and by the self-generated
field in the dynamo model.

While the onset of slow magnetostrophic waves
for $|\omega_C| > |\omega_M| \approx |\omega_A| > |\omega_\eta|$ 
is known to
produce the axial dipole from a chaotic multipolar
state (see figures \ref{freqcg2}(a,b) \& \ref{brcmb}), 
the suppression of slow waves
for the same condition (figure \ref{compare_ra}(c))
leads to polarity
reversals in strongly driven dynamos.
 This study differentiates between polarity-reversing
and multipolar solutions, both of which onset at
the same Rayleigh number. Polarity reversals
result from the suppression of slow waves that already
exist in a strong field state of mean square
intensity $O(1)$, whereas the multipolar solution forms
from a seed field when the 
forcing is sufficiently strong that
the slow waves are never excited. Multipolar
solutions are also obtained at stronger forcing, regardless
of whether one begins from a seed field or a 
strong field.

The critical 
ratios of forces and energies in the dynamo have been
instrumental in setting criteria for polarity transitions.
Earlier studies \citep{olson2006dipole,chraub2006}
have proposed a critical value $\approx 0.1$
for the local Rossby number $Ro_\ell$, above which
 the dipole--multipole
transition occurs. However, this condition relies on
the finiteness of nonlinear inertia
on the scale of the
buoyancy perturbations, which we consider unlikely in 
rapidly rotating planetary cores. 
 While the
condition of the energy ratio $E_k/E_m >1$ 
(table \ref{parameters}) by itself
is admissible as a 
criterion for the dipole--multipole transition 
\citep{kutzner2002,tassin2021},
the increase of this ratio is likely a consequence
of the transition rather than a cause for it. 
Polarity reversals are never obtained for $E_k/E_m <1$,
so it is not certain that the reversals in the present
study are Earth-like.

The self-similarity of polarity reversals 
in the inertia-free regime can place
a useful constraint on the Rayleigh number that
admits reversals in the Earth. For $Ra_\ell \sim 10^4$,
the classical Rayleigh number $R=Ra/E  \, \sim 10^{17}$,
taking a turbulent viscosity $\nu \approx \eta$ and a 
plausible ratio of core depth to the convective length scale, 
$L/\delta \sim 10^2$. { That having been said,
the Earth's core is thought to convect in
response to large lateral variations in lower-mantle
heat flux \citep[e.g.][]{olson2015,mound2019}, which
can potentially reduce 
the value of $Ra$ for reversals.}  Our understanding
of the convective regime for polarity reversals is
not complete, but the idea that
 polarity transitions are self-similar
 would eventually lead to improved parameter constraints 
for reversals.

\section*{Acknowledgments}
 This study was supported by Research Grant
 MoE-STARS/STARS-1/504 under Scheme for
 Transformational and Advanced Research 
in Sciences awarded by the Ministry of Education, India. 
DM's doctoral studentship is granted by
the Council of Scientific and Industrial Research, India.
The computations were performed on SahasraT and
Param Pravega, the supercomputers at the 
Indian Institute of Science, Bangalore. The
authors thank three anonymous reviewers for their
thoughtful comments and suggestions.

\section*{Declaration of interests}
The authors report no conflict of interest.

\appendix

\section{Calculation of the initial wavenumber, $k_0$}
\label{k0}
The Fourier transform of the initial condition \eqref{pert} is,
\begin{equation}
	\hat{\Theta}_0 = \pi^{3/2} \delta^3 \exp{\left(-\frac{k^2\delta^2}{4} \right)},
\end{equation}
where $k=\sqrt{k_x^2+k_y^2+k_z^2}$. 

The initial wavenumber is defined by,
\begin{equation}
	k_0=\left[\dfrac{\displaystyle \int_{\!\!-\infty}^{\infty} \displaystyle \int_{\!\!-\infty}^{\infty}
\displaystyle \int_{\!\!-\infty}^{\infty} k^2 |\hat{\Theta}_0|^2
\,\mbox{d}k_x\,\mbox{d}k_y\,\mbox{d}k_z}{\displaystyle \int_{\!\!-\infty}^{\infty}
 \displaystyle\int_{\!\!-\infty}^{\infty}
\displaystyle \int_{\!\!-\infty}^{\infty}|\hat{\Theta}_0|^2\, \mbox{d}k_x\, \mbox{d}k_y\, \mbox{d}k_z}\right]^{1/2}.
\end{equation}
Letting $k_x=k \sin\phi \cos\theta$, $k_y= k \sin\phi \sin\theta$, $k_z = k \cos\phi$, 
we obtain
\begin{eqnarray}
	k_0 &=& \left[\frac{\displaystyle \int_{0}^{2 \pi} \displaystyle \int_{0}^{\pi}
\displaystyle \int_{0}^{\infty} k^2 |\pi^{3/2}\delta^3\exp{\left(-k^2\delta^2/4 \right)}|^2
k^2 \sin\phi \,\mbox{d}k\, \mbox{d}\theta\, \mbox{d}\phi}{\displaystyle \int_{0}^{2 \pi}
\displaystyle \int_{0}^{\pi} \displaystyle \int_{0}^{\infty}|\pi^{3/2}\delta^3  \exp{\left(-k^2\delta^2/4 \right)}|^2k^2 
\sin\phi\, \mbox{d}k\, \mbox{d}\theta\, \mbox{d}\phi}\right]^{1/2}, \\
 &=& \sqrt{3}/\delta,
\end{eqnarray}
on evaluation of the integrals.

\section{Equations for nonmagnetic convection}
\label{nmeqns}
For $Pr=Pm$, the convection-driven dynamo given by equations
\eqref{momentum}--\eqref{div} can be compared with
nonmagnetic convection given by the equations
\begin{align}
	E Pr^{-1}  \Bigl(\frac{\partial {\bm u}}{\partial t} + 
	(\nabla \times {\bm u}) \times {\bm u}
	\Bigr)+  {\hat{\bm{z}}} \times {\bm u} = - \nabla p^\star +
	Ra \, T \, {\bm r} \, + E\nabla^2 {\bm u}, \label{mom1} \\
	\frac{\partial T}{\partial t} +({\bm u} \cdot \nabla) T =  \nabla^2 T,  \label{h1}\\
	\nabla \cdot {\bm u} = 0.  \label{div1}
\end{align}
Here, lengths are scaled by the thickness of the spherical shell $L$, 
time is scaled by $L^2/\kappa$, the velocity $\bm{u}$ is scaled by $\kappa/L$
and $p^\star= \frac1{2} E Pr^{-1} |\bm{u}|^2$.

For a magnetic (dynamo) calculation with the parameters
 $E=6 \times 10^{-5}$, $Pm=Pr=5$, $Ra=1000$, 
 the equivalent nonmagnetic calculation
has the parameters $E=6 \times 10^{-5}$, $Pr=5$, $Ra=1000$.

\clearpage

\newpage


\bibliography{library3}

\begin{thebibliography}{47}
\expandafter\ifx\csname natexlab\endcsname\relax\def\natexlab#1{#1}\fi
\def\au#1{#1} \def\ed#1{#1} \def\yr#1{#1}\def\at#1{#1}\def\jt#1{\textit{#1}}
  \def\bt#1{#1}\def\bvol#1{\textbf{#1}} \def\vol#1{#1} \def\pg#1{#1}
  \def\publ#1{#1}\def\arxiv#1{#1}\def\org#1{#1}\def\st#1{\textit{#1}}

\bibitem[Bardsley \& Davidson(2017)]{bardsley2017}
{\sc \au{Bardsley, O.~P.} \& \au{Davidson, P.~A.}} \yr{2017}  \at{The
  dispersion of magnetic-coriolis waves in planetary cores}.  \jt{Geophys. J.
  Int.}  \bvol{210}~(1),  \pg{18--26}.

\bibitem[Braginsky(1967)]{brag1967}
{\sc \au{Braginsky, S.~I.}} \yr{1967}  \at{Magnetic waves in the {E}arth's
  core}.  \jt{Geomagn. Aeron.}  \bvol{7},  \pg{851--859}.

\bibitem[Braginsky \& Roberts(1995)]{95brag}
{\sc \au{Braginsky, S.~I.} \& \au{Roberts, P.~H.}} \yr{1995}  \at{Equations
  governing convection in {E}arth's core and the geodynamo}.  \jt{Geophys.
  Astrophys. Fluid Dyn.}  \bvol{79},  \pg{1--97}.

\bibitem[Buffett \& Bloxham(2002)]{buffett2002energetics}
{\sc \au{Buffett, B.~A.} \& \au{Bloxham, J}} \yr{2002}  \at{Energetics of
  numerical geodynamo models}.  \jt{Geophys. J. Int.}  \bvol{149}~(1),
  \pg{211--224}.

\bibitem[Bullard \& Gellman(1954)]{bullard1954}
{\sc \au{Bullard, E.~C.} \& \au{Gellman, H.}} \yr{1954}  \at{Homogeneous
  dynamos and terrestrial magnetism}.  \jt{Philos. Trans. R. Soc. London, Ser.
  A}  \bvol{247}~(928),  \pg{213--278}.

\bibitem[Busse {\em et~al.\/}(2007)Busse, Dormy, Simitev \&
  Soward]{07bussechapter}
{\sc \au{Busse, F.}, \au{Dormy, E.}, \au{Simitev, R.} \& \au{Soward, A.}}
  \yr{2007}  \at{Dynamics of rotating fluids}.  \bt{In {\em Mathematical
  Aspects of Natural Dynamos\/} (ed. \ed{E.~Dormy \& A.~M. Soward})},  \st{The
  Fluid Mechanics of Astrophysics and Geophysics},  \vol{vol.~13},  \pg{pp.
  165--168}.  \publ{CRC Press}.

\bibitem[Christensen \& Aubert(2006)]{chraub2006}
{\sc \au{Christensen, U.~R.} \& \au{Aubert, J.}} \yr{2006}  \at{Scaling
  properties of convection-driven dynamos in rotating spherical shells and
  application to planetary magnetic fields}.  \jt{Geophys. J. Int.}
  \bvol{166}~(1),  \pg{97--114}.

\bibitem[Davidson \& Ranjan(2018)]{davidson2018spatial}
{\sc \au{Davidson, P.~A.} \& \au{Ranjan, A.}} \yr{2018}  \at{On the spatial
  segregation of helicity by inertial waves in dynamo simulations and planetary
  cores}.  \jt{J. Fluid Mech.}  \bvol{851},  \pg{268--287}.

\bibitem[Davidson {\em et~al.\/}(2006)Davidson, Staplehurst \&
  Dalziel]{06davidstaple}
{\sc \au{Davidson, P.~A.}, \au{Staplehurst, P.~J.} \& \au{Dalziel, S.~B.}}
  \yr{2006}  \at{On the evolution of eddies in a rapidly rotating system}.
  \jt{J. Fluid Mech.}  \bvol{557},  \pg{135--144}.

\bibitem[Dormy(2016)]{16dormy}
{\sc \au{Dormy, E.}} \yr{2016}  \at{Strong-field spherical dynamos}.  \jt{J.
  Fluid Mech.}  \bvol{789},  \pg{500--513}.

\bibitem[Glatzmaier(2013)]{glatz2013}
{\sc \au{Glatzmaier, G.~A.}} \yr{2013} {\em Introduction to Modeling Convection
  in Planets and Stars: Magnetic Field, Density Stratification, Rotation\/}.
  \publ{Princeton University Press}.

\bibitem[Glatzmaier {\em et~al.\/}(1995)Glatzmaier, Coe, Hongre \&
  Roberts]{glatz1999}
{\sc \au{Glatzmaier, G.~A.}, \au{Coe, R.~S.}, \au{Hongre, L.} \& \au{Roberts,
  P.~H.}} \yr{1995}  \at{The role of the {E}arth’s mantle in controlling the
  frequency of geomagnetic reversals}.  \jt{Nature}  \bvol{401},
  \pg{885--890}.

\bibitem[Glatzmaier \& Roberts(1995{\natexlab{{\em a\/}}})]{glatz1995a}
{\sc \au{Glatzmaier, G.~A.} \& \au{Roberts, P.~H.}} \yr{1995{\natexlab{{\em
  a\/}}}}  \at{A three-dimensional convective dynamo solution with rotating and
  finitely conducting inner core and mantle}.  \jt{Phys. Earth Planet. Inter.}
  \bvol{91}~(1-3),  \pg{63--75}.

\bibitem[Glatzmaier \& Roberts(1995{\natexlab{{\em b\/}}})]{glatz1995b}
{\sc \au{Glatzmaier, G.~A.} \& \au{Roberts, P.~H.}} \yr{1995{\natexlab{{\em
  b\/}}}}  \at{A three-dimensional self-consistent computer simulation of a
  geomagnetic field reversal}.  \jt{Nature}  \bvol{377},  \pg{203--209}.

\bibitem[Gubbins(1999)]{gubbins1999}
{\sc \au{Gubbins, D.}} \yr{1999}  \at{The distinction between geomagnetic
  excursions and reversals}.  \jt{Geophys. J. Int.}  \bvol{137}~(1),
  \pg{F1--F3}.

\bibitem[Kutzner \& Christensen(2002)]{kutzner2002}
{\sc \au{Kutzner, C.} \& \au{Christensen, U.~R.}} \yr{2002}  \at{From stable
  dipolar towards reversing numerical dynamos}.  \jt{Phys. Earth Planet.
  Inter.}  \bvol{131}~(1),  \pg{29--45}.

\bibitem[Loper {\em et~al.\/}(2003)Loper, Chulliat \& Shimizu]{loper2003}
{\sc \au{Loper, D.~E.}, \au{Chulliat, A.} \& \au{Shimizu, H.}} \yr{2003}
  \at{Buoyancy-driven perturbations in a rapidly rotating, electrically
  conducting fluid: Part {I} -- {F}low and magnetic field}.  \jt{Geophys.
  Astrophys. Fluid Dyn.}  \bvol{97},  \pg{429--469}.

\bibitem[McDermott \& Davidson(2019)]{mcdermott2019}
{\sc \au{McDermott, B.~R.} \& \au{Davidson, P.~A.}} \yr{2019}  \at{A physical
  conjecture for the dipolar--multipolar dynamo transition}.  \jt{J. Fluid
  Mech.}  \bvol{874},  \pg{995--1020}.

\bibitem[Merrill(2011)]{merrill2011}
{\sc \au{Merrill, R.~T.}} \yr{2011} {\em Our {M}agnetic {E}arth: {T}he
  {S}cience of {G}eomagnetism\/}.  \publ{University of Chicago Press}.

\bibitem[Moffatt(1978)]{moffatt1978}
{\sc \au{Moffatt, H.~K.}} \yr{1978} {\em Magnetic Field Generation in
  Electrically Conducting Fluids\/}.  \publ{Cambridge University Press}.

\bibitem[Monchaux {\em et~al.\/}(2009)Monchaux, Berhanu, Auma{\^\i}tre,
  Chiffaudel, Daviaud, Dubrulle, Ravelet, Fauve, Mordant, P{\'e}tr{\'e}lis {\em
  et~al.\/}]{monchaux2009}
{\sc \au{Monchaux, R.}, \au{Berhanu, M.}, \au{Auma{\^\i}tre, S.},
  \au{Chiffaudel, A.}, \au{Daviaud, F.}, \au{Dubrulle, B.}, \au{Ravelet, F.},
  \au{Fauve, S.}, \au{Mordant, N.}, \au{P{\'e}tr{\'e}lis, F.} \& \au{others}}
  \yr{2009}  \at{The von {K}{\'a}rm{\'a}n {So}dium experiment: {T}urbulent
  dynamical dynamos}.  \jt{Phys. Fluids}  \bvol{21}~(3),  \pg{035108}.

\bibitem[Mound {\em et~al.\/}(2019)Mound, Davies, Rost \& Aurnou]{mound2019}
{\sc \au{Mound, J.}, \au{Davies, C.}, \au{Rost, S.} \& \au{Aurnou, J.}}
  \yr{2019}  \at{Regional stratification at the top of {E}arth's core due to
  core--mantle boundary heat flux variations}.  \jt{Nat. Geosci.}
  \bvol{12}~(7),  \pg{575--580}.

\bibitem[Nicolas \& Buffett(2023)]{nicolas2023}
{\sc \au{Nicolas, Q.} \& \au{Buffett, B.}} \yr{2023}  \at{Excitation of
  high-latitude {MAC} waves in {E}arth’s core}.  \jt{Geophys. J. Int.}
  \bvol{233}~(3),  \pg{1961--1973}.

\bibitem[Olson {\em et~al.\/}(1999)Olson, Christensen \&
  Glatzmaier]{olson1999numerical}
{\sc \au{Olson, P.}, \au{Christensen, U.} \& \au{Glatzmaier, G.~A}} \yr{1999}
  \at{Numerical modeling of the geodynamo: mechanisms of field generation and
  equilibration}.  \jt{J. Geophs. Res. Solid Earth}  \bvol{104}~(B5),
  \pg{10383--10404}.

\bibitem[Olson \& Christensen(2006)]{olson2006dipole}
{\sc \au{Olson, P.} \& \au{Christensen, U.~R}} \yr{2006}  \at{Dipole moment
  scaling for convection-driven planetary dynamos}.  \jt{Earth Planet. Sci.
  Lett.}  \bvol{250}~(3-4),  \pg{561--571}.

\bibitem[Olson {\em et~al.\/}(2015)Olson, Deguen, Rudolph \& Zhong]{olson2015}
{\sc \au{Olson, P.}, \au{Deguen, R.}, \au{Rudolph, M.L.} \& \au{Zhong, S.}}
  \yr{2015}  \at{Core evolution driven by mantle global circulation}.
  \jt{Phys. Earth Planet. Inter.}  \bvol{243},  \pg{44--55}.

\bibitem[Olson {\em et~al.\/}(2009)Olson, Driscoll \& Amit]{olsdris2009}
{\sc \au{Olson, P.}, \au{Driscoll, P.} \& \au{Amit, H.}} \yr{2009}  \at{Dipole
  collapse and rever- sal precursors in a numerical dynamo}.  \jt{Phys. Earth
  Planet. Inter.}  \bvol{173},  \pg{121--140}.

\bibitem[Petitdemange(2018)]{petitdemange2018}
{\sc \au{Petitdemange, L.}} \yr{2018}  \at{Systematic parameter study of dynamo
  bifurcations in geodynamo simulations}.  \jt{Phys. Earth Planet. Inter.}
  \bvol{277},  \pg{113--132}.

\bibitem[Ranjan {\em et~al.\/}(2020)Ranjan, Davidson, Christensen \&
  Wicht]{ranjan2020etal}
{\sc \au{Ranjan, A}, \au{Davidson, P.~A.}, \au{Christensen, U.~R.} \&
  \au{Wicht, J.}} \yr{2020}  \at{On the generation and segregation of helicity
  in geodynamo simulations}.  \jt{Geophys. J. Int.}  \bvol{221}~(2),
  \pg{741--757}.

\bibitem[Sarson \& Jones(1999)]{sarson1999}
{\sc \au{Sarson, G.~R.} \& \au{Jones, C.~A.}} \yr{1999}  \at{A convection
  driven geodynamo reversal model}.  \jt{Phys. Earth Planet. Inter.}
  \bvol{111}~(1-2),  \pg{3--20}.

\bibitem[Schaeffer {\em et~al.\/}(2017)Schaeffer, Jault, Nataf \&
  Fournier]{schaeff2017}
{\sc \au{Schaeffer, N.}, \au{Jault, D.}, \au{Nataf, H-C.} \& \au{Fournier, A.}}
  \yr{2017}  \at{Turbulent geodynamo simulations: a leap towards {E}arth's
  core}.  \jt{Geophys. J. Int.}  \bvol{211},  \pg{1--29}.

\bibitem[Schwaiger {\em et~al.\/}(2019)Schwaiger, Gastine \&
  Aubert]{schwaiger2019}
{\sc \au{Schwaiger, T.}, \au{Gastine, T.} \& \au{Aubert, J.}} \yr{2019}
  \at{Force balance in numerical geodynamo simulations: a systematic study}.
  \jt{Geophys. J. Int.}  \bvol{219}~(Supplement\_1),  \pg{S101--S114}.

\bibitem[Soderlund {\em et~al.\/}(2012)Soderlund, King \&
  Aurnou]{soderlund2012}
{\sc \au{Soderlund, K.~M.}, \au{King, E.~M.} \& \au{Aurnou, J.~M.}} \yr{2012}
  \at{The influence of magnetic fields in planetary dynamo models}.  \jt{Earth
  Planet. Sci. Lett.}  \bvol{333},  \pg{9--20}.

\bibitem[Sreenivasan \& Jones(2006)]{sreenivasan2006}
{\sc \au{Sreenivasan, B} \& \au{Jones, C.~A}} \yr{2006}  \at{The role of
  inertia in the evolution of spherical dynamos}.  \jt{Geophys. J. Int.}
  \bvol{164}~(2),  \pg{467--476}.

\bibitem[Sreenivasan \& Jones(2011)]{jfm11}
{\sc \au{Sreenivasan, B.} \& \au{Jones, C.~A.}} \yr{2011}  \at{Helicity
  generation and subcritical behaviour in rapidly rotating dynamos}.  \jt{J.
  Fluid Mech.}  \bvol{688},  \pg{5}.

\bibitem[Sreenivasan \& Kar(2018)]{prf18}
{\sc \au{Sreenivasan, B.} \& \au{Kar, S.}} \yr{2018}  \at{Scale dependence of
  kinetic helicity and selection of the axial dipole in rapidly rotating
  dynamos}.  \jt{Phys. Rev. Fluids}  \bvol{3}~(9),  \pg{093801}.

\bibitem[Sreenivasan \& Maurya(2021)]{jfm21}
{\sc \au{Sreenivasan, B.} \& \au{Maurya, G.}} \yr{2021}  \at{Evolution of
  forced magnetohydrodynamic waves in a stratified fluid}.  \jt{J. Fluid Mech.}
   \bvol{922}.

\bibitem[Sreenivasan \& Narasimhan(2017)]{jfm17b}
{\sc \au{Sreenivasan, B.} \& \au{Narasimhan, G.}} \yr{2017}  \at{Damping of
  magnetohydrodynamic waves in a rotating fluid}.  \jt{J. Fluid Mech.}
  \bvol{828},  \pg{867--905}.

\bibitem[Sreenivasan {\em et~al.\/}(2014)Sreenivasan, Sahoo \&
  Dhama]{sreeni2014}
{\sc \au{Sreenivasan, B.}, \au{Sahoo, S.} \& \au{Dhama, G.}} \yr{2014}  \at{The
  role of buoyancy in polarity reversals of the geodynamo}.  \jt{Geophys. J.
  Int.}  \bvol{199}~(3),  \pg{1698--1708}.

\bibitem[Starchenko \& Jones(2002)]{starjones2002}
{\sc \au{Starchenko, S.} \& \au{Jones, C.~A.}} \yr{2002}  \at{Typical
  velocities and magnetic fields in planetary interiors}.  \jt{Icarus}
  \bvol{157},  \pg{426--435}.

\bibitem[Tassin {\em et~al.\/}(2021)Tassin, Gastine \& Fournier]{tassin2021}
{\sc \au{Tassin, T.}, \au{Gastine, T.} \& \au{Fournier, A.}} \yr{2021}
  \at{Geomagnetic semblance and dipolar--multipolar transition in top-heavy
  double-diffusive geodynamo models}.  \jt{Geophys. J. Int.}  \bvol{226}~(3),
  \pg{1897--1919}.

\bibitem[Teed {\em et~al.\/}(2015)Teed, Jones \& Tobias]{teed2015}
{\sc \au{Teed, R.~J.}, \au{Jones, C.~A.} \& \au{Tobias, S.~M.}} \yr{2015}
  \at{The transition to {E}arth-like torsional oscillations in
  magnetoconvection simulations}.  \jt{Earth Planet. Sci. Lett.}  \bvol{419},
  \pg{22--31}.

\bibitem[Valet {\em et~al.\/}(2005)Valet, Meynadier \& Guyodo]{valet2005}
{\sc \au{Valet, J-P.}, \au{Meynadier, L.} \& \au{Guyodo, Y.}} \yr{2005}
  \at{Geomagnetic dipole strength and reversal rate over the past two million
  years}.  \jt{Nature}  \bvol{435}~(7043),  \pg{802--805}.

\bibitem[Varma \& Sreenivasan(2022)]{aditya2022}
{\sc \au{Varma, A.} \& \au{Sreenivasan, B.}} \yr{2022}  \at{The role of slow
  magnetostrophic waves in the formation of the axial dipole in planetary
  dynamos}.  \jt{Phys. Earth Planet. Inter.}  \bvol{333},  \pg{106944}.

\bibitem[Wicht \& Olson(2004)]{wicht2004}
{\sc \au{Wicht, J.} \& \au{Olson, P.}} \yr{2004}  \at{A detailed study of the
  polarity reversal mechanism in a numerical dynamo model}.  \jt{Geochem.
  Geophys. Geosyst.}  \bvol{5}~(3).

\bibitem[Willis {\em et~al.\/}(2007)Willis, Sreenivasan \& Gubbins]{07willis}
{\sc \au{Willis, A.~P.}, \au{Sreenivasan, B.} \& \au{Gubbins, D.}} \yr{2007}
  \at{Thermal core-mantle interaction: {E}xploring regimes for \lq locked'
  dynamo action}.  \jt{Phys. Earth Planet. Inter.}  \bvol{165},  \pg{83--92}.

\bibitem[Zaire {\em et~al.\/}(2022)Zaire, Jouve, Gastine, Donati, Morin, Landin
  \& Folsom]{zaire2022}
{\sc \au{Zaire, B.}, \au{Jouve, L.}, \au{Gastine, T.}, \au{Donati, J.~F.},
  \au{Morin, J.}, \au{Landin, N.} \& \au{Folsom, C.~P.}} \yr{2022}
  \at{Transition from multipolar to dipolar dynamos in stratified systems}.
  \jt{Mon. Not. R. Astron. Soc.}  \bvol{517}~(3),  \pg{3392--3406}.

\end{thebibliography}
\bibliographystyle{jfm}

\clearpage
\newpage
\setcounter{figure}{0}
\renewcommand{\figurename}{Figure}
\renewcommand{\thefigure}{S\arabic{figure}}
\begin{figure}
	\centering
	\section*{SUPPLEMENTARY MATERIAL}
	\includegraphics[width=0.6\linewidth]{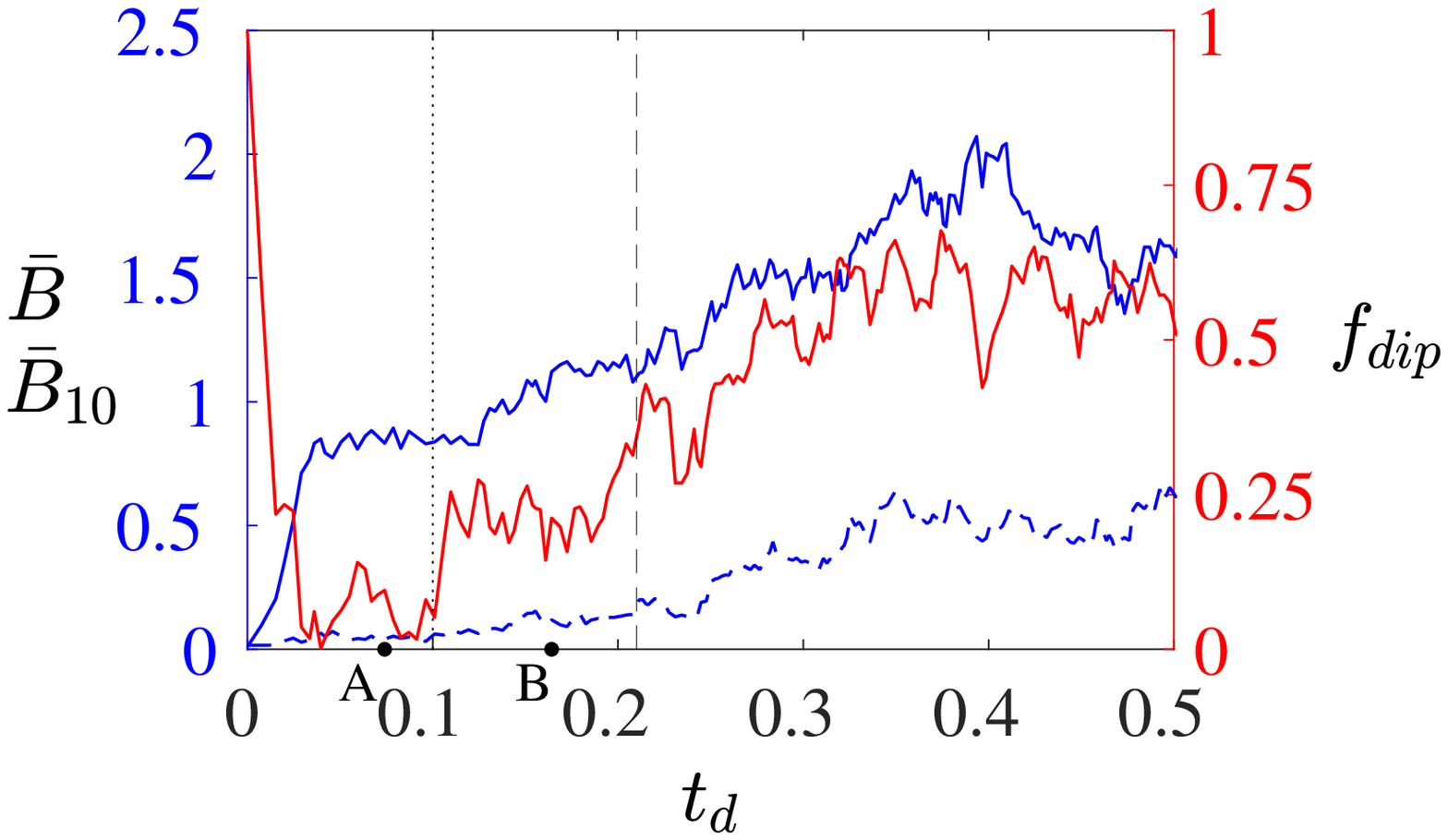}
	\caption{
		The figure shows the evolution in time (measured in units
		of the magnetic diffusion time) of 
		the root mean square value of 
		the magnetic field intensity, $\bar{B}$ (solid blue line),
		the axial dipole part of the field, $\bar{B}_{10}$
		(dashed blue line), and the relative
		dipole field strength at the outer boundary, $f_{dip}$
		(red line). The run starts from a 
		dipole  seed magnetic field. The vertical 
		dotted line marks the emergence of 
		the slow MAC waves, which happens when
		$|\omega_M| \approx |\omega_A|$ \citep{aditya2022}; the dashed line 
		denotes $f_{dip}=0.35$, which is suggested as the lower
		bound for the existence of
		dipole-dominated numerical dynamos \citep{chraub2006}. 
		Point `A' ($f_{dip}=0.095$) is a multipolar
		state before the 
		appearance of the slow MAC waves and 
		point `B' ($f_{dip}=0.212$) is a multipolar
		state where slow waves are present. 
		The dynamo parameters are 
		$E=6 \times 10^{-5}$, $Pm=Pr=5$ and $Ra=18000$.}
	\label{fdip}
\end{figure}
\begin{figure}
	\centering
	\hspace{-2.25 in}(a)  \hspace{2.25 in} (b)  \\
	\includegraphics[width=0.33\linewidth]{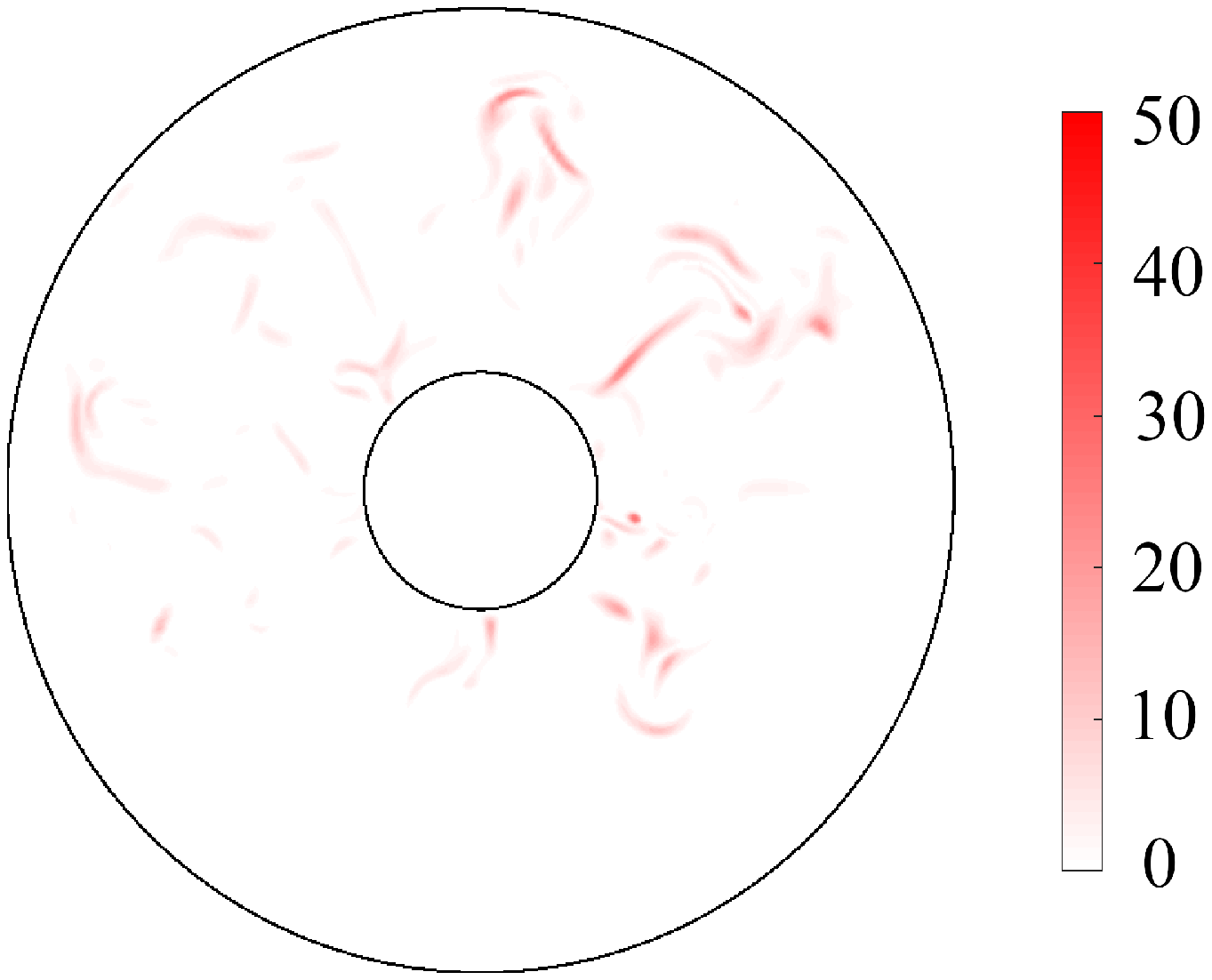}
	\includegraphics[width=0.33\linewidth]{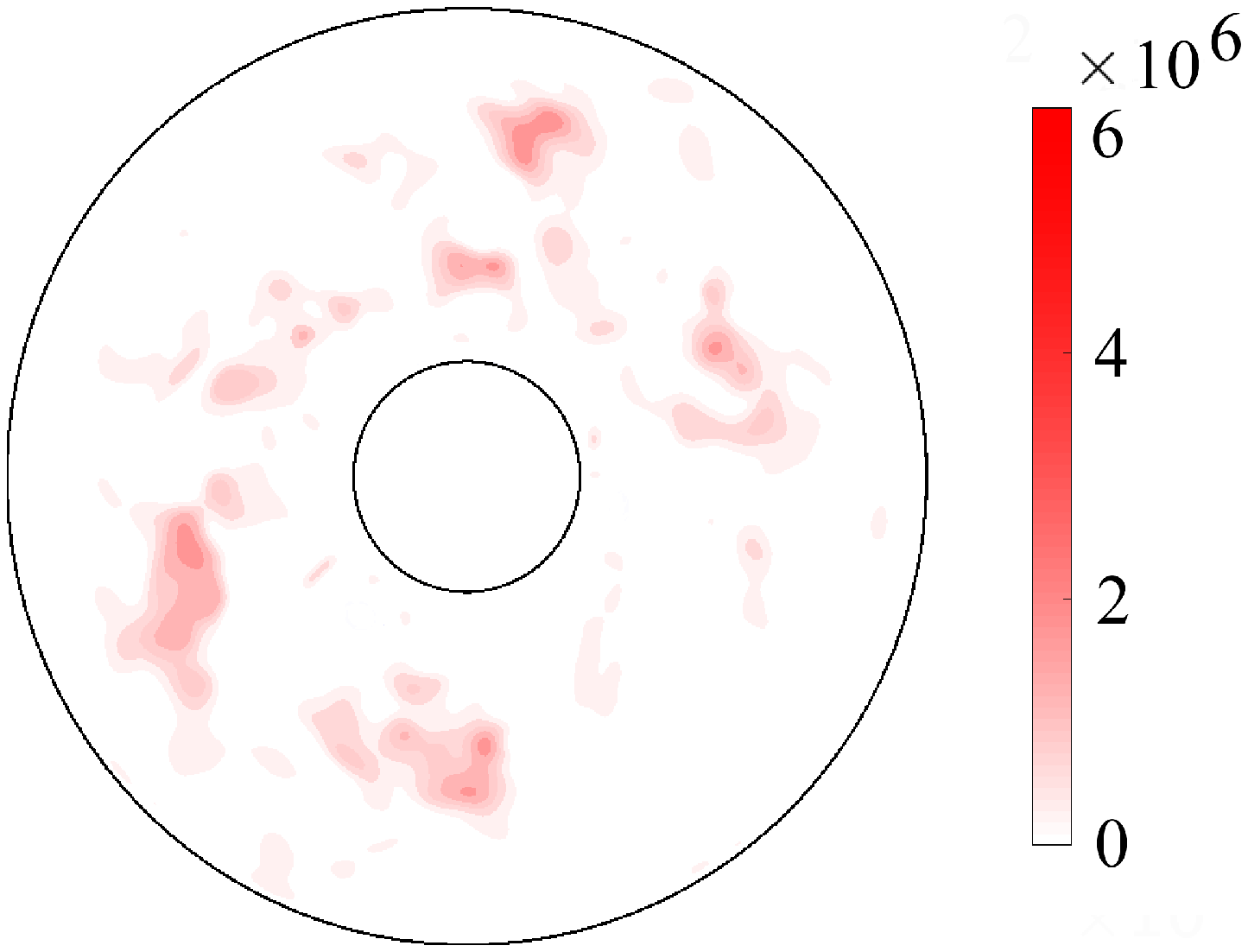}\\
	\hspace{-2.25 in}(c)  \hspace{2.25 in} (d)  \\
	\includegraphics[width=0.33\linewidth]{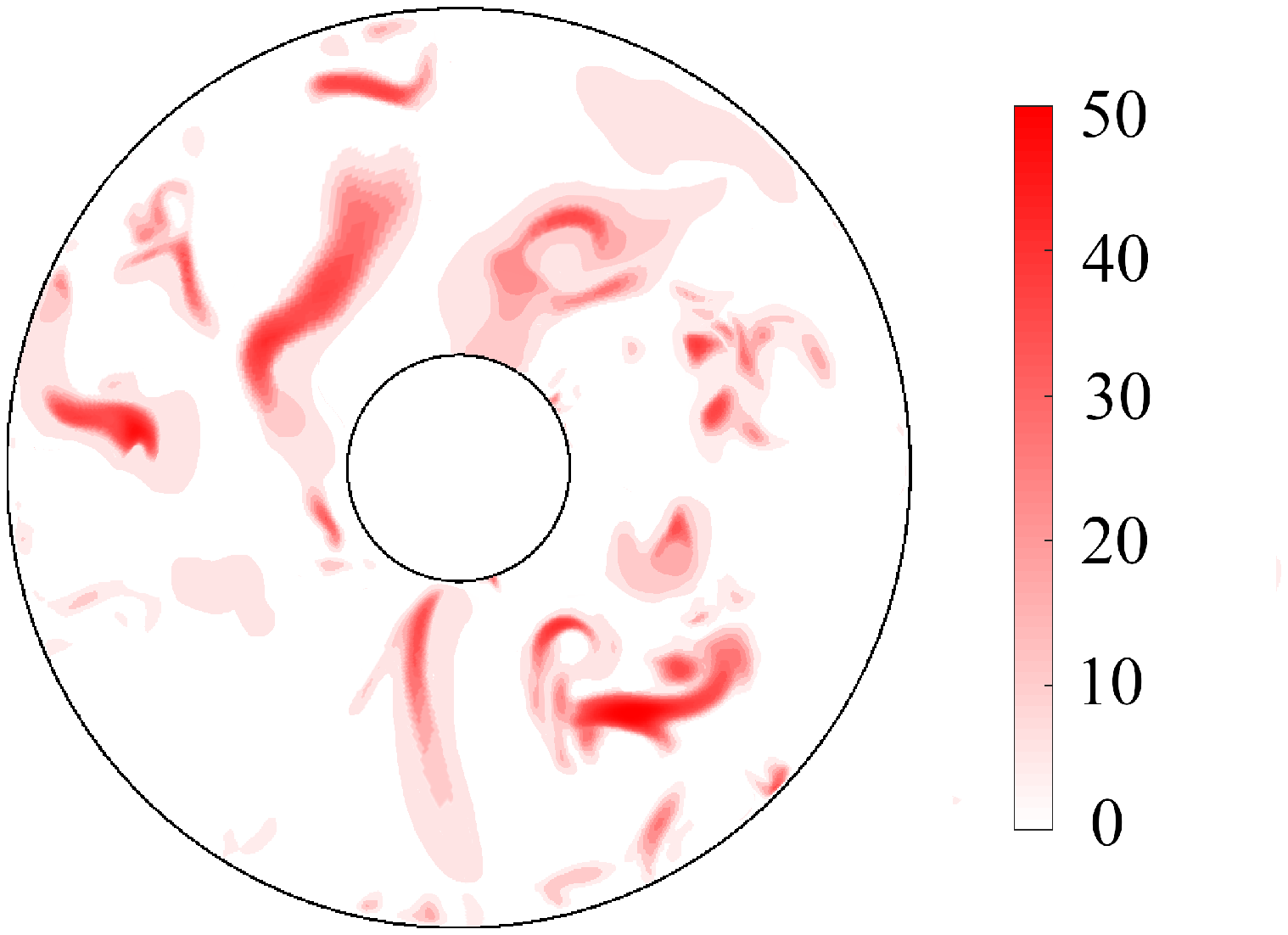}
	\includegraphics[width=0.33\linewidth]{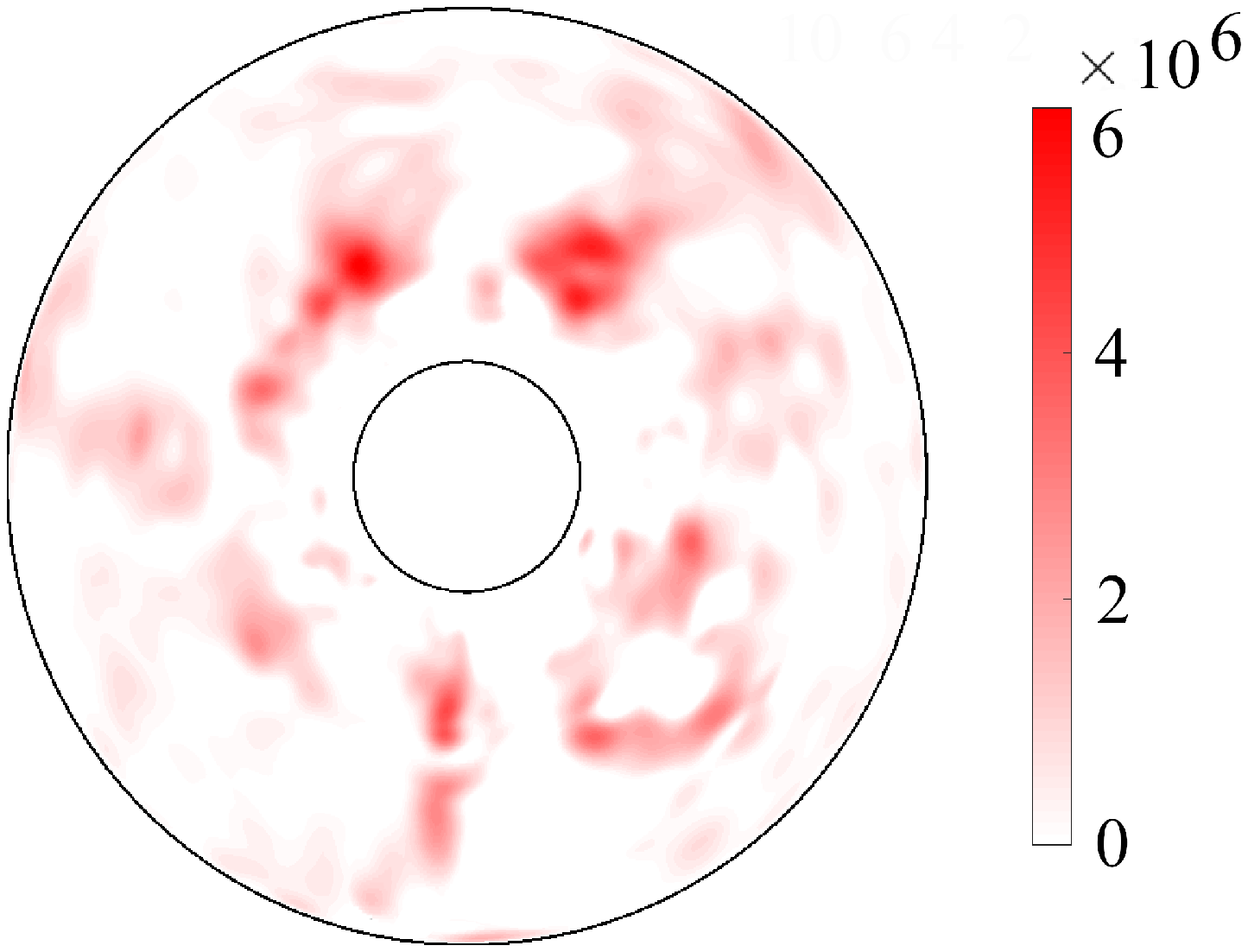}\\
	\caption{Horizontal section plots at a 
		distance $z=0.4$ below the equator of 
		the square of the total magnetic
		field, $B^2$ (left panels) and the 
		kinetic helicity $H$ (right panels) 
		for the energy-containing range
		of spherical harmonic degrees 
		$l\le 25$. Figures (a) and (b) 
		represent point `A' in figure \ref{fdip} 
		while figures (c) and (d) represent point `B' 
		in figure \ref{fdip}.
		The minimum and maximum values of the contours in
		(a), (b), (c) and (d) are $[0, 17]$, 
		$[0, 2.6 \times 10^6]$, $[0, 46]$ and $[0, 5.8 \times
		10^6]$ respectively. 
		The dynamo parameters 
		are $E=6 \times 10^{-5}$, $Pm=Pr=5$ and $Ra=18000$.  
	}
	\label{bhplots}
\end{figure}

\end{document}